\documentclass[11pt,onecolumn,twoside,a4paper,openany]{article}
\usepackage[dvips]{graphicx}
\usepackage{bbm}
\usepackage{amsfonts,amsmath,amssymb,graphics,psfrag}
\usepackage[arrow,matrix]{xy}
\makeatletter
\@addtoreset{equation}{section}
\makeatother

\newcommand{\lp}{\ell_p}

\newcommand{\sst}{\scriptscriptstyle}
\newcommand{\st}{\scriptstyle}
\newcommand{\op}[1]{\widehat{#1}}
\newcommand{\wt}[1]{\widetilde{#1}}
\newcommand{\lb}{\left}
\newcommand{\rb}{\right}
\newcommand{\half}{\frac{1}{2}}
\newcommand{\ti}{\tilde}

\newcommand{\bx}{{\sst \Box}}
\newcommand{\vt}{\tilde{v}}

\newcommand{\mn}{m_{\sst 0}}
\newcommand{\mnt}{\wt{m}_{\sst 0}}
\newcommand{\nn}{n_{\sst 0}}
\newcommand{\nnt}{\wt{n}_{\sst 0}}
\newcommand{\elln}{\ell_{\sst 0}}

\newcommand{\Jn}{J_{\sst 0}}
\newcommand{\Jnt}{\wt{J}_{\sst 0}}
\newcommand{\Kn}{K_{\sst 0}}
\newcommand{\Knt}{\wt{K}_{\sst 0}}
\newcommand{\In}{I_{\sst 0}}
\newcommand{\Int}{\wt{I}_{\sst 0}}
\newcommand{\sn}{\sigma_{\sst 0}}
\newcommand{\snt}{\wt{\sigma}_{\sst 0}}

\newcommand{\e}[3]{e^{#1}_{#2}(#3)}
\newcommand{\n}[4]{n_{#1 #2 #3 #4}}
\newcommand{\hol}[4]{\op{h}_{#1 #2 #3 #4}}
\newcommand{\holin}[4]{\op{h}^{-1}_{#1 #2 #3 #4}}
\newcommand{\chol}[4]{h_{#1 #2 #3 #4}}
\newcommand{\cholin}[4]{h^{-1}_{#1 #2 #3 #4}}
\newcommand{\holloop}[5]{\op{h}_{\beta_{#1 #2 #3 #4 #5}}}
\newcommand{\holloopin}[5]{\op{h}^{-1}_{\beta_{ #1 #2 #3 #4 #5}}}
\newcommand{\cholloop}[5]{h_{\beta_{#1 #2 #3 #4 #5}}}
\newcommand{\cholloopin}[5]{h^{-1}_{\beta_{#1 #2 #3 #4 #5}}}

\newcommand{\Psigen}{\Psi^{t}_{\{g,J,\sigma,j,L\}}}
\newcommand{\Psigenedge}{\Psi^{t}_{g J\sigma j\tilde{v}}}

\newcommand{\Psigeng}{\Psi^{t}_{\alpha,m}}

\newcommand{\hgen}{\hat{h}_{J \sigma j v}}
\newcommand{\hgencl}{h_{J \sigma j v}}
\newcommand{\pgen}{p_{J \sigma j\vt}}

\newcommand{\pgenv}{\hat{p}_{J \sigma j v}}
\newcommand{\ppgen}{(p_{J j v})^+}
\newcommand{\pmgen}{(p_{J j v})^-}
\newcommand{\pmm}{(p)^-}
\newcommand{\ggen}{g_{J \sigma j v}}
\newcommand{\ngen}{n_{J \sigma j \tilde{v}}}
\newcommand{\ngent}{\wt{n}_{J \sigma j \tilde{v}}}
\newcommand{\ngenv}{n_{J \sigma j v}}
\newcommand{\ngentv}{\wt{n}_{J \sigma j v}}
\newcommand{\npgen}{(n_{J j v})^+}
\newcommand{\nmgen}{(n_{J j v})^-}

\newcommand{\nmm}{(n)^-}
\newcommand{\xgen}{x_{J \sigma j\tilde{v}}}
\newcommand{\xgenv}{x_{J \sigma j v}}
\newcommand{\xpgen}{(x_{J j v})^+}
\newcommand{\xmgen}{(x_{J j v})^-}
\newcommand{\xmgenn}{(\tilde{x}_{J j v})^-}
\newcommand{\xmm}{(\tilde{x})^-}
\newcommand{\qmm}{(q^{-1})^-}
\newcommand{\qmgen}{(q_{J j v})^-}
\newcommand{\kgen}{k_{J \sigma j \vt}}
\newcommand{\kgenv}{k_{J \sigma j v}}
\newcommand{\phigen}{\varphi_{J \sigma j \tilde{v}}}
\newcommand{\phigenv}{\varphi_{J \sigma j v}}

\newcommand{\phipgen}{(\varphi_{Jjv})^+}
\newcommand{\phimgen}{(\varphi_{Jjv})^-}
\newcommand{\tsgen}{\vartheta_{J \sigma j \tilde{v}}}
\newcommand{\tsgenv}{\vartheta_{J \sigma j v}}
\newcommand{\tgen}{t_{J \sigma j \tilde{v}}}
\newcommand{\kdel}[8]{\delta_{\sst (#1,#2, #3,#4),(#5,#6,#7,#8)}}
\newcommand{\Bdel}{\Delta(\In,\Jn,\sn,\mn,v,J,\sigma,j,\vt)}
\newcommand{\Bdelt}{\Delta(\Int,\Jnt,\snt,\mnt,v,J,\sigma,j,\vt)}
\newcommand{\Bdelv}{\Delta(\In,\Jn,\sn,\mn,v,J,\sigma,j,v)}
\newcommand{\Bdeltv}{\Delta(\In,\Jn,\sn,\mn,v,J,\sigma,j,v)}
\newcommand{\Bdelvm}{(\Delta)^{-}(\In,\Jn,v,J,j,v)}
\newcommand{\Bdelvp}{(\Delta)^{+}(\In,\Jn,v,J,j,v)}
\newcommand{\Bdelvtm}{(\Delta)^{-}(\Int,\Jnt,v,J,j,v)}
\newcommand{\Bdelvtp}{(\Delta)^{+}(\Int,\Jnt,v,J,j,v)}
\newcommand{\Del}{\Delta(\In,\Int,\Jn,\Jnt,\sn,\snt,\mn,\mnt,v,J,\sigma,j,\vt)}
\newcommand{\Delv}{\Delta(\In,\Int,\Jn,\Jnt,\sn,\snt,\mn,\mnt,v,J,\sigma,j,v)}

\newcommand{\Dtm}{(\Delta)^-}
\newcommand{\Dtmt}{(\wt{\Delta})^-}

\newcommand{\dpgen}{(\delta)^+_{\sst (J,j,v),(\Kn,\nn,v)}}
\newcommand{\dpgent}{(\delta)^+_{\sst (J,j,v),(\Knt,\nnt,v)}}
\newcommand{\dmgen}{(\delta)^-_{\sst (J,j,v),(\Kn,\nn,v)}}
\newcommand{\dmgent}{(\delta)^-_{\sst (J,j,v),(\Knt,\nnt,v)}}
\newcommand{\Ygen}{\op{X}_{J\sigma j v}}
\newcommand{\Y}[4]{\op{X}_{#1 #2 #3 #4}}

\newcommand{\MeO}{\frac{\langle\Omnt\Psigen\,\Big|\,\Omn\Psigen\rangle}{||\Psigen||^2}}
\newcommand{\Omn}{\op{O}^{\mn,\nn}_{\In\Jn\Kn\sn v}}
\newcommand{\Omnt}{\op{O}^{\mnt,\nnt}_{\Int\Jnt\Knt\snt v}}
\def\be{\begin{equation}}
\def\ee{\end{equation}}
\def\ba{\begin{eqnarray}}
\def\ea{\end{eqnarray}}
\def\Nl{{\mathchoice
{\setbox0=\hbox{$\displaystyle\rm N$}\hbox{\hbox to0pt
{\kern0.4\wd0\vrule height0.9\ht0\hss}\box0}}
{\setbox0=\hbox{$\textstyle\rm N$}\hbox{\hbox to0pt
{\kern0.4\wd0\vrule height0.9\ht0\hss}\box0}}
{\setbox0=\hbox{$\scriptstyle\rm N$}\hbox{\hbox to0pt
{\kern0.4\wd0\vrule height0.9\ht0\hss}\box0}}
{\setbox0=\hbox{$\scriptscriptstyle\rm N$}\hbox{\hbox to0pt
{\kern0.4\wd0\vrule height0.9\ht0\hss}\box0}}}}
\def\Zl{{\mathchoice
{\setbox0=\hbox{$\displaystyle\rm Z$}\hbox{\hbox to0pt
{\kern0.4\wd0\vrule height0.9\ht0\hss}\box0}}
{\setbox0=\hbox{$\textstyle\rm Z$}\hbox{\hbox to0pt
{\kern0.4\wd0\vrule height0.9\ht0\hss}\box0}}
{\setbox0=\hbox{$\scriptstyle\rm Z$}\hbox{\hbox to0pt
{\kern0.4\wd0\vrule height0.9\ht0\hss}\box0}}
{\setbox0=\hbox{$\scriptscriptstyle\rm Z$}\hbox{\hbox to0pt
{\kern0.4\wd0\vrule height0.9\ht0\hss}\box0}}}}
\def\Ql{{\mathchoice
{\setbox0=\hbox{$\displaystyle\rm Q$}\hbox{\hbox to0pt
{\kern0.4\wd0\vrule height0.9\ht0\hss}\box0}}
{\setbox0=\hbox{$\textstyle\rm Q$}\hbox{\hbox to0pt
{\kern0.4\wd0\vrule height0.9\ht0\hss}\box0}}
{\setbox0=\hbox{$\scriptstyle\rm Q$}\hbox{\hbox to0pt
{\kern0.4\wd0\vrule height0.9\ht0\hss}\box0}}
{\setbox0=\hbox{$\scriptscriptstyle\rm Q$}\hbox{\hbox to0pt
{\kern0.4\wd0\vrule height0.9\ht0\hss}\box0}}}}
\def\Rl{{\mathchoice
{\setbox0=\hbox{$\displaystyle\rm R$}\hbox{\hbox to0pt
{\kern0.4\wd0\vrule height0.9\ht0\hss}\box0}}
{\setbox0=\hbox{$\textstyle\rm R$}\hbox{\hbox to0pt
{\kern0.4\wd0\vrule height0.9\ht0\hss}\box0}}
{\setbox0=\hbox{$\scriptstyle\rm R$}\hbox{\hbox to0pt
{\kern0.4\wd0\vrule height0.9\ht0\hss}\box0}}
{\setbox0=\hbox{$\scriptscriptstyle\rm R$}\hbox{\hbox to0pt
{\kern0.4\wd0\vrule height0.9\ht0\hss}\box0}}}}
\def\Cl{{\mathchoice
{\setbox0=\hbox{$\displaystyle\rm C$}\hbox{\hbox to0pt
{\kern0.4\wd0\vrule height0.9\ht0\hss}\box0}}
{\setbox0=\hbox{$\textstyle\rm C$}\hbox{\hbox to0pt
{\kern0.4\wd0\vrule height0.9\ht0\hss}\box0}}
{\setbox0=\hbox{$\scriptstyle\rm C$}\hbox{\hbox to0pt
{\kern0.4\wd0\vrule height0.9\ht0\hss}\box0}}
{\setbox0=\hbox{$\scriptscriptstyle\rm C$}\hbox{\hbox to0pt
{\kern0.4\wd0\vrule height0.9\ht0\hss}\box0}}}}
\def\Co{{\mathchoice
{\setbox0=\hbox{$\displaystyle\rm C$}\hbox{\hbox to0pt
{\kern0.4\wd0\vrule height0.9\ht0\hss}\box0}}
{\setbox0=\hbox{$\textstyle\rm C$}\hbox{\hbox to0pt
{\kern0.4\wd0\vrule height0.9\ht0\hss}\box0}}
{\setbox0=\hbox{$\scriptstyle\rm C$}\hbox{\hbox to0pt
{\kern0.4\wd0\vrule height0.9\ht0\hss}\box0}}
{\setbox0=\hbox{$\scriptscriptstyle\rm C$}\hbox{\hbox to0pt
{\kern0.4\wd0\vrule height0.9\ht0\hss}\box0}}}}
\def\Hl{{\mathchoice
{\setbox0=\hbox{$\displaystyle\rm H$}\hbox{\hbox to0pt
{\kern0.4\wd0\vrule height0.9\ht0\hss}\box0}}
{\setbox0=\hbox{$\textstyle\rm H$}\hbox{\hbox to0pt
{\kern0.4\wd0\vrule height0.9\ht0\hss}\box0}}
{\setbox0=\hbox{$\scriptstyle\rm H$}\hbox{\hbox to0pt
{\kern0.4\wd0\vrule height0.9\ht0\hss}\box0}}
{\setbox0=\hbox{$\scriptscriptstyle\rm H$}\hbox{\hbox to0pt
{\kern0.4\wd0\vrule height0.9\ht0\hss}\box0}}}}
\def\Ol{{\mathchoice
{\setbox0=\hbox{$\displaystyle\rm O$}\hbox{\hbox to0pt
{\kern0.4\wd0\vrule height0.9\ht0\hss}\box0}}
{\setbox0=\hbox{$\textstyle\rm O$}\hbox{\hbox to0pt
{\kern0.4\wd0\vrule height0.9\ht0\hss}\box0}}
{\setbox0=\hbox{$\scriptstyle\rm O$}\hbox{\hbox to0pt
{\kern0.4\wd0\vrule height0.9\ht0\hss}\box0}}
{\setbox0=\hbox{$\scriptscriptstyle\rm O$}\hbox{\hbox to0pt
{\kern0.4\wd0\vrule height0.9\ht0\hss}\box0}}}}
\DeclareMathOperator{\MCO}{\boldsymbol{\widehat{\mathsf{M}}}}%
\DeclareMathOperator{\MC}{\boldsymbol{\mathsf{M}}}
\DeclareMathOperator{\sgn}{sgn}
\DeclareMathOperator{\tr}{Tr}
\title{\sf Algebraic Quantum Gravity (AQG)\\ II. Semiclassical Analysis}
\author{\sf
K. 
Giesel\thanks{\sf Kristina.Giesel@aei.mpg.de, kgiesel@perimeterinstitute.ca}
~~ and ~
T. 
Thiemann\thanks{\sf Thomas.Thiemann@aei.mpg.de, tthiemann@perimeterinstitute.ca}\\
\\
{\sf MPI f. Gravitationsphysik, Albert-Einstein-Institut,} \\
          {\sf Am M\"uhlenberg 1, 14476 Potsdam, Germany}\\
\\
{\sf and}\\
\\
{\sf Perimeter Institute for Theoretical Physics,}\\ 
{\sf 31 Caroline Street N, Waterloo, ON N2L 2Y5, Canada}}
\date{\sf {\small Preprint AEI-2006-059}}

\begin{document}
\maketitle
\begin{abstract}
{\sf In the previous article \cite{I} a new combinatorial and thus purely algebraical approach to quantum gravity, called Algebraic Quantum Gravity (AQG), was introduced. In the framework of AQG existing semiclassical tools can be applied to operators that encode the dynamics of AQG such as the Master constraint operator. In this article we will analyse the semiclassical limit of the (extended) algebraic Master constraint operator and show that it reproduces the correct infinitesimal generators of General Relativity. 
Therefore the question whether General Relativity is included in the semiclassical sector of the theory, which  is still an open problem in LQG, can be significantly improved in the framework of AQG.
For the calculations we will substitute $SU(2)$ by $U(1)^3$. That this substitution is justified will be demonstrated in the third article \cite{III} of this series}
\end{abstract}
\newpage
\section{Introduction}
In the previous companion paper \cite{I} of this series we introduced a new top down approach to quantum gravity, called Algebraic Quantum Gravity (AQG). This  combinatorial approach is very much inspired by the ideas and concepts of LQG \cite{books,reviews}. However, it departs in a crucial way from LQG by discarding the notion of embedded graphs and considering algebraic graphs instead. Since these graphs are algebraic, we lose information such as topology and the differential structure of the spatial manifold that are  fundamental for LQG. Nevertheless, we showed that all physical (gauge invariant) operators such as the Master constraint operator can be formulated in an algebraic (i.e. embedding independent) way and thus be lifted from LQG to AQG.
In this sense AQG offers a technical simpler approach since one just has to deal with one fundamental infinite algebraic graph, while within LQG one considers an infinite number of finite embedded graphs. The missing information in AQG about the topology and the differential structure of the spacetime manifold as well as the background metric to be approximated is encoded in the coherent states and thus only of interest in the semiclassical limit.
As pointed out in \cite{I} the analysis of the semiclassical limit of the dynamics of LQG could not been performed so far, because existing semiclassical tools fail to be applied  to graph-changing operators such as the Hamiltonian or the graph-changing version of the Master constraint operator\cite{QSD,M1,M2,Test}. The reason for the failure in the case of the Hamiltonian constraint operator is that in order to quantise this operators without anomalies, it has to be formulated in a graph-changing fashion. The action of an graph-changing operator on coherent states will necessarily add degrees of freedom to the coherent states under consideration. The fluctuation of these additional degrees of freedom are not well suppressed by the coherent states leading to an unacceptable semiclassical approximation of the Hamilton constraint operator. The graph-changing version of the Master constraint operator is spatially diffemorphism invariant. In \cite{ALMMT} it was shown that such operators have to be defined directly on the spatially diffeomorphism invariant Hilbert space. Hence, we would need spatially diffeomorphism invariant coherent states, that so far have not been defined in LQG.
\\ 
In contrast within the framework of AQG we work with the (extended) Master constraint operator, which is quantised in a graph-non-changing formulation. Therefore the dynamics will not change the degrees of freedom and thus existing semiclassical tools can be used to analyse the semiclassical behaviour of the AQG-dynamics. Furthermore, since we have only one fundamental or maximal graph in AQG, we are able to remove the graph-dependence that is present in the semiclassical tools of LQG.\\ \\
In this paper we will display the semiclassical analysis of the (extended) algebraic Master constraint operator associated with an algebraic graph of cubic symmetry and show that AQG reproduces the correct infinitesimal generators of General Relativity in the semiclassical limit. We will use the semiclassical tools developed in \cite{U1hoch3,Complexifier,STW}. Since, we are working on the algebraic level, the restriction to an algebraic graph of cubic symmetry incorporates all graphs of valence six or lower\footnote{The graphs with valence $(6-n)$ with $3\le n<6$ can be obtained by simply not exiting $n$ edges at each vertex of the algebraic graph.}. We will substitute $SU(2)$ by $U(1)^3$, because this will simplify the calculation enormously. That this substitution is satisfied was already shown in \cite{U1hoch3} where it was proven that the electric fluxes and holonomies for $SU(2)$ are well approximated in the semiclassical limit. Additionally, we will prove in our companion paper \cite{III} that the $U(1)^3$-substitution is also satisfied for operators such as the (extended) Master constraint operator.
Here we will only consider the gravitational sector. However, the techniques used here carry over to all standard matter coupling. Since the Gauss constraint consists of a linear combination of flux operators and for those the correct semiclassical limit has been already demonstrated in \cite{U1hoch3}, we neglect the Gauss constraint in our analysis. \\
Due to the fact that the relation $\{H_E^{(1)},V\}=\int_\sigma d^3x K_a^j E^a_j$ for $SU(2)$ on which eqn
(2.18) in \cite{I} relies fails to hold, we cannot approximate the Lorentzian part of the Hamiltonian by $U(1)^3$ correctly. Hence, we will only consider the Euclidean part here. However, the discussion in \cite{III} show that the correct $SU(2)$ calculation reproduces the correct semiclassical limit.
\\
\\
This article is organised as follows:\\
In section 2 we introduce the necessary technical tools in order to perform the semiclassical analysis. We discuss the notion of the infinite algebraic graph of cubic symmetry as well as explain in detail how the $U(1)$-approximation is performed. Afterwards we introduce the (extended) algebraic Master constraint operator in the $U(1)^3$-approximation. Here we follow the ideas of \cite{Hanno} and generalise them to our case. Note, that, in contrast to the Master constraint operator, the operators considered in \cite{Hanno} contained no loop operators.  We will display certain details of the calculation in section 2 in the appendix and just referring to them in the main text. We decided to present this calculation very detailed, because as pointed out in our companion paper \cite{III}, this is the first time semiclassical perturbation theory wihtin AQG allows to compute expectation values of dynamical operators.
In section 3 we discuss the leading order (LO) contrinution of the expectation value of the Master constraint operator.
In section 4 we analyse in detail how this LO-contributions is related to the classical master constraint.
In section 5 we demonstrate the next-to-leading order (NLO) term of the expectation value of the Master constraint operator.
In section 6 we discuss our result and finally conclude.

\section{The Master Constraint Operator for an Algebraic Infinite Graph of Cubic Topology and within the $U(1)^3$ Approximation}
\subsection{The Infinite Algebraic Graph of Cubic Topology}
\begin{figure}[hbt]
\begin{center}
\psfrag{e1p}{$e_1^{+}(v)$}
\psfrag{e1m}{$e_1^{-}(v)$}
\psfrag{e2p}{$e_2^{+}(v)$}
\psfrag{e2m}{$e_2^{-}(v)$}
\psfrag{e3p}{$e_3^{+}(v)$}
\psfrag{e3m}{$e_3^{-}(v)$}
\psfrag{Vp1}{$v+\hat{1}$}
\psfrag{Vm1}{$v-\hat{1}$}
\psfrag{Vp2}{$v+\hat{2}$}
\psfrag{Vm2}{$v-\hat{2}$}
\psfrag{Vp3}{$v+\hat{3}$}
\psfrag{Vm3}{$v-\hat{3}$}
\psfrag{v}{$v$}
\psfrag{cubicgraph}{cubic graph}
\includegraphics[width=12cm]{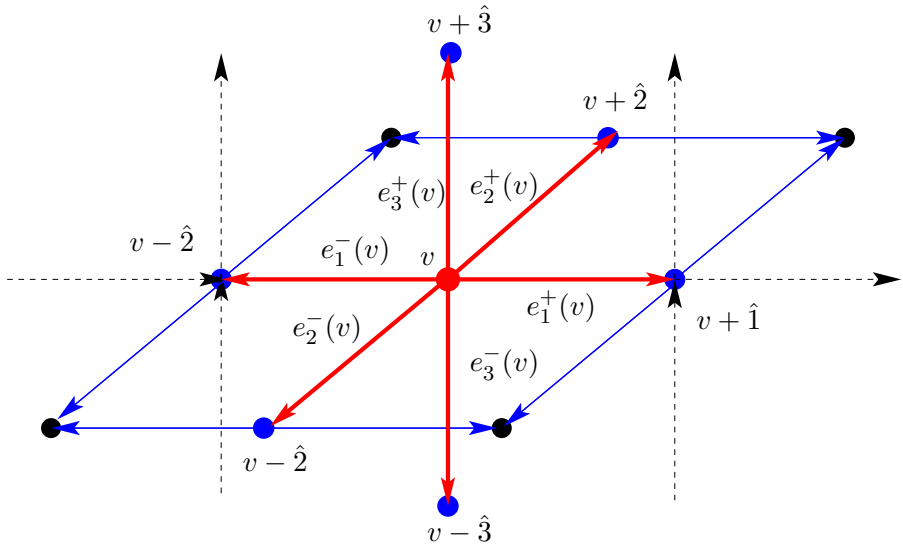}
\caption{Sketch of a graph of cubic topology}
\label{cubicgraph}
\end{center}
\end{figure}
We will consider an algebraic graph with cubic topology, sketched in figure \ref{cubicgraph}. Each vertex is six-valent with three edges going out and three edges going in. We will choose the embedding such that for a given vertex $v$ all six edges are outgoing, as shown in figure \ref{cubicgraph}. Since the algebraic Master constraint operator acts on vertices only and moreover consists of a sum of the contributions at each vertex, it is always possible to restrict attention to one vertex only. 
For a given vertex $v$ we will label the six edges with $e_J^{\sigma}(v)$, whereby $\sigma=\{+,-\}$ and $J=\{1,2,3\}$. We will label the outgoing edges by $e_J(v)$ and choose an ordering such that the triple $\{e_1,e_3,e_3\}$ is  right handed  with respect to the given orienteation of $\Sigma$. We use the notation $e^{+}_J(v):=e_J(v)$ and $e^{-}_J(v):=e_J(v-\hat{J})$ where $v-\hat{J}$ denotes the point translated one unit along the $\hat{J}$ axis while the other two directions do not change. The dual surfaces associated with $e^{+}_J(v)$ is $S_{e_J}$ while the one belonging to $e^{-}_J(v)$ is $S_{e_J(v-{\sst \hat{J}})}$ with its orientation reversed.
Beside the six edges directly connected to the vertex $v$ (the red or thick ones respectively  in figure \ref{cubicgraph}) the action of the algebraic Master constraint involves additionally edges of next neighbouring vertices. In figure \ref{cubicgraph} these edges are the blue or thinner ones respectively that are not dashed.  The next neighbouring vertices are also blue or thinner respectively.. We will choose the orientation of these edges such that, when embedding the graph via coherent states, the orientation of the edge $e^{\sigma}_{J}(v+\sigma^{\prime}\hat{I})$ agrees with the orientation of $e^{\sigma}_J$.
\subsection{The $U(1)^3$-Approximation}
In our calculations we will use the approximation that $SU(2)$ is replaced by $U(1)^3$. Former work \cite{U1hoch3} showed that although replacing $SU(2)$ by $U(1)^3$ is incorrect the  results are reproduced qualitatively. Moreover the main advantage of this approximation is that the $U(1)^3$ volume operator counterpart diagonalizes the $U(1)^3$ counterparts of the SNF. often called charge network functions (CNFs). Thus calculations involving the volume operator, as it is the case for the Master constraint operator, become enormously easier. Since we are mainly interested in the question whether the zeroth order of the expectation value of the Master constraint operator with respect to to coherent states reproduces the correct classical expression. 
the approximation should be appropriate for our purpose. In \cite{III} we will justify this approximation rigorously.\\
Let us denote the $U(1)^3$-holonomy by $h_e$ and the dimensionless electrical flux by $p^e$. Note, that in order emphasize the difference between $SU(2)$ and $U(1)^3$, we will not choose the letters $A(e)$ and $E(e)$ here. The $U(1)^3$-approximation includes then the following replacements
\ba
A(e)&\to& h_e:=(h^1_e,h^2_e,h^3_e)\nonumber\\
E(e)&\to& p^e:=(p^e_1,p^e_2,p^e_3)
\ea
where
\be
h^j_e(m):=\exp(i\int\limits_e A^j)\quad\mathrm{and}\quad
p^e_j(m):=\frac{1}{a_e^2}\int\limits_{S_e}(*E)_j
\ee
The Poisson algebra of $h^j_e$ and $p^e_j$ is given by
\be
\{p^e_j,h^k_{e^{\prime}}\}=i\frac{\kappa}{a^2_e}h^k_{e^{\prime}}\delta^k_j\delta^e_{e^{\prime}}\quad\quad
\{p^e_j,p^e_k\}=-\frac{\kappa}{a_e^2}\epsilon{jk\ell}p^{e^{\prime}}_{\ell}\delta_{e,e^{\prime}}
\ee
leads to the following commutator relations
\be
[\hat{p}^e_j,\hat{h}^k_{e^{\prime}}]=-\frac{\lp^2}{a_e^2}\hat{h}^j_e\delta^k_j\delta^e_{e^{\prime}}\quad [\hat{p}^e_j,\hat{p}^{e}_k]=[\hat{h}^j_e,\hat{h}^k_e]=0
\ee
Here, we introduced a parameter  $a_e$ with dimension of length in order to work with dimensionless fluxes. Its relation with the classicality parameter $t_e$ of the coherent states is $t_e=\lp a_e^2$. Working with dimensionless fluxes will convenient for the later discussion of the quantum fluctuations.\\
For the holonomies and fluxes of our cubic graph we use the following abbreviations in order to keep our notation as simple as possible
\be
 \op{h}_{e^{\sigma}_J}^j:=\hgen,\quad \op{p}^{e^{\sigma}_J}_j:=\pgenv
\ee
\subsection{The Algebraic (Extended) Master Constraint Operator for an Algebraic Graph of Cubic Topology}
\begin{figure}[hbt]
\begin{center}
\psfrag{Al1p2p}{$\op{h}_{\beta_{\{1,+,2,+,\mn,v\}}}$}
\psfrag{Al1m2p}{$\op{h}_{\beta_{\{1,-,2,+,\mn,v\}}}$}
\psfrag{Al1m2m}{$\op{h}^{-1}_{\beta_{\{1,-,2,-,\mn,v\}}}$}
\psfrag{Al2m1p}{$\op{h}^{-1}_{\beta_{\{1,+,2,-,\mn,v\}}}$}
\psfrag{e1p}{$e^{+}_1$}
\psfrag{e2p}{$e^{+}_2$}
\psfrag{e3p}{$e^{+}_3$}
\psfrag{e1m}{$e^{-}_1$}
\psfrag{e2m}{$e^{-}_2$}
\psfrag{e3m}{$e^{-}_3$}
\includegraphics[width=8cm]{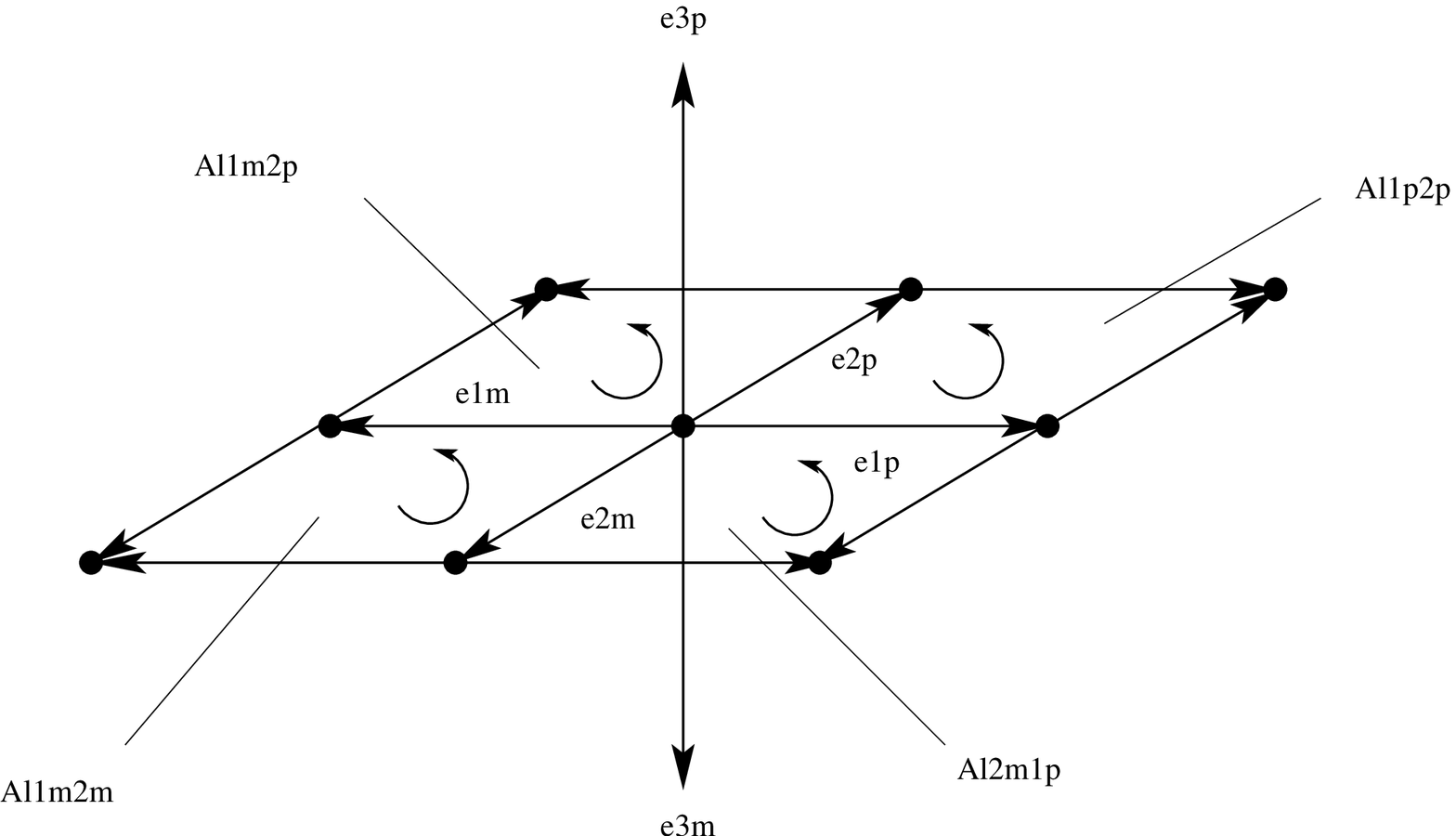}
\caption{The four possible minimal loops $\op{h}_{\beta_{1 + 2,+ \mn v}}$, $\op{h}_{\beta_{1 - 2 + \mn v}}$, $\op{h}^{-1}_{\beta_{1 - 2 - \mn v}}$ and $\op{h}^{-1}_{\beta_{1 + 2 - \mn v}}$ for $\Kn=3,\sn=+$.}
\label{4loops}
\end{center}
\end{figure}

The algebraic extended $SU(2)$ Master constraint reads
\be 
\MC=\sum_v [\sum_a {\rm Tr}(A(\beta^a_v) A(e^a_v) 
[A(e^a_v)^{-1},\sqrt{V_v}])]^2+\sum\limits_{\ell=1}^3{\rm Tr}(\tau_\ell A(\beta^a_v) A(e^a_v) 
[A(e^a_v)^{-1},\sqrt{V_v}])]^2
\ee
where $\beta^a_v$ denotes the minimal plaquettes loop in the $x^a=const$ direction. The substitution of $U(1)^3$ for $SU(2)$ replaces 
\ba
{\rm Tr}(A(\beta^a_v) A(e^a_v)[A(e^a_v)^{-1},\sqrt{V_v}])]^2&\rightarrow&
\sum\limits_{\In\Jn\Kn}\epsilon^{\In\Jn\Jn}h^{\nn}_{\beta_{\In\Jn}}h^{\nn}_{e_{\Kn}(v)}\{(h^{\nn}_{e_{\Kn})^{-1}(v)},\sqrt{V_v}\}\nonumber\\
\sum\limits_{\ell=1}^3{\rm Tr}(\tau_\ell A(\beta^a_v) A(e^a_v) 
[A(e^a_v)^{-1},\sqrt{V_v}])]^2&\rightarrow&
\sum\limits_{\elln=1}^3\sum\limits_{\In\Jn\Kn}\epsilon^{\In\Jn\Jn}\epsilon_{\elln\mn\nn}h^{\mn}_{\beta_{\In\Jn}}(h^{\nn}_{e_{\Kn})^{-1}(v)}\{h^{-1}_{e_{\Kn}(v)},\sqrt{V_v}\}
\ea
where $h^{\nn}_{\beta_{\In\Jn}}$ denotes a minimal loop of $U(1)^3$-holonomies along the edges $e_{\In},e_{\Jn}$.
Let us parametrise the minimal loops by the parameters $\In,\sn,\Jn,\sn^{\prime}$, then any possible loop
loop can be written as
\be
\beta_{\{\In,\sn,\Jn,\sn^{\prime},v\}}=e^{\sn}_{\In}(v)\circ e^{\sn^{\prime}}_{\Jn}(v+\sn\hat{\In})\circ(e^{\sn}_{\In})^{-1}(v+\sn^{\prime}\hat{\Jn})\circ (e^{\sn^{\prime}}_{\Jn})^{-1}(v)
\ee
and hence the $U(1)^3$-loop is given by
\be
\holloop{\In,\sn}{\Jn}{\sn^{\prime}}{\mn}{v}=\hol{\In}{\sn}{\mn}{v}\circ\hol{\Jn}{\sn^{\prime}}{\mn}{v+\sn\hat{\In}}\circ\holin{\In}{\sn}{\mn}{v+\sn^{\prime}\hat{\Jn}}\circ\holin{\Jn}{\sn^{\prime}}{\mn}{v}
\ee
The summation over all possible minimal loops $\beta_{\{\In,\sn,\Jn,\sn^{\prime},v\}}$ can be expressed in terms of $\epsilon_{ijk}$-tensors such that the
algebraic Master contraint operator associated with a graph having cubic topology for $U(1)^3$ is given by
 \be
\MCO_{v}=\sum\limits_{\mu=0}^{3}\op{C}_{\mu,v}^{\dagger}\op{C}_{\mu,v}
\ee 
where
\ba
\op{C}_{0,v}&=&\sum\limits_{\In\Jn\Kn}\sum\limits_{\sn=+,-}\sum\limits_{\sn^{\prime}=+,-}\sum\limits_{\sn^{\prime\prime}=+,-}\frac{4}{\kappa}\epsilon^{\In\Jn\Kn}
\holloop{\In \sn^{\prime}}{\Jn}{\sn^{\prime\prime}}{\elln}{v}\hol{\Kn}{\sn}{\elln}{v}\frac{1}{i\hbar}\left[\holin{\Kn}{\sn}{\elln}{v},\op{V}^{\frac{1}{2}}_{\alpha,v}\right]\\
\op{C}_{\elln,v}&=&\sum\limits_{\In\Jn\Kn}\sum\limits_{\sn=+,-}\sum\limits_{\sn^{\prime}=+,-}\sum\limits_{\sn^{\prime\prime}=+,-}
\frac{4}{\kappa}\epsilon^{\In\Jn\Kn}\epsilon_{\elln \mn\nn}
\holloop{\In \sn^{\prime}}{\Jn}{\sn^{\prime\prime}}{\mn}{v}\hol{\Kn}{\sn}{\nn}{v}\frac{1}{i\hbar}\left[\holin{\Kn}{\nn}{\sn}{v},\op{V}^{\frac{1}{2}}_{\alpha,v}\right]\nonumber\\
\ea
When considering the Master constraint operator, we realise that for a fixed values of $\Kn,\sn$, we have four possible minimal loops. These loops are shown in figure \ref{4loops} for the case $\Kn=3,\sn=+$. 
Through out our calculation we want to use the simplification $\sn=\sn^{\prime}=\sn^{\prime\prime}$. This assumption will not affect our final semiclassical result\footnote{When considering four loops we have to divide by a factor of four and hence semiclassically this factor is cancelled.}, but has the advantage that the four loops reduce to only one loop. For instance in figure \ref{4loops} only the loop $\op{h}_{\beta_{\{1,+,2,+,\mn,v\}}}$ fullfills $\sn=\sn^{\prime}=\sn^{\prime\prime}$. Hence, in total we will have less edges involved in the action of the Master constraint, whose $\op{C}_{\mu,v}$ operators, inserting our assumption $\sn=\sn^{\prime}=\sn^{\prime\prime}$, are
\ba
\op{C}_{0,v}&=&\sum\limits_{\In\Jn\Kn}\sum\limits_{\sn=+,-}\frac{4}{\kappa}\epsilon^{\In\Jn\Kn}
\holloop{\In}{\Jn}{\sn}{\elln}{v}\hol{\Kn}{\sn}{\elln}{v}\frac{1}{i\hbar}\left[\holin{\Kn}{\sn}{\elln}{v},\op{V}^{\frac{1}{2}}_{\alpha,v}\right]\nonumber\\
\op{C}_{\elln,v}&=&\sum\limits_{\In\Jn\Kn}\sum\limits_{\sn=+,-}\frac{4}{\kappa}\epsilon^{\In\Jn\Kn}\epsilon_{\elln \mn\nn}
\holloop{\In}{\Jn}{\sn}{\mn}{v}\hol{\Kn}{\sn}{\nn}{v}\frac{1}{i\hbar}\left[\holin{\Kn}{\nn}{\sn}{v},\op{V}^{\frac{1}{2}}_{\alpha,v}\right]
\ea
where we introduced $\holloop{\In}{\Jn}{\sn}{\mn}{v}:=\holloop{\In}{\Jn \sn}{\sn}{\mn}{v}$ which we will use as the notation for the loops from now on, because we always have $\sn^{\prime}=\sn^{\prime\prime}$ and do not have to carry a separated $\sn$-label for $\In$ and $\Jn$.
By introducing the operators $\op{X}^{e^{\sigma}_J}_j:=\Ygen=i\hgen\partial/\partial\hgen$ and taking advantage of the cubic symmetry of $\alpha$, the volume operator $\op{V}_{\alpha,v}$ can be rewritten as
\be
\op{V}_{\alpha,v}=\lp^3\sqrt{\left|\epsilon^{jkl}\left[\frac{\Y{1}{+}{j}{v}-\Y{1}{-}{j}{v}}{2}\right]\left[\frac{\Y{2}{+}{k}{v}-\Y{2}{-}{k}{v}}{2}\right]\left[\frac{\Y{3}{+}{l}{v}-\Y{3}{-}{l}{v}}{2}\right]\right|}
\ee
The eigenvalue of $\op{V}_{\alpha,v}$ is given by
\be
\lambda^{\frac{1}{2}}(\{\ngen\})=\lp^3\left(\sqrt{\left|\epsilon^{jkl}\left[\frac{\n{1}{+}{j}{v}-\n{1}{-}{j}{v}}{2}\right]\left[\frac{\n{2}{+}{k}{v}-\n{2}{-}{k}{v}}{2}\right]\left[\frac{\n{3}{+}{l}{v}-\n{3}{-}{l}{v}}{2}\right]\right|}\right)^{\frac{1}{2}}
\ee
Note, that we use the embedding dependend operator introduced in \cite{VAL}, because the embedding independent one \cite{VRS} has been ruled out by a recent analysis \cite{GT}.
\subsection{$U(1)^3$ Coherent States associated with a Graph of Cubic Symmetry}
The $U(1)^3$ coherent states are given by
\be
\label{cohstate}
\Psi^t_{\alpha,m}=\prod\limits_{e\in E(\alpha)}\prod\limits_{j=1,2,3}\Psi^{t_e}_{g_e(m)},
\ee
where
\be
\Psi^{t_e}_{g(m)}(h)=\sum\limits_{n\in\Zl}e^{-t_e n^2/2}\lb(g_e h^{-1}\rb)^n
\ee
and $g_{e^{\sigma}_J}^j:=e^{\pgen}\hgencl=\ggen$ and $t_e:=\lp^2/a_e^2$ is the so called classically parameter.
Now we want to calculate expectation values of $\MCO_{v}$ for coherent $U(1)^3$ states
\ba
\label{expvalueM}
\frac{\langle\Psigeng\,|\,\MCO_{v}\,|\,\Psigeng\rangle}{||\Psigeng||^2}&=&
\frac{\sum\limits_{\mu=0}^{3}\langle\Psigeng\,|\,\op{C}_{\mu,v}^{\dagger}\op{C}_{\mu,v}\,|\,\Psigeng\rangle}
{||\Psigeng||^2}\nonumber\\
&=&\frac{\sum\limits_{\mu=0}^{3}\langle\op{C}_{\mu,v}\Psigeng\,|\,\op{C}_{\mu,v}\Psigeng\rangle}
{||\Psigeng||^2}
\ea
\begin{figure}[htb]
\begin{center}
\psfrag{Al12p}{$\holloop{1}{2}{+}{\mn}{v}$}
\psfrag{Al23p}{$\holloop{2}{3}{+}{\mn}{v}$}
\psfrag{Al31p}{$\holloop{3}{1}{+}{\mn}{v}$}
\psfrag{Al12m}{$\holloopin{1}{2}{-}{\mn}{v}$}
\psfrag{Al23m}{$\holloopin{2}{3}{-}{\mn}{v}$}
\psfrag{Al31m}{$\holloopin{3}{1}{-}{\mn}{v}$}
\psfrag{e1p}{$e^{+}_1$}
\psfrag{e2p}{$e^{+}_2$}
\psfrag{e3p}{$e^{+}_3$}
\includegraphics[width=8cm]{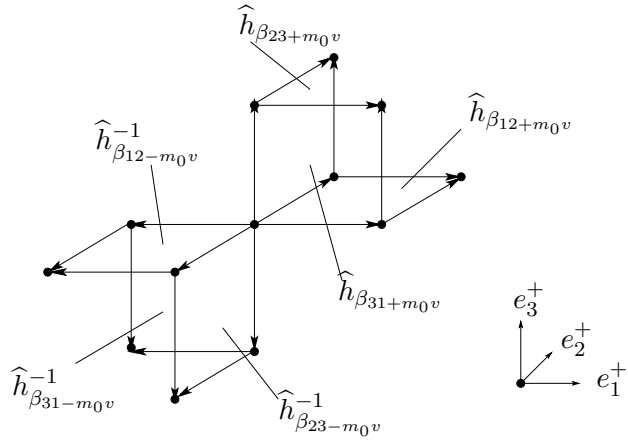}
\caption{The Eighteen edges that are involved in the action of the Master constraint at a given vertex and the six corresponding minimal loops.}
\label{18edges}
\end{center}
\end{figure}
Let us discuss a bit more in detail how many and which edges precisely are involved in the action of $\MCO_{v}$ at a given vertex $v$.
Since we chose $\sn=\sn^{\prime}=\sn^{\prime\prime}$ for simplicity, for each chosen $(\Kn,\sn)\in(\{1,2,3\},\{+.-\})$ there is only one possible minimal loop $\holloop{\In}{\Jn}{\sn}{\mn}{v}$. Hence, when summing over all possible in total we have $6+3\cdot4=18$ edges which are involved in the action of $\MCO_v$, whereby the additional $12$ edges are not directly connected to the vertex $v$. The volume operator considers only the six edges $\{e^{\sigma}_{J}\,|\, \sigma=+,-\, ;\, J=1,2,3\}$ that are directly connected to $v$, hence we do not get any additional edges to consider from the commutator term. These 18 edges are shown in figure \ref{18edges}.
\\
Eqn (\ref{cohstate}) states that a coherent state associated to a graph $\alpha$ can be written in terms of the product of the coherent states associated with each edge $e\in E(\alpha)$ of the graph. Consequently, when considering expectation values of the form in eqn (\ref{expvalueM}), all edges that are not involved in the operator action, will simply be cancelled by their corresponding norm in the denominator. Hence, the expectation value of $\MCO_{v}$ with respect to $\Psigeng$ is equivalent to the expectation value of $\MCO_{v}$ with respect to $\Psi^{t}_{\alpha_{18},m}$, where $\Psi^{t}_{\alpha_{18},m}$ denotes the coherent state associated to the graph with 18 edges shown in figure \ref{18edges}.
\subsection{The basic building blocks of the expectation value of the Master constraint operator}
Let us introduce the following shorthand
\be
\Omn:=\holloop{\In}{\Jn}{\sn}{\mn}{v}\hol{\Kn}{\sn}{\nn}{v}\frac{1}{i\hbar}\left[\holin{\Kn}{\sn}{\nn}{v},\op{V}^{\frac{1}{2}}_{\alpha,v}\right]
\ee
then the basic building block of the Master constraint operator is given by
\be
\label{bbb}
\frac{\langle\Omnt\Psigeng\,\Big|\,\Omn\Psigeng\rangle}{||\Psigeng||^2}
\ee
If we know the explicit value of the expectation value of $(\Omnt)^{\dagger}\Omn$ in eqn (\ref{bbb}) for general $\In,\Jn,\Kn,\sn,\mn,\nn$ and $\Int,\Jnt,\Knt,\snt,\mnt,\nnt$ respectively, the expectation value of the Master constraint can be expressed in terms of a sum of expectation values of $(\Omnt)^{\dagger}\Omn$.
\begin{figure}[hbt]
\begin{center}
\psfrag{PsiL}{$\Psi^t_{\{g,J,\sigma,j,L\}}$}
\psfrag{Al12p}{$\holloop{1}{2}{+}{\mn}{v}$}
\psfrag{Al32m}{$\holloopin{2}{3}{-}{\mn}{v}$}
\psfrag{v}{$v$}
\psfrag{vp1}{$v+\hat{1}$}
\psfrag{vp2}{$v+\hat{2}$}
\psfrag{vm3}{$v-\hat{3}$}
\psfrag{vm2}{$v-\hat{2}$}
\includegraphics[width=5cm]{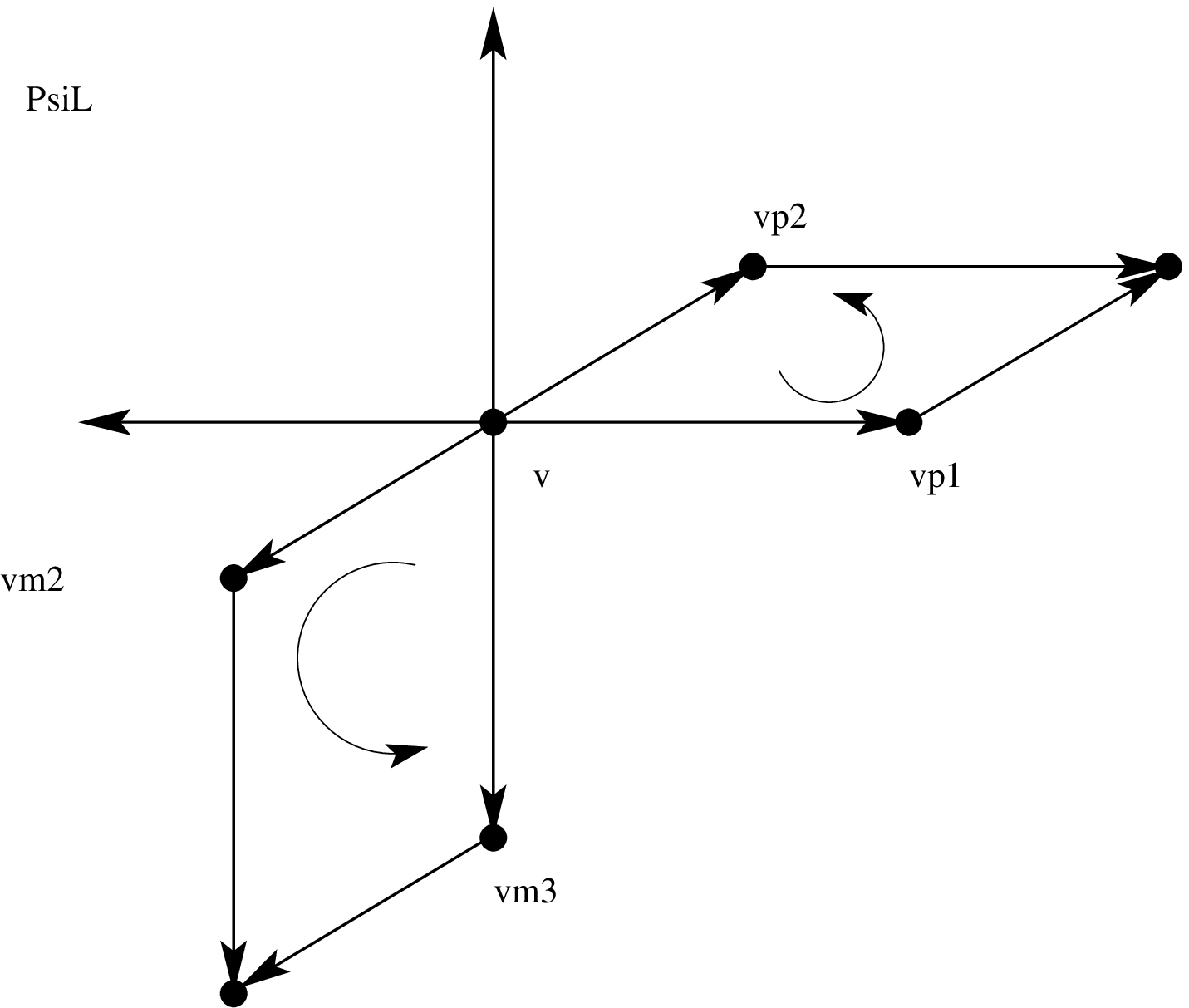}
\caption{$\Psigen$  with $L=L(3,2,1,-,1,2,3,+,v)$ for the loops $\holloop{3}{2}{-}{\mn}{v}=\holloopin{2}{3}{-}{\mn}{v}$ and $\holloop{1}{2}{+}{\mn}{v}$}
\label{PsiL}
\end{center}
\end{figure}
Due to the fact that we have fixed but general $\In,\Jn,\Kn,\sn,\mn,\nn$ and $\Int,\Jnt,\Knt,\snt,\mnt,\nnt$ here only ten out of the 18  different edges of $\Psi^{t}_{\alpha_{18},m}$ are considered by the operator $(\Omnt)^{\dagger}\Omn$.
Let us introduce the set $L(\Int,\Jnt,\Knt,\snt,\In,\Jn,\Kn,\sn,v)$ that contains these ten edges and use the following notation
\ba
\label{DefL}
L_v&:=&\{e^{\sigma}_J(v)\,|\,\sigma=+.-\, ,\, J=1,2,3\}\nonumber\\
L:=L(\Int,\Jnt,\Knt,\snt,\In,\Jn,\Kn,\sn,v)&:=&L_v
\cup\{e^{\sn}_{\Jn}(v+\sn\hat{\In}),(e^{\sn}_{\In})^{-1}(v+\sn\hat{\Jn}),e^{\snt}_{\Jnt}(v+\snt\hat{\Int}),(e^{\snt}_{\Int})^{-1}(v+\snt\hat{\Jnt})\}\nonumber\\
\ea
Apart from the six edges that are directly connected to the vertex $v$ at most four different additional edges that are not connected to $v$ are modified by the action of $(\Omnt)^{\dagger}\Omn$. An example of such a graph is shown in figure \ref{PsiL}. When, defining the coherent state associated to this graph  consisting of at most 10 edges, we have to take the product of the coherent states associated to each edge, see eqn (\ref{cohstate}). Introducing the set
of vertices $V:=\{v,v+\sn\hat{\In},v+\sn\hat{\Jn},v+\snt\hat{\Int},v+\snt\hat{\Jnt}\}$, we can parametrise these 10 edges by the labels $J,\sigma,j,\tilde{v}$, whereby $j,J\in\{1,2,3\}$, $\sigma\in\{+,-\}$ and $\tilde{v}\in V$. Denoting the coherent state associated with this graph by $\Psigen$ we can express it as
\ba
\label{Psigen}
\Psigen&=&\prod\limits_{\tilde{v}\in V}\prod\limits_{(J,\sigma,j)\atop \in L}\Psigenedge\nonumber\\
&=&\prod\limits_{\tilde{v}\in V}\prod\limits_{(J,\sigma,j)\atop\in L}\sum\limits_{\ngen\atop \in\Zl}e^{-\frac{\tgen\ngen^2}{2}}e^{+\pgen\ngen}e^{+i\phigen\ngen}e^{-i\tsgen\ngen}
\ea
Consequently, as explained already in the case of the 18-edges graph, when discussing the action of\\ $(\Omnt)^{\dagger}\Omn$ on $\Psi_{\alpha_{18},m}$ it is enough to know the action on $\Psigen$, thus we have
\be
\frac{\langle\Omnt\Psi_{\alpha_{18},m}\,\Big|\,\Omn\Psi_{\alpha_{18},m}\rangle}{||\Psigeng||^2}=\frac{\langle\Omnt\Psigen\,\Big|\,\Omn\Psigen\rangle}{||\Psigen||^2}
\ee
Note, that most generally the classicality parameter $\tgen$ can be different for each single edge. Hence, we would have to take 10 different limits $\tgen\to 0$ when actually calculating our expectation values. Since we need already a lot of notation through out our calculation and the final result will not be affected in general when we chose $t:=\tgen$ for all $g,Jm\sigma,\vt$, we will do this in the following discussion.
\subsection{The Action of the Operator $\Omn$}
For the benefit of the reader we will discuss the explicit action of $\Omn$ on coherent states in detail. First we will analyse the action of the loop operator contained in $\Omn$, then the action of the remaining holonomy and commutator term and afterwards combining both into the total action of $\Omn$.
\subsubsection{The Action of the Loop Operator $\holloop{\In}{\Jn}{\sn}{\mn}{v}$}
The loop $\holloop{\In}{\Jn}{\sn}{\mn}{v}$ expressed in terms of four single holonomies reads 
\ba
\holloop{\In}{\Jn}{\sn}{\mn}{v}
&=&\hol{\In}{\sn}{\mn}{v}\circ\hol{\Jn}{\sn}{\mn}{v+\sn\hat{\In}}\circ
\holin{\In}{\mn}{\sn}{v+\sn\hat{\Jn}}\circ\holin{\Jn}{\mn}{\sn}{v}
\ea
Hence, the action of $\holloop{\In}{\Jn}{\sn}{\mn}{v}$ is given by
\ba
\lefteqn{\holloop{\In}{\Jn}{\sn}{\mn}{v}\frac{\Psigen}{||\Psigen||}}\nonumber\\
&=&\hol{\In}{\sn}{\mn}{v}\circ\hol{\Jn}{\sn}{\mn}{v+\sn\hat{\In}}\circ
\holin{\In}{\mn}{\sn}{v+\sn\hat{\Jn}}\circ\holin{\Jn}{\mn}{\sn}{v}\frac{\Psigen}{||\Psigen||}\nonumber\\
&=&\frac{1}{||\Psigen||}\prod\limits_{\sst \vt\in V}\prod\limits_{\sst (J,\sigma,j)\atop\in L}\sum\limits_{\ngen\atop \in\Zl}
e^{-\frac{1}{2}\left(t  (\ngen)^2\right)}
e^{+\left(\pgen\ngen\right)}
e^{+i\left(\phigen\ngen\right)}\nonumber\\
&&
e^{-i\tsgen\left(\ngen
+\kdel{\Jn}{\sn}{\mn}{v}{J}{\sigma}{j}{\vt}
+\kdel{\In}{\sn}{v+\sn\hat{\Jn}}{\mn}{J}{\sigma}{j}{\vt}
-\kdel{\In}{\sn}{v}{\mn}{J}{\sigma}{j}{\vt}
-\kdel{\Jn}{\sn}{v+\sn\hat{\In}}{\mn}{J}{\sigma}{j}{\vt}\right)}
\ea
where 
\be
\kdel{\Jn}{\sn}{\mn}{v}{J}{\sigma}{j}{\vt}=\delta_{\Jn,J}\delta_{\sn,\sigma}\delta_{\mn,j}\delta_{v,\vt}
\ee
In order to get succinct expression for the $\delta-$functions, we introduce the abbreviation
\ba
\lefteqn{\Bdel}\nonumber\\
&:=&\left(
+\kdel{\Jn}{\sn}{\mn}{v}{J}{\sigma}{j}{\vt}
+\kdel{\In}{\sn}{\mn}{v+\sn\hat{\Jn}}{J}{\sigma}{j}{\vt}
-\kdel{\In}{\sn}{\mn}{v}{J}{\sigma}{j}{\vt}
-\kdel{\Jn}{\sn}{\mn}{v+\sn\hat{\In}}{J}{\sigma}{j}{\vt}
\right)
\ea
\subsubsection{The Action of $\hol{\Kn}{\sn}{\nn}{v}\frac{1}{i\hbar}\left[\holin{\nn}{\Kn}{\sn}{v},\op{V}^{\frac{1}{2}}_{\alpha,v}\right]$}
The action of  $\hol{\Kn}{\sn}{\nn}{v}\left[\holin{\Kn}{\sn}{\nn}{v},\op{V}^{\frac{1}{2}}_{\alpha,v}\right]$ involves for a given vertex $v$  only the six edges that are directly connected to the vertex $v$. 
\ba
\lefteqn{\frac{\hol{\Kn}{\sn}{\nn}{v}\frac{1}{i\hbar}\left[\holin{\Kn}{\sn}{\nn}{v},\op{V}^{\frac{1}{2}}_{\alpha,v}\right]\Psigen}{||\Psigen||}}\nonumber\\
&=&\frac{\frac{1}{i\hbar}\left(\op{V}^{\frac{1}{2}}-\hol{\Kn}{\sn}{\nn}{v}\op{V}{\frac{1}{2}}_{\alpha,v}\holin{\Kn}{\sn}{\nn}{v}\right)\Psigen}{||\Psigen||}\nonumber\\
&=&
\left(\prod\limits_{\vt\neq v\atop\in V}\prod\limits_{(J,\sigma,j)\atop\in L\setminus L_v}\sum\limits_{\ngen\atop\in\Zl}e^{-\frac{1}{2}\left(t  (\ngen)^2\right)}
e^{+\left(\pgen\ngen\right)}
e^{+i\left(\phigen\ngen\right)}\right)\nonumber\\
&&\frac{\prod\limits_{(J,\sigma,j)\atop\in L_v}\sum\limits_{\ngenv\atop\in\Zl}
\left(\lambda^{\frac{1}{2}}(\{\ngenv\})-\lambda^{\frac{1}{2}}(\{\ngenv+\kdel{J}{\sigma}{j}{v}{\Kn}{\sn}{\nn}{v}\})\right)e^{-\frac{1}{2}\left(t  (\ngenv)^2\right)}
e^{+\left(\pgenv\ngenv\right)}
e^{+i\left(\phigenv\ngenv\right)}}{||\Psigen||}\nonumber\\
\ea
where
\be
\lambda^{\frac{1}{2}}(\{\ngenv\})=a^{\frac{3}{2}}\frac{t^{\frac{3}{4}}}{i\hbar}\left(\sqrt{\left|\epsilon^{jkl}\left[\frac{\n{1}{+}{j}{v}-\n{1}{-}{j}{v}}{2}\right]\left[\frac{\n{2}{+}{k}{v}-\n{2}{-}{k}{v}}{2}\right]\left[\frac{\n{3}{+}{l}{v}-\n{3}{-}{l}{v}}{2}\right]\right|}\right)^{\frac{1}{2}}
\ee
Note we inserted a factor of one into the eigenvalue $\lambda^{\half}$ by multiplying and by the same time dividing the whole term by a factor of $a^{\frac{3}{2}}$ since $t=\lp/a^2$. This will be convenient for our later notation.\\
The volume operator acts on edges directly connected to the vertex $v$ only. Therefore the parts of the coherent state  associated with edges at $\vt\neq v$ commute with the volume operator and can therefore be moved to the lefthand side of the holonomy-commutator term. Recall from eqn (\ref{DefL}) that the set $L_v=\{e^{\sigma}_I\,|\,\sigma=+,-\,;\,I=1,2,3\}$.
Combining together the separate action of the loop and the commutator term, we end up with the following action of $\Omn$
\ba
\lefteqn{\frac{1}{||\Psigen||}\Omn\Psigen}\nonumber\\
&=&\frac{1}{||\Psigen||}
\holloop{\In}{\Jn}{\sn}{\mn}{v}\hol{\Kn}{\sn}{\nn}{v}\frac{1}{i\hbar}
\left[(\holin{\Kn}{\sn}{\nn}{v},\op{V}^{\frac{1}{2}}_{\alpha,v}\right]\Psigen\nonumber\\
&=&
\frac{1}{||\Psigen||}\left(\prod\limits_{\vt\neq v\atop\in V}\prod\limits_{\sst (J,\sigma,j)\atop\in L\setminus L_v}\sum\limits_{\ngen\atop\in\Zl}e^{-\frac{1}{2}\left(t  (\ngen)^2\right)}
e^{+\left(\pgen\ngen\right)}
e^{+i\left(\phigen\ngen\right)}
e^{+i\tsgen\left(-\ngen-\Bdel\right)}\right)\nonumber\\
&&\hspace{2.2cm}
\left(\prod\limits_{(J,\sigma,j)\atop\in L_v}\sum\limits_{\ngenv\atop\in\Zl}
\left(\lambda^{\frac{1}{2}}(\{\ngenv\})-\lambda^{\frac{1}{2}}(\{\ngenv+\kdel{J}{\sigma}{j}{v}{\Kn}{\sn}{\nn}{v}\})\right)e^{-\frac{1}{2}\left(t  (\ngenv)^2\right)}
e^{+\left(\pgenv\ngenv\right)}
\right.\nonumber\\
&&\hspace{4.3cm}\left.
e^{+i\left(\phigenv\ngenv\right)}e^{+i\tsgenv\left(-\ngenv-\Bdelv\right)}\right)
\ea
Hence, we are able to give an expression for the expectation value of $\Omn$
\ba
\lefteqn{\frac{\langle\Omnt\Psigen\,\Big|\,\Omn\Psigen\rangle}{||\Psigen||^2}}\nonumber\\
&=&\frac{1}{|||\Psigen||^2}\nonumber\\
&&\left(\prod\limits_{\vt\neq v\atop\in V}\prod\limits_{\sst (J,\sigma,j)\atop\in L\setminus L_v}\sum\limits_{\ngen\atop\in\Zl}\sum\limits_{\ngent\atop \in\Zl}
e^{-\frac{1}{2}\left(t  \left(\lb(\ngen\rb)^2+\lb(\ngent\rb)^2\right)\right)}
e^{+\left(\pgen\left(\ngen+\ngent\right)\right)}
e^{+i\left(\phigen\left(\ngen-\ngent\right)\right)}\right.\nonumber\\
&&\hspace{3cm}
\underbrace{\left.\int\,d\tsgen e^{+i\tsgen\left(-\ngen-\Bdel+\ngent+\Bdelt\right)}\right)}_{\delta\left(\ngent+\Bdelt-\ngen-\Bdel\right)}\nonumber\\
&&\left(\prod\limits_{\sst (J,\sigma,j)\atop\in L_v}\sum\limits_{\ngen\atop\in\Zl}\sum\limits_{\ngentv\in\Zl}
e^{-\frac{1}{2}\left(t  \left(\lb(\ngenv\rb)^2+\lb(\ngentv\rb)^2\right)\right)}
e^{+\left(\pgenv\left(\ngenv+\ngentv\right)\right)}
e^{+i\left(\phigenv\left(\ngenv-\ngentv\right)\right)}\right.\nonumber\\
&&
\left[\lambda^{\frac{1}{2}}(\{\ngenv\})-\lambda^{\frac{1}{2}}(\{\ngenv+\kdel{J}{\sigma}{j}{v}{\Kn}{\sn}{\nn}{v}\})\right]
\left[\lambda^{\frac{1}{2}}(\{\ngentv\})-\lambda^{\frac{1}{2}}(\{\ngentv+\kdel{J}{\sigma}{j}{v}{\Knt}{\snt}{\nnt}{v}\})\right]\nonumber\\
&&\hspace{3.3cm}
\underbrace{\left.\int\,d\tsgenv e^{+i\tsgenv\left(-\ngenv-\Bdelv+\ngentv+\Bdeltv\right)}\right)}_{\delta\left(\ngentv+\Bdeltv-\ngenv-\Bdelv\right)}
\ea
The $\delta$-function forces the following condition on $\ngen,\ngenv$ and $\ngent,\ngentv$ respectively
\ba
\ngent&=&\ngen+\Bdel-\Bdelt\nonumber\\
\ngentv&=&\ngenv+\Bdelv-\Bdeltv
\ea
Introducing
\ba
\label{Bdelta}
\Del&:=&\Bdelt-\Bdel\nonumber\\
\Delv&:=&\Bdeltv-\Bdelv\nonumber\\
\ea
the condition for $\ngen,\ngenv$ ca be rewritten as
\ba
\ngent&=&\ngen-\Del\nonumber\\
\ngentv&=&\ngenv-\Delv
\ea
Reinserting this condition into the expectation value of $\Omn$, we obtain:
\ba
\label{vorPR}
\lefteqn{\frac{\langle\Omnt\Psigen\,\Big|\,\Omn\Psigen\rangle}{||\Psigen||^2}}\nonumber\\
&=&\frac{1}{||\Psigen||^2}\nonumber\\
&&\left(\prod\limits_{\vt\neq v\atop\in V}\prod\limits_{\sst (J,\sigma,j)\atop\in L\setminus L_v}
e^{-\pgen\Del}e^{+i\phigen\Del}\right.\nonumber\\
&&e^{-\frac{t}{2}\left(\Del\right)^2}\nonumber\\
&&\left.\sum\limits_{\ngen\atop\in\Zl}
e^{-t\left(\left(\ngen\right)^2 - 2\ngen\Del\right)}
e^{+2\pgen\ngen}\right)\nonumber\\
&&\left(\prod\limits_{\sst (J,\sigma,j)}
e^{-\pgenv\Delv}e^{+i\phigenv\Delv}\right.\nonumber\\
&&e^{-\frac{t}{2}\left(\Delv\right)^2}\nonumber\\
&&\sum\limits_{\ngenv\atop\in\Zl}
e^{-t\left(\left(\ngenv\right)^2 - 2\ngenv\Delv\right)}
e^{+2\pgenv\ngenv}\nonumber\\
&&
\left[\lambda^{\frac{1}{2}}\lb(\{\ngenv\}\rb)-\lambda^{\frac{1}{2}}\lb(\{\ngenv+\kdel{J}{\sigma}{j}{v}{\Kn}{\sn}{\nn}{v}\}\rb)\right]\nonumber\\
&&
\left[+\lambda^{\frac{1}{2}}\lb(\{\ngenv-\Delv\}\rb)\right.\nonumber\\
&&\hspace{2.2cm}\left.
\left.-\lambda^{\frac{1}{2}}\lb(\{\ngenv-\Delv+\kdel{J}{\sigma}{j}{v}{\Kn}{\snt}{\nnt}{v}\}\rb)\right]\right)\nonumber\\
\ea
Note that $\Delv$ is the special case where $\vt=v$ since only edges that are directly connected to $v$ are considered. Thus  four out of the eight Kronecker-deltas can be neglected and we have
\ba
\lefteqn{\Delv}\nonumber\\
&=&\left(
\kdel{\Jnt}{\sn}{\mnt}{v}{J}{\sigma}{j}{v}
-\kdel{\Int}{\snt}{\mnt}{v}{J}{\sigma}{j}{v}
-\kdel{\Jn}{\sn}{\mn}{v}{J}{\sigma}{j}{v}
+\kdel{\In}{\sn}{\mn}{v}{J}{\sigma}{j}{v}
\right)
\ea
\subsection{Application of the Poisson Re-summation Theorem}
 The aim of this work is to discuss the semiclassical behaviour of the algebraic Master constraint, thus we are mainly interested in the properties of the expectation value in eqn (\ref{vorPR}) for tiny values of the classicality parameter $t$. Looking at eqn (\ref{vorPR}), tiny values of $t$ will correspond to a slow convergence behaviour when considering the sum over $\ngen$. Therefore we will perform a Poisson resummation in which $t$ gets replaced by $1/t$. Then the series converges rapidly when considering small, tiny values of the classicality parameter.
 Let us introduce the following quantities
\be
T:=\sqrt{t}\quad \xgen:=T\ngen\quad \xgenv:=T\ngenv
\ee
with the help of whose all quantities can be expressed in terms of $\xgenv$.
\be
\lambda^{\half}(\{\ngenv\})=T^{-\frac{3}{4}}\lambda^{\half}(\{T\ngenv\})=T^{-\frac{3}{4}}\lambda^{\half}(\{\xgenv\})
\ee
and the expectation value can be rewritten in terms of $\xgen$ as
\ba
\lefteqn{\MeO}\nonumber\\
&=&\frac{T^{-\frac{3}{2}}}{||\Psigen||^2}\nonumber\\
&&\left(e^{-\sum\limits_{\vt\neq v\atop\in V}\sum\limits_{(J,\sigma,j)\atop\in L\setminus L_v}\pgen\Del}
e^{+i\sum\limits_{\vt\neq v\atop\in V}\sum\limits_{((J,\sigma,j)\atop\in L\setminus L_v}\phigen\Del}\right.\nonumber\\
&&
e^{-\frac{t}{2}\sum\limits_{\vt\neq v\atop\in V}\sum\limits_{ (J,\sigma,j)\atop\in L\setminus L_v}\left(\Del\right)^2}\nonumber\\
&&\sum\limits_{\xgen\in\Zl T}
e^{-\sum\limits_{\vt\neq v\atop\in V}\sum\limits_{(J,\sigma,j)\atop\in L\setminus L_v}\left(\xgen\right)^2-\xgen\lb(\frac{2}{T}\pgen+T\Del\rb)}\nonumber\\
&&\hspace{1cm}
\left(e^{-\sum\limits_{(J,\sigma,j)\atop\in L_v}\pgenv\Delv}e^{+i\sum\limits_{((J,\sigma,j)\atop\in L_v}\phigenv\Delv}\right.\nonumber\\
&&
e^{-\frac{t}{2}\sum\limits_{ (J,\sigma,j)\atop\in L_v}\left(\Delv\right)^2}\nonumber\\
&&\sum\limits_{\xgenv\in\Zl T}
e^{-\sum\limits_{(J,\sigma,j)\atop\in L_v}\left(\xgenv\right)^2-\xgenv\lb(\frac{2}{T}\pgenv+T\Delv\rb)}\nonumber\\
&&\left.
\Lambda^{\half}\lb(\{\xgenv\},\e{\sn}{\Kn}{v},\nn\rb)
\Lambda^{\half}\lb(\{\xgenv-T\Delv\},\e{\snt}{\Knt}{v},\nnt\rb)\right)\nonumber\\
\ea
where we introduced  
\be
\label{Lambda}
\Lambda^{\half}\lb(\{\xgenv\},\e{\sn}{\Kn}{v},\nn\rb):=T^{-\frac{3}{4}}\lb[\lambda^{\frac{1}{2}}\lb(\{\xgenv\}\rb)-\lambda^{\frac{1}{2}}\lb(\{\xgenv+T\kdel{J}{\sigma}{j}{v}{\Kn}{\sn}{\nn}{v}\}\rb)\rb]
\ee
in order to keep the expression as short as possible. Moreover, the denominator can be reexpressed as
\be
||\Psigen||^2=\prod\limits_{(J,\sigma,j)}\sqrt{\frac{\pi}{t}}e^{-\frac{1}{t}\lb(\pgen\rb)^2}\lb[1+K_t(p)\rb]
=\lb(\sqrt{\frac{\pi}{t}}\rb)^{30}e^{+\frac{1}{t}\sum\limits_{\vt\in V}\sum\limits_{(J,\sigma,j)\atop\in L}\lb(\pgen\rb)^2}\lb[1+K_t(p)\rb]^{30}
\ee
The application of
the Poisson resummation formula leads therefore to the following expectation value
\ba
\lefteqn{\MeO}\nonumber\\
&=&\frac{1}
{(\sqrt{\frac{\pi}{t}})^{30}[1+K_t(p)]^{30}}\frac{(\sqrt{\pi})^{12}}{T^{30}}\nonumber\\
&&
\left(
e^{+i\sum\limits_{\vt\neq v\atop\in V}\sum\limits_{(J,\sigma,j)\atop\in L\setminus L_v}\phigen\Del}e^{-\frac{t}{4}\sum\limits_{\vt\neq v\atop\in V}\sum\limits_{(J,\sigma,j)\atop\in L\setminus L_v}\lb(\Del\rb)^2}\right.\nonumber\\
&&\left.
\sum\limits_{\ngen\in\Zl}e^{-\frac{\pi^2}{t}\sum\limits_{\vt\neq v\atop\in V}\sum\limits_{(J,\sigma,j)\atop\in L\setminus L_v}\lb(\ngen\rb)^2}e^{-2i\frac{\pi}{t}\sum\limits_{\vt\neq v\atop\in V}\sum\limits_{(J,\sigma,j)\atop\in L\setminus L_v}\ngen\lb(\pgen+\frac{T^2}{2}\Del\rb)}\right)\nonumber\\
&&\left(e^{-\sum\limits_{(J,\sigma,j)\atop\in L_v}\pgenv\Delv}
e^{+i\sum\limits_{(J,\sigma,j)\atop\in L_v}\phigenv\Delv}\right.\nonumber\\
&&e^{-\frac{t}{2}\sum\limits_{(J,\sigma,j)\atop\in L_v}\lb(\Delv\rb)^2}
e^{-\frac{1}{t}\sum\limits_{(J,\sigma,j)\atop\in L_v}\lb(\pgenv\rb)^2}\nonumber\\
&&\sum\limits_{\ngenv\in\Zl}e^{-\frac{\pi^2}{t}\sum\limits_{(J,\sigma,j)\atop\in L_v}\lb(\ngenv\rb)^2}e^{-2i\frac{\pi}{t}\sum\limits_{(J,\sigma,j)\atop\in L_v}\ngenv\lb(\pgenv+\frac{T^2}{2}\Del\rb)}\nonumber\\
&&\int\limits_{\Rl^{18}}\,d^{18}\xgenv e^{-\sum\limits_{(J,\sigma,j)\atop\in L_v}\left(\xgenv\right)^2-\frac{2}{T}\xgenv\lb(\pgenv+\frac{T^2}{2}\Delv-i\pi\ngenv\rb)}\nonumber
\\
&&\left.
\Lambda^{\half}\lb(\{\xgenv\},\e{\sn}{\Kn}{v},\nn\rb)
\Lambda^{\half}\lb(\{\xgenv-T\Delv\},\e{\snt}{\Knt}{v},\nnt\rb)\right)\nonumber\\
\ea
Similar to \cite{Hanno} we introduce new $\xgenv$ variables denoted by $\xpgen:=\half(x_{J+jv}+x_{J-jv})$ and $\xmgen:=\half(x_{J+jv}-x_{J-jv})$. These variables have the advantage that the $\Lambda^{\half}$ are functions on $\xmgen$ only. Hence, the 9-dimensional integral over $\xpgen$ contains no $\Lambda^{\half}$ anymore and can be easily computed, because it has become a usual complex Gaussian integral. Additionally all the other quantities as $\ngenv,\pgenv$ undergo analogous transformations. The transformation for the terms involving $\delta$-functions will depend on the sign of $\sn$ and $\snt$ respectively and will be of the general form
\ba
\frac{1}{2}\sgn(\sn)\dmgen&:=&\frac{\kdel{J}{+}{j}{v}{\Kn}{\sn}{\nn}{v}-\kdel{J}{-}{j}{v}{\Kn}{\sn}{\nn}{v}}{2}\nonumber\\
\half\dpgen&:=&\frac{\kdel{J}{+}{j}{v}{K}{\sn}{\nn}{v}+\kdel{J}{-}{j}{v}{\Kn}{\sn}{\nn}{v}}{2}
\ea
Note, that it was necessary to reexpress $\Delv$ in terms of the difference of $\Bdeltv-\Bdelv$ (see eqn (\ref{Bdelta})) since there is no global $\sgn$-term to factor out in this case. 
\ba
\half\sgn(\sn)\Bdelvm&:=&\frac{\Delta(\In,\Jn,\sn,\mn,v,J,+,j,v)-\Delta(\In,\Jn,\sn,\mn,v,J,-,j,v)}{2}\nonumber\\
\half\Bdelvp&:=&\frac{\Delta(\In,\Jn,\sn,\mn,v,J,+,j,v)+\Delta(\In ,\Jn,\sn,\mn,v,J,-,j,v)}{2}\nonumber\\
\ea
For the details of this transformation, see \ref{Trafo} in appendix.
The change of variables and the performation of the Gaussian integral simplify the expectation value to
\ba
 \label{x9Int}
\lefteqn{\MeO}\nonumber\\
&=&
\frac{e^{+i\sum\limits_{\vt\neq v\atop\in V}\sum\limits_{(J,\sigma,j)\atop\in L\setminus L_v}\phigen\Del}e^{-\frac{t}{4}\sum\limits_{\vt\neq v\atop\in V}\sum\limits_{(J,\sigma,j)\atop\in L\setminus L_v}\lb(\Del\rb)^2}}
{(\sqrt{\frac{\pi}{t}})^{30}[1+K_t(p)]^{30}}\nonumber\\
&&\frac{(\sqrt{\pi})^{12}}{T^{30}}2^9\lb(\sqrt{\frac{\pi}{2}}\rb)^9\nonumber\\
&&\left(
\sum\limits_{\ngen\in\Zl}e^{-\frac{\pi^2}{t}\sum\limits_{\vt\neq v\atop\in V}\sum\limits_{(J,\sigma,j)\atop\in L\setminus L_v}\lb(\ngen\rb)^2}e^{-2i\frac{\pi}{t}\sum\limits_{\vt\neq v\atop\in V}\sum\limits_{(J,\sigma,j)\atop\in L\setminus L_v}\ngen\lb(\pgen+\frac{T^2}{2}\Del\rb)}\right)\nonumber\\
&&\left(
e^{-\frac{t}{8}\sum\limits_{(J,j)\atop\in L_v}\lb(\lb(\Bdelvtp-\Bdelvp\rb)^2\rb)}
e^{+\frac{i}{2}\sum\limits_{(J,j)\atop\in L_v}\lb(\phipgen(\Bdelvtp-\Bdelvp)\rb)}\rb.\nonumber\\
&&\sum\limits_{\npgen\in\Zl}\lb.
e^{-2\frac{\pi^2}{t}\sum\limits_{(J,j)\atop\in L_v}\lb(\npgen\rb)^2}
e^{-4i\frac{\pi}{t}\sum\limits_{(J,j)\atop\in L_v}\npgen\lb(\ppgen+\frac{T^2}{4}(\Bdelvtp-\Bdelvp)\rb)}\rb)\nonumber\\
&&\lb(
e^{-\half\sum\limits_{(J,j)\atop\in L_v}\lb(\pmgen(\sgn(\snt)\Bdelvtm-\sgn(\sn)\Bdelvm)\rb)}\rb.\nonumber\\
&&e^{+\frac{i}{2}\sum\limits_{(J,j)\atop\in L_v}\lb(\phimgen(\sgn(\snt)\Bdelvtm-\sgn(\sn)\Bdelvm)\rb)}\nonumber\\
&&
e^{-\frac{2}{t}\sum\limits_{(J,j)\atop\in L_v}\lb(\lb(\pmgen\rb)^2\rb)}
e^{-\frac{t}{4}\sum\limits_{(J,j)\atop\in L_v}\lb(\lb(\sgn(\snt)\Bdelvtm-\sgn(\snt)\Bdelvm\rb)^2\rb)}\nonumber\\
&&\sum\limits_{\nmgen\in\Zl}
\int\limits_{\Rl^9}d^9\xmgen\nonumber\\
&&
e^{-2\sum\limits_{(J,j,v)\atop\in L_v}\left((\xmgen)^2-\frac{2}{T}\lb(\pmgen+\frac{T^2}{2}\lb[\half\sgn(\snt)\Bdelvtm-\half\sgn(\sn)\Bdelvm\rb]-i\pi\nmgen\rb)\right)}\nonumber\\ 
&&
\Lambda^{\half}\lb(\lb\{\xmgen-\frac{T}{2}\lb[\sgn(\snt)\Bdelvtm-\sgn(\sn)\Bdelvm\rb]\rb\},\half\sgn(\snt)\e{}{\Knt}{v},\nnt\rb)\nonumber\\
&&\lb.
\Lambda^{\half}\lb(\lb\{\xmgen\rb\},\half\sgn(\sn)\e{}{\Kn}{v},\nn\rb)\rb)
\ea
\subsection{Only the Term with $\ngen=0$ matters}
The remaining integral in eqn (\ref{x9Int}) cannot be calculated in a closed form
 for the reason that the $\Lambda^{\half}$-functions prevent it from being a usual Gaussian integral. Moreover, we have an infinite summation over $\nmgen$  occurring in the argument of the exponential function. Therefore we also have to discuss which terms in this $\ngen$-summation have to be considered and which can be neglected. The problem with completing the square in the exponent is, that we will have to continue the $\Lambda^{\half}$-function into the complex plane. Since $\Lambda^{\half}$ is not analytic in $\Cl^9$, we cannot use a simple contour argument in order to estimate the remaining integral. In order to get the integrand univalent, we express $\Lambda$ in terms of squares of determinants by simply squaring the usual expression of the determinant in $\Lambda^{\half}$ in eqn (\ref{Lambda}) and at the same time taking the square root out of the squared determinant
 \ba
 \label{lambdadet}
\lefteqn{\st \Lambda^{\half}\lb(\lb\{\xmgenn+\frac{1}{T}\lb(\pmgen-i\pi\nmgen\rb)+\frac{T}{4}\lb[\sgn(\snt)\Dtmt-\sgn(\sn)\Dtm\rb]\rb\},\half\sgn(\sn)\e{}{\Kn}{v},\nn\rb)}\nonumber\\
&=&{\st t^{\frac{3}{8}}\lb(
\lb[\lb(\det\lb(\lb\{\xmgenn+\frac{1}{T}\lb(\pmgen-i\pi\nmgen\rb)+\frac{T}{4}\lb[\sgn(\snt)\Dtmt-\sgn(\sn)\Dtm\rb]\rb\}\rb)\rb)^2\rb]^{\frac{1}{8}}\right.}\nonumber\\
&&\hspace{-0.5cm}{\st \left.
 -\lb[\lb(\det\lb(\lb\{\xmgenn+\frac{1}{T}\lb(\pmgen-i\pi\nmgen\rb)+\frac{T}{4}\lb[\sgn(\snt)\Dtmt-\sgn(\sn)\Dtm\rb]-\frac{T}{2}\sgn(\sn)\dmgen\rb)\rb\}\rb)^2\rb]^{\frac{1}{8}}\rb)}\nonumber\\
\ea
and furthermore using the trivial identity
\be
\label{lndet}
\lb(\lb[\det(\{\xmgenn+B^{\Kn}_{Jjv}\})\rb]^2\rb)^{\frac{1}{8}}=\exp\lb(\frac{1}{8}\ln\lb(\lb[\det(\{\xmgenn+B^{\Kn}_{Jjv}\})\rb]^2\rb)\rb)
\ee
 whereby $B^{\Kn}_{Jjv}$ denotes symbolically all the additional terms to $\xmgenn$ that occur in the argument of the determinant and we have to use the branch of the $\ln(z)=\ln(|z|e^{i\phi})$ for any complex number $z=|z|e^{i\phi}$ with $\phi\in[0,2\pi)$. With this branch in mind, the integrand becomes indeed univalent on the entire complex plane $\Cl^9$ except at the points where $\det(\{\xmgenn+B^{\Kn}_{Jjv}\})=0$. With having a univalent integrand now, a contour argument can be found that allows us to move the integration path away from the real hyperplane $\Rl^9$ in $\Cl^9$ without changing the result. Consequently, we can complete the square in the exponent now and obtain by using
\be
\xmgenn
:=\xmgen-\frac{1}{T}\pmgen+\frac{T^2}{4}\lb(\sgn(\snt)\Dtmt-\sgn(\sn)\Dtm\rb)-\frac{i\pi}{T}\nmgen\nonumber\\
\ee
where the shorthands
\be
\Dtm:=\Bdelvm\quad\quad
\Dtmt:=\Bdelvtm
\ee
were introduced. This results in the following form of the expectation value
  \ba
 \lefteqn{\MeO}\nonumber\\
&&\frac{e^{+i\sum\limits_{\vt\in V}\sum\limits_{(J,\sigma,j)\atop\in L}\phigen\Del}e^{-\frac{t}{4}\sum\limits_{\vt\in V}\sum\limits_{(J,\sigma,j)\atop\in L}\lb(\Del\rb)^2}}
{(\sqrt{\frac{\pi}{t}})^{30}[1+K_t(p)]^{30}}\nonumber\\
&&\frac{(\sqrt{\pi})^{12}}{T^{30}}2^9\lb(\sqrt{\frac{\pi}{2}}\rb)^9\nonumber\\
&&\left(
\sum\limits_{\ngen\in\Zl}e^{-\frac{\pi^2}{t}\sum\limits_{\vt\neq v\atop\in V}\sum\limits_{(J,\sigma,j)\atop\in L\setminus L_v}\lb(\ngen\rb)^2}e^{-2i\frac{\pi}{t}\sum\limits_{\vt\neq v\atop\in V}\sum\limits_{(J,\sigma,j)\atop\in L\setminus L_v}\ngen\lb(\pgen+\frac{T^2}{2}\Del\rb)}\right)\nonumber\\
&&\left(\sum\limits_{\npgen\in\Zl}
e^{-2\frac{\pi^2}{t}\sum\limits_{(J,j)\atop\in L_v}\lb(\npgen\rb)^2}
e^{-4i\frac{\pi}{t}\sum\limits_{(J,j)\atop\in L_v}\npgen\lb(\ppgen+\frac{T^2}{4}(\Bdelvtp-\Bdelvp)\rb)}\rb)\nonumber\\
&&\left(\sum\limits_{\nmgen\in\Zl}
e^{-2\frac{\pi^2}{t}\sum\limits_{(J,j)\atop\in L_v}\lb(\nmgen\rb)^2}
e^{-4i\frac{\pi}{t}\sum\limits_{(J,j)\atop\in L_v}\nmgen\lb(\pmgen+\frac{T^2}{4}(\sgn(\snt)\Bdelvtm-\sgn(\sn)\Bdelvm)\rb)}\rb.\nonumber\\
&&\int\limits_{\Rl^9}d^9\xmgenn
e^{-2\sum\limits_{(J,j)}\left(\xmgenn\right)^2}\nonumber\\ 
&&\hspace{-0.75cm}
\Lambda^{\half}\lb(\lb\{\xmgenn+\frac{1}{T}\lb(\pmgen-i\pi\nmgen\rb)-\frac{T}{4}\lb[\sgn(\snt)\Dtmt-\sgn(\sn)\Dtm\rb]\rb\},\half\sgn(\snt)\e{}{\Knt}{v},\nnt\rb)\nonumber\\
&&\hspace{-0.75cm}\lb.
\Lambda^{\half}\lb(\lb\{\xmgenn+\frac{1}{T}\lb(\pmgen-i\pi\nmgen\rb)+\frac{T}{4}\lb[\sgn(\snt)\Dtmt-\sgn(\sn)\Dtm\rb]\rb\},\half\sgn(\sn)\e{}{\Kn}{v},\nn\rb)\rb)\nonumber\\
\ea
Here we combined all the exponentials that do not depend on $\ngen$ of the various edges to a compact form summing over $\vt$ and $J,j,\sigma$ again.\\ 
In section \ref{nNull} in appendix we show that the only term that contributes to the infinite sum over $\nmgen$ is the term with $\nmgen=0$ all other term are of the order $O(t^{\infty})$. Hence, up to order $O(t^{\infty})$ the expectation value is given by
\ba
\label{ExpnNull}
\lefteqn{\MeO}\nonumber\\
&=&
\frac{e^{+i\sum\limits_{\vt\in V}\sum\limits_{(J,\sigma,j)\atop\in L}\phigen\Del}e^{-\frac{t}{4}\sum\limits_{\vt\in V}\sum\limits_{(J,\sigma,j)\atop\in L}\lb(\Del\rb)^2}}
{(\sqrt{\frac{\pi}{t}})^{30}[1+K_t(p)]^{30}}\nonumber\\
&&\frac{(\sqrt{\pi})^{12}}{T^{30}}2^9\lb(\sqrt{\frac{\pi}{2}}\rb)^9
\int\limits_{\Rl^9}d^9\xmgenn
e^{-2\sum\limits_{(J,j)\atop\in L_v}\left(\xmgenn\right)^2}\nonumber\\ 
&&\hspace{-0.75cm}
\Lambda^{\half}\lb(\lb\{\xmgenn+\frac{1}{T}\lb(\pmgen\rb)-\frac{T}{4}\lb[\sgn(\snt)\Dtmt-\sgn(\sn)\Dtm\rb]\rb\},\half\sgn(\snt)\e{}{\Knt}{v},\nnt\rb)\nonumber\\
&&\hspace{-0.75cm}
\Lambda^{\half}\lb(\lb\{\xmgenn+\frac{1}{T}\lb(\pmgen\rb)+\frac{T}{4}\lb[\sgn(\snt)\Dtmt-\sgn(\sn)\Dtm\rb]\rb\},\half\sgn(\sn)\e{}{\Kn}{v},\nn\rb)\nonumber\\
\ea
\subsection{Expansion of the $\Lambda^{\half}$-Functions}
Although the expectation value simplifies a lot when considerung the $\ngen=0$ tern only, the integral in eqn (\ref{ExpnNull}) cannot be performed analytically due to the occurring $\Lambda^{\half}$-functions. The way out of this problem is to expand these functions in terms of powers of $\xmgenn$ which yields integrals of the form $\int\limits_{\Rl^9}d^9\xmgenn e^{-2(\xmgenn)^2}(\xmgenn)^k$ with $k\in\Nl$ that can be solved analytically. Here we will use the same technique that was introduced in \cite{Hanno}.
Recalling again the definition of $\Lambda^{\half}$ in terms of the determinants
and furthermore introducing the dimensionless matrix
\be
\qmgen:=\pmgen t^{-\alpha},\quad\quad\mathrm{with}\quad s=t^{\half-\alpha}
\ee
where $t^{\alpha}=\frac{\epsilon^2}{a^2}$. This relation takes its origin in the analysis of the estimation of the disretisation as well as the quantisation error. Roughly speaking, on the one hand the discretisation error will be proportional to $(\frac{\epsilon}{a})^n$ where $n>0$. On the other hand we have the quantum fluctuation that are proportional to $\frac{t}{(\frac{\epsilon}{a})^m}$ with $m>0$. Thus, the discretisation error decreases when $\epsilon$ gets smaller, while the quantum fluctuation error increases and might even diverge in the limit $\epsilon\to 0$. Therefore, we can conclude that the total error will be minimised for $\frac{\epsilon}{a}\propto t^{\frac{\alpha}{2}}$ for some $\alpha>0$. The concrete value of  the optimal $\alpha$ will strongly depend on the fact whether one is interested in a very small classical error and accepts larger, but still finite quantum fluctuation, or one wants to keep the quantum fluctuations as small as possible and deals with a larger discretisation error, or one takes the point of view that both errors should be treated with equal weight. In \cite{ITP} a value of $\alpha=\frac{1}{6}$ was proposed by having made a rough estimate. We do not want to fix the value of $\alpha$ here, but rahter keep it as long as possible as general as possible.\\
In any case, with the help of $\qmgen$, we can rewrite $\Lambda^{\half}$ as
\ba
\lefteqn{\Lambda^{\half}\lb(\lb\{\xmgenn+\frac{1}{T}\lb(\pmgen\rb)+\frac{T}{4}\lb[\sgn(\snt)\Dtmt-\sgn(\sn)\Dtm\rb]\rb\},\half\sgn(\sn)\e{}{\Kn}{v},\nn\rb)}\nonumber\\
&=&\frac{t^{\frac{3}{8}}}{T^{\frac{3}{4}}}\lb(\frac{a^{\frac{3}{2}}}{i\hbar}\rb)\lb(|\det(\pmm)|^{\frac{1}{4}}\rb)\nonumber\\
&&\lb[\lb|\det\lb(1+s\qmm\xmm+\frac{T}{4}s\qmm\lb[\sgn(\snt)\Dtmt-\sgn(\sn)\Dtm\rb]\rb)\rb|^{\frac{1}{4}}\rb.\nonumber\\
&&\lb. 
-\lb|\det\lb(1+s\qmm\xmm+\frac{T}{4}s\qmm\lb[\sgn(\snt)\Dtmt-\sgn(\sn)\Dtm\rb]+\frac{T}{2}s\qmm\sgn(\sn)\dmgen\rb)\rb|^{\frac{1}{4}}\rb]\nonumber\\
\ea
The terms involving determinants can be further expanded by using the following identity:
\be
\label{Expdet}
\det(1+A)^2=1+\underbrace{2z^{\prime}_A+(z^{\prime}_A)^2}_{\displaystyle =:z_A}=:1+z_A,
\quad\mathrm{with}\quad
z^{\prime}_A=\tr(A)+\half\lb(\lb[\tr(A)\rb]^2-\tr(A^2)\rb)+\det(A)
\ee
In our case we have to consider four different $A$ matrices. Let us denote them by $A_1:=sq^{-1}x+\Delta$, $A_2:=sq^{-1}x-\Delta$, $A_3:=sq^{-1}x+\Delta+\delta$ and  $A_4:=sq^{-1}x-\Delta+\delta$. Thus, we need the explicit expressions for $z_{sq^{-1}x+\Delta}$, $z_{sq^{-1}x-\Delta}$, $z_{sq^{-1}x+\Delta+\delta}$ and $z_{sq^{-1}x-\Delta+\delta}$ in order to expand all four determinants contained in the two $\Lambda^{\half}$-functions. These explicit expressions are derived in section \ref{zA} in appendix.\\
We then define
\ba
y&:=&1+z_{sq^{-1}x+\Delta}\quad\quad y_1:=1+z_{sq^{-1}x+\Delta+\delta}\nonumber\\
\wt{y}&:=&1+z_{sq^{-1}x-\Delta}\quad\quad \wt{y}_1:=1+z_{sq^{-1}x-\Delta+\tilde{\delta}}
\ea
and due to eqn (\ref{Expdet}), we can express the $\Lambda^{\half}$-functions as
\ba
\lefteqn{\Lambda^{\half}\lb(\lb\{\xmgenn+\frac{1}{T}\lb(\pmgen\rb)+\frac{T}{4}\lb[\sgn(\snt)\Dtmt-\sgn(\sn)\Dtm\rb]\rb\},\half\sgn(\sn)
\e{}{\Kn}{v},\nn\rb)}\nonumber\\
&=&
\frac{a^{\frac{3}{2}}\lb|\det(\pmm)\rb|^{\frac{1}{4}}}{i\hbar}\lb(y^{\frac{1}{8}}-y_1^{\frac{1}{8}}\rb)
\hspace{9.5cm}
\ea
and
\ba
\lefteqn{\Lambda^{\half}\lb(\lb\{\xmgenn+\frac{1}{T}\lb(\pmgen\rb)-\frac{T}{4}\lb[\sgn(\snt)\Dtmt-\sgn(\sn)\Dtm\rb]\rb\},\half\sgn(\snt)\e{}{\Knt}{v},\nnt\rb)}\nonumber\\
&=&
\frac{a^{\frac{3}{2}}\lb|\det(\pmm)\rb|^{\frac{1}{4}}}{(-i\hbar)}\lb(\wt{y}^{\frac{1}{8}}-\wt{y}_1^{\frac{1}{8}}\rb)\hspace{9.5cm}
\ea
In order to continue the calculation we will expand $y^{\frac{1}{8}},y_1^{\frac{1}{8}},\wt{y}^{\frac{1}{8}},\wt{y}_1^{\frac{1}{8}}$ around $y=y_1=\wt{y}=\wt{y}_1=1$. Here we follow the guideline given in \cite{Hanno}. The general expansion yields
\ba
\label{GenExp}
\lefteqn{\Lambda^{\half}\lb(\lb\{\xmgenn+\frac{1}{T}\lb(\pmgen\rb)+\frac{T}{4}\lb[\sgn(\snt)\Dtmt-\sgn(\sn)\Dtm\rb]\rb\},\half\sgn(\sn)\e{}{\Kn}{v},\nn\rb)}\nonumber\\
&=&\frac{a^{\frac{3}{2}}\lb|\det(\pmm)\rb|^{\frac{1}{4}}}{i\hbar}\nonumber\\
&&\Big\{(y-y_1)\lb(\sum\limits_{k=1}^n f^{(k)}_{\frac{1}{8}}(1)
\sum\limits_{l=0}^{k-1}(y-1)^l(y_1-1)^{k-1-l}\rb)
+\underbrace{\Big[f^{(n+1)}_{\frac{1}{8}}(y)(y-1)^{n+1} - f^{(n+1)}_{\frac{1}{8}}(y_1)(y_1-1)^{(n+1)}\Big]}_{\displaystyle\mathrm{remainder}}\Big\}\nonumber\\
\ea
and similar for the second $\Lambda^{\half}$-functions where $y,y_1$ are replaced by $\wt{y},\wt{y}_1$. 
Here, we will be  only interested in the semiclassical limit (leading order) and the first quantum correction (next-to-leading order). Noting that $t=\lp^2/a^2$ and therefore $\hbar=t/\kappa a^2$, we obtain the following power counting in $s$ for each single $\Lambda^{\half}$-function
\ba
\label{powerc}
\lefteqn{\Lambda^{\half}\lb(\lb\{\xmgenn+\frac{1}{T}\lb(\pmgen\rb)+\frac{T}{4}\lb[\sgn(\snt)\Dtmt-\sgn(\sn)\Dtm\rb]\rb\},\half\sgn(\sn)\e{}{\Kn}{v},\nn\rb)}\nonumber\\
&=&\frac{sT}{t}\lb(1+ s\xmgen +s^2(\xmgen)^2 +O(sT)\rb)\hspace{5cm}
\ea
The fact whether the $s^3$- or the $sT$-contribution is the next-to-next-to leading order term depends on the value of $\alpha$. The quotient $sT/s^3=t^{\half-2\alpha}$ is small aslong as $\alpha\le\frac{1}{4}$. When $\alpha$ passes the value of $\frac{1}{4}$ the $s^3$ contribution becomes larger than the corresponding one coming from $sT$. Similar to \cite{Hanno} we consider the $sT$-contribution as the next-to-next-to-leading order (NNLO) term. If one wants to work with an $\alpha$ being greater than $\frac{1}{4}$
 one should replace $O(sT)$ by $O(s^3)$ in the power counting above. Since the expectation value contains a product of two $\Lambda^{\half}$-functions, we will expand the expectation value up to order $(\frac{sT}{t})^2s^2$.
In section \ref{yExp} in appendix is shown that as long as $n\le n_0$ the integral over the remainder when the expansion is reinserted into the expectation value is smaller than the $s^{n}$ contribution. As mentioned in section \ref{yExp} the precise value of $n_0$ will depend on the chosen value of $\alpha$. In our case, we have to ensure that when expanding up to order $s^{n^{\prime}}$ with $n^{\prime}>2$ that $s^{n^{\prime}+1}\ll sTs^2$. This is equivalent to the condition 
$s^{n^{\prime}-2}\ll T=t^{\frac{1}{2}}$ from which the minimal value of $n^{\prime}$ can be computed. The result reads
$n^{\prime}>\frac{\half+1-2\alpha}{\half-\alpha}$. For instance for the suggested value of $\alpha=\frac{1}{6}$ in \cite{ITP}, the minimal value of $n^{\prime}$ is $n^{\prime}=4$ which is well below the value of $n_0\gg 1$. Thus the error of neglecting the remainder is indeed of higher order in $s$ than $(\frac{sT}{t})^2s^2$. Moreover, when $\alpha$ is coming closer to the value of $\frac{1}{2}$, the value of $n^{\prime}$ increases strongly and a carefull analysis whether $n^{\prime}<n_0$ is necessary.
\\ \\
In appendix in section \ref{Expy} we derive the explicit forms of the $y,y_1,\wt{y}$ and $\wt{y}_1$ up to the necessary orders. It turns out that the lowest contribution in the terms $(y-y_1)$ and $(\wt{y}-\wt{y}_1)$ respectively is already of the order $sT$. The highest order we want to consider is the next-to-leading order term of order $(s^3T/t)$. Since all other terms occurring in the $\Lambda^{\half}$-expansion are multiplied by $(y-y_1)$ and $(\wt{y}-\wt{y}_1)$ respectively, we expand $(y-y_1)$ and $(\wt{y}-\wt{y}_1)$ respectively up to order $(s^3T/t)$ and all the other terms occurring in the expansion up to order $O(s^3)$\footnote{Note, that the proposed value of $\alpha=\frac{1}{6}\le\frac{1}{4}$ in \cite{ITP} corresponds to a NNLO-term  of order $sT$.}. \\
The expansion of the $\Lambda^{\half}$-functions up to $O((sT)^2/t)$ yields
\ba
\label{ExpL1}
\lefteqn{\Lambda^{\half}\lb(\lb\{\xmgenn+\frac{1}{T}\lb(\pmgen\rb)+\frac{T}{4}\lb[\sgn(\snt)\Dtmt-\sgn(\sn)\Dtm\rb]\rb\},\half\sgn(\sn)\e{}{\Kn}{v},\nn\rb)}\nonumber\\
&=&\frac{a^{\frac{3}{2}}|\det(\pmm)|^{\frac{1}{4}}}{i\hbar}
(y-y_1)\lb\{\lb(f^{(1)}_{\frac{1}{8}}(1)+f^{(2)}_{\frac{1}{8}}(1)\lb(2(y-1)+(y_1-y)\rb)\rb.\rb.\hspace{2cm}\nonumber\\
&&\lb.\lb.
+f^{(3)}_{\frac{1}{8}}(1)\lb(3(y_1-1)^2+3(y-1)(y_1-1)+(y_1-y)^2\rb)\rb)\rb\}+O((sT)^2/t)
\ea
and
\ba
\label{ExpL2}
\lefteqn{\Lambda^{\half}\lb(\lb\{\xmgenn+\frac{1}{T}\lb(\pmgen\rb)-\frac{T}{4}\lb[\sgn(\snt)\Dtmt-\sgn(\sn)\Dtm\rb]\rb\},\half\sgn(\snt)\e{}{\Knt}{v},\nnt\rb)}\nonumber\\
&=&\frac{a^{\frac{3}{2}}|\det(\pmm)|^{\frac{1}{4}}}{(-i\hbar)}
(\wt{y}-\wt{y}_1)\lb\{\lb(f^{(1)}_{\frac{1}{8}}(1)+f^{(2)}_{\frac{1}{8}}(1)\lb(2(\wt{y}-1)+(\wt{y}_1-\wt{y})\rb)\rb.\rb.\hspace{2cm}\nonumber\\
&&\lb.\lb.
+f^{(3)}_{\frac{1}{8}}(1)\lb(3(\wt{y}_1-1)^2+3(\wt{y}-1)(\wt{y}_1-1)+(\wt{y}-\wt{y})^2\rb)\rb)\rb\}+O((sT)^2/t)
\ea
\section{Leading Order of the Expectation Value}
Let us summarise the structure of our calculation. The aim is to calculate the expectation value of the algebraic Master constraint operator $\MCO_v$ with respect to certain coherent states. We have introduced an operator $\Omn$ which has the advantage that the expectation value of $\MCO_v$ can be expressed in terms of a sum over $\In,\Jn\,\Kn,\Int,\Jnt,\Knt$ of expectation values of $(\Omnt)^{\dagger}\Omn$. By actually analysing the expectation value of $(\Omnt)^{\dagger}\Omn$ we saw that the so called $\Lambda^{\half}$-function occurring in the expectation value cannot be integrated analytically. Therefore, we are forced to expand these functions in terms of powers as $sT/t(1+s\xmm+s^2(\xmm)^2+O(sT))$. \\
In this section we are interested in the leading order of the expectation value of $\MCO_v$. 
Consequently, we need the leading order of $\Lambda^{\half}$ in order to calculate the leading order of the expectation value of $(\Omnt)^{\dagger}\Omn$. With the knowledge of the result of the expectation value of $(\Omnt)^{\dagger}\Omn$, finally we are able to give an expression for the leading order of the expectation value of $\MCO_v$.
\\ \\
 The detailed analysis of the explicit expressions for $y,y_1,\wt{y}$ and $\wt{y}_1$ which are derived in section \ref{Expy} in appendix shows that the leading of $\Lambda^{\half}$ is of order $sT/t$. This is due to the fact that any term in the expansion is multiplied by a term of the form  $(y-y_1)$ and $(\wt{y}-\wt{y}_1)$ respectively. The lowest order of this terms is $sT$ which together with the $1/\hbar\propto1/t$ in eqn (\ref{ExpL1}) and (\ref{ExpL2}) respectively combines into terms of the order $sT/t$. The explicit expressions are given by
\ba
y-y_1\Big|_{sT}&=&-sT\sgn(\sn)\qmm_{\Kn\nn}\nonumber\\
\wt{y}-\wt{y}_1\Big|_{sT}&=&-sT\sgn(\snt)\qmm_{\Knt\nnt}
\ea 
Hence, the leading order expansion for the $\Lambda^{\half}$-functions are
\ba
\lefteqn{\Lambda^{\half}\lb(\lb\{\xmgenn+\frac{1}{T}\lb(\pmgen\rb)+\frac{T}{4}\lb[\sgn(\snt)\Dtmt-\sgn(\sn)\Dtm\rb]\rb\},\half\sgn(\sn)\e{}{\Kn}{v},\nn\rb)}\nonumber\\
&=&\frac{a^{\frac{3}{2}}|\det(\pmm)|^{\frac{1}{4}}}{i\hbar}f^{(1)}_{\frac{1}{8}}(1)\lb(-sT\sgn(\sn)\qmm_{\Kn\nn}\rb)\hspace{6cm}
\ea
and
\ba
\lefteqn{\Lambda^{\half}\lb(\lb\{\xmgenn+\frac{1}{T}\lb(\pmgen\rb)-\frac{T}{4}\lb[\sgn(\snt)\Dtmt-\sgn(\sn)\Dtm\rb]\rb\},\half\sgn(\snt)\e{}{\Knt}{v},\nnt\rb)}\nonumber\\
&=&\frac{a^{\frac{3}{2}}|\det(\pmm)|^{\frac{1}{4}}}{(-i\hbar)}f^{(1)}_{\frac{1}{8}}(1)\lb(-sT\sgn(\snt)\qmm_{\Knt\nnt}\rb)\hspace{6cm}
\ea
We know that the expectation value of $(\Omnt)^{\dagger}\Omn$ contains a product of these two leading order $\Lambda^{\half}$-functions. Consequently, the leading order of the expectation value will be of the order $O((sT/t)^2$.
Reinserting the leading order $\Lambda^{\half}$ into the expression of the expectation value of the $(\Omnt)^{\dagger}\Omn$ leads to
\ba
\lefteqn{\MeO}\nonumber\\
&=&\frac{e^{+i\sum\limits_{\vt\in V}\sum\limits_{(J,\sigma,j)\atop\in L}\phigen\Del}e^{-\frac{t}{4}\sum\limits_{\vt\in V}\sum\limits_{(J,\sigma,j)\atop\in L}\lb(\Del\rb)^2}}
{(\sqrt{\frac{\pi}{t}})^{30}[1+K_t(p)]^{30}}\nonumber\\
&&\frac{(\sqrt{\pi})^{12}}{T^{30}}2^9\lb(\sqrt{\frac{\pi}{2}}\rb)^9
\int\limits_{\Rl^9}d^9\xmgenn
e^{-2\sum\limits_{(J,j)\atop\in L_v}\left(\xmgen\right)^2}\nonumber\\
&&
\lb(\frac{a^{\frac{3}{2}}|\det(\pmm)|^{\frac{1}{4}}}{\hbar}\rb)^2\lb(f^{(1)}_{\frac{1}{8}}(1)\rb)^2\lb(-sT\sgn(\sn)\qmm_{\Kn\nn}\rb)
\lb(-sT\sgn(\snt)\qmm_{\Knt\nnt}\rb)\nonumber\\
&=&e^{+i\sum\limits_{\vt\in V}\sum\limits_{(J,\sigma,j)\atop\in L}\phigen\Del}\nonumber\\
&&
\lb(\frac{a^{\frac{3}{2}}|\det(\pmm)|^{\frac{1}{2}}}{\hbar}\rb)^2\lb(sT\rb)^2\lb(f^{(1)}_{\frac{1}{8}}(1)\rb)^2
\lb(\sgn(\sn)\qmm_{\Kn\nn}\rb)\lb(\sgn(\snt)\qmm_{\Knt\nnt}\rb)\nonumber\\
\ea
In the last step we expanded the exp-function
\ba
\lefteqn{e^{-\frac{t}{4}\sum\limits_{\vt\in V}\sum\limits_{(J,\sigma,j)\atop\in L}\lb(\Del\rb)^2}}\nonumber\\
&=&1-\frac{t}{4}\sum\limits_{\vt\in V}\sum\limits_{(J,\sigma,j)\atop\in L}\lb(\Del\rb)^2 +...\hspace{5cm}
\ea
and took only the first $1$ in the expansion since we are collecting terms of order $(sT/t)^2$ only. Finally, we use the above result in order to build the expectation value of $\MCO_v$ out of it. This yields
\ba
\lefteqn{\frac{\langle\Psigen\,|\,\MCO_{v}\,|\,\Psigen\rangle}{||\Psigen||^2}}\nonumber\\
&=&
\sum\limits_{\In\Jn\Kn}\sum\limits_{\Int\Jnt\Knt}\sum\limits_{\sn=+,-}\sum\limits_{\snt=+,-}\epsilon^{\In\Jn\Kn}\epsilon^{\Int\Jnt\Knt}\lb(\frac{4}{\kappa}\rb)^2\nonumber\\
&&
\Big\{
\lb(\delta_{\mn,\nn}\delta_{\mnt,\nnt}
+\sum\limits_{\elln=1}^3 \epsilon_{\elln \mn\nn}\epsilon_{\elln \mnt\nnt}\rb)
\lb(\frac{a^{\frac{3}{2}}|\det(\pmm)|^{\frac{1}{4}}}{\hbar}\rb)^2\lb(sT\rb)^2e^{+i\sum\limits_{(J,\sigma,j)}\phigen\Del}\nonumber\\
&&\lb(f^{(1)}_{\frac{1}{8}}(1)\rb)^2\lb(\sgn(\sn)\qmm_{\Kn\nn}\rb)\lb(\sgn(\snt)\qmm_{\Knt\nnt}\rb)\Big\}+O((sT/t)^2)
\ea
\section{(One) Semiclassical Limit of Algebraic Quantum Gravity}
Recall the philosophy of Algebraic Quantum Gravity: On the algebraic level we have an algebraic Master constraint operator $\MCO$ which acts on algebraic graphs $\alpha$. Furthermore we can define coherent states associated with an embedded image $\gamma=X(\alpha)$ of the algebraic graph. Hence, when the coherent states enter the game, the missing information such as the topology, the differential structure and the background metric to be approximated are encoded in these coherent states. This has the consequence, that Algebraic Quantum Gravity has not one single semiclassical limit, rather for each set of coherent states that represent a different topology, a different differential structure and a different background metric to approximate the semiclassical limit will be different in general. 
\subsection{Comparison of the Leading Order Expectation Value with the Classical (Discretised) Master Constraint}
In this section we want to show that the leading order of the expectation value of the algebraic Master constraint with respect to the coherent states used in the calculations can indeed be interpreted as the classical Master constraint of General Relativity. Hence, we can demonstrate that there exists coherent states such that the semiclassical limit of Algebraic Quantum Gravity reproduces the infinitesimal generators of General Relativity. Thus, the problem whether the semiclassical sector includes General Relativity, that is still unsolved within the framework of Loop Quantum Gravity, is significantly improved  in the context of Algebraic Quantum Gravity.\\ \\
In order to show that the semiclassical limit of the expectation value of the algebraic Master constraint with respect to these coherent states associated to a cubic graph $\alpha$ is the classical Master constraint (associated with a cubic graph), we will use three steps.
\begin{itemize}
\item[1.)] 
Recall that the coherent states are labelled by the so called classicality parameter $t$. Thus we will take the limit $\lim_{t\to 0}< \MCO>_{t}$ in order to substract the semiclassical limit out of the expectation value.
\item[2.)] We will show that  $\lim_{t\to 0}< \MCO>_{t}=\sum\limits_{v\in E(\alpha)}\MC^{cubic}_v=:\MC^{cubic}$ whereby  $\MC^{cubic}$ can be interpreted as a discretised version of the classical Master constraint on a lattice with cubic symmetry, i.e. a cubic graph with infinitely many edges.
\item[3.)] We will investigate the limit of $\MC^{cubic}$ in which we shrink the lattice length $\epsilon$ to zero and show that this limit is exactly the continuum expression $M$, i.e. the classical Master constraint. Note, that since $\MC^{cubic}$ does not depend on the lattice length explicitly due to the fundamental background independence of the theory, this limit is rather easy to take.
\end{itemize}
\underline{Step1:}\\
This was already done in the last section when actually calculating the leading order contribution of the expectation value of the algebraic Master constraint operator $\MCO$. We therefore take the result of the last sections as our starting point and proceed with step 2.
\\
\underline{Step 2:}
\\
The discretisation of the classical Master constraint on a cubic graph is given by
\be
\MC^{cubic}=\sum\limits_{v\in V(\alpha)} \MC^{cubic}_v\quad\quad \MC^{cubic}_v=\sum\limits_{\mu=0}^3 \lb[C^{cubic}_{\mu,v}\rb]^2
\ee
where 
\ba
C^{cubic}_{0,v}&=&\sum\limits_{\In\Jn\Kn}\sum\limits_{\sn=+,-}\frac{4}{\kappa}a^{\frac{3}{2}}\epsilon^{\In\Jn\Kn}\cholloop{\In}{\Jn}{\sn}{\mn}{v}\chol{\Kn}{\sn}{\mn}{v}
\{\cholin{\Kn}{\sn}{\mn}{v},V^{\half}_{\alpha,v}\}\nonumber\\
C^{cubic}_{\elln,v}&=&\sum\limits_{\In\Jn\Kn}\sum\limits_{\sn=+,-}\frac{4}{\kappa}a^{\frac{3}{2}}\epsilon_{\elln\mn\nn}\epsilon^{\In\Jn\Kn}\cholloop{\In}{\Jn}{\sn}{\mn}{v}\chol{\Kn}{\sn}{\nn}{v}
\{\cholin{\Kn}{\sn}{\nn}{v},V^{\half}_{\alpha,v}\}
\ea
where $V_{\alpha,v}$ denotes the dimensionless volume of a cube centered around the vertex $v$ with edge parameter length $\epsilon$.
These cube can be parameterised by an embedding $X^a_v:[-\frac{\epsilon}{2},+\frac{\epsilon}{2}]^3\to\sigma$ with $(t^1,t^2,t^3)\mapsto X^a_v(t^1,t^2,t^3)$ and $X^a_v(0)=v$. The dimensionless volume 
can be expressed  in terms of the dimensionless fluxes $p_{J\sigma jv}$
\be
V_{\alpha,v}=\sqrt{\lb|\epsilon^{jkl}\lb(\frac{p_{1+jv}-p_{1-jv}}{2}\rb)\lb(\frac{p_{2+kv}-p_{2-kv}}{2}\rb)\lb(\frac{p_{3+lv}-p_{3-lv}}{2}\rb)\rb|}
\ee
Recall that in our notation $p_{1+jv}$ denotes the dimensionless the $j$ component of the flux through the surface $S_{e^{+}_1(v)}$ etc.
Introducing $(p_{Jjv})^{-}:=\frac{1}{2}\lb(p_{J+jv}-p_{J-jv}\rb)$ the volume can be reexpressed as
\be
V_{\alpha,v}=\sqrt{\lb|\det((p_{Jjv})^{-})\rb|}
\ee
We will show:
\ba
\cholloop{\In}{\Jn}{\sn}{\mn}{v}h_{\Kn\sn\nn v}\{(h_{\Kn\sn\nn v})^{-1},V^{\half}_{\alpha,v}\}
&=&
i\frac{\kappa}{8a^2}\cholloop{\In}{\Jn}{\mn}{\sn}{v}\lb|\det((p_{J\sigma j v})^{-})\rb|^{\frac{1}{4}}\lb[(p_{\Kn\sn\nn v})^{-}\rb]^{-1}
\ea,
where
\be
\cholloop{\In}{\Jn}{\sn}{\mn}{v}=e^{+i\lb(\varphi_{\In\sn\mn v}+\varphi_{\Jn\sn\mn  v+\sn\hat{\In}}-\varphi_{\Jn\sn\mn v}-\varphi_{\In\sn\mn v+\sn\hat{\Jn})}\rb)}
\ee
The classical Poisson bracket reads
\ba
h_{\Kn\sn\nn v}\{(h_{\Kn\sn\nn v})^{-1},V^{\half}_{\alpha,v}\}&=&\kappa h_{\Kn\sn\nn v}\int d^3z\lb(\frac{\delta\lb(h_{\Kn\sn\nn v}\rb)^{-1}}{\delta A^k_b(z)}\rb)\lb(\frac{\delta V^{\half}{\alpha,v}((p_{Jjv})^{-})}{\delta E^b_k(z)}\rb)
\ea
We have
\ba
h_{\Kn\sn\nn v}\lb(\frac{\delta\lb(h_{\Kn\sn\nn v}\rb)^{-1}}{\delta A^k_b(z)}\rb)&=&
i\int\limits_0^1dt\dot{e}^{\sn}_{\Kn}(t)\delta^{k}_{\nn}\delta^{b}_a\delta(t,z)\nonumber\\
\lb(\frac{\delta V^{\half}_{\alpha,v}((p_{J\sigma j v})^{-})}{\delta E^b_k(z)}\rb)
&=&
\frac{1}{4}\lb|\det((p_{Jjv})^{-})\rb|^{-\frac{3}{4}}\sgn(\det((p_{Jjv})^{-}))\frac{\delta \det((p_{Jjv})^{-})}{\delta E^b_k(z)}
\ea
Recalling
\be
(p_{Jjv})^{-}=\frac{1}{2a^2}\Big(\int\limits_{S_{e^{+}_{J}}} n^{S_{e^{+}_{J}}}_a E^a_j -\int\limits_{S_{e^{-}_{J}}} n^{S_{e^{-}_{J}}}_a E^a_j\Big)
\ee
we obtain
\ba
\lefteqn{\frac{\delta \det((p_{Jjv})^{-})}{\delta E^b_k(z)}=
\frac{\det((p_{Jjv})^{-})}{2a^2}(p^{-1}_{Iiv})^{-}}\nonumber\\
&&
\Big(\int\limits_{S_{e^{+}_{I}}}d^2u\, n^{S_{e^{+}_{I}}}(u_1,u_2)\delta^a_{b}\delta^{k}_i\delta(x(u),z)
-\int\limits_{S_{e^{-}_{I}}}d^2u\, n^{S_{e^{-}_{I}}}(u_1,u_2)\delta^a_{b}\delta^{k}_i\delta(x(u),z)\Big)\nonumber\\
\ea
Thus, we get
\ba
\lefteqn{\cholloop{\In}{\Jn}{\sn}{\mn}{v}h_{\Kn\nn\sn v}\{h^{-1}_{\Kn\nn\sn v},V^{\half}_{\alpha,v}\}}\nonumber\\
&=&
\frac{i\kappa}{8a^2}\cholloop{\In}{\Jn}{\sn}{\mn}{v}\lb|\det((p_{Jjv})^{-})\rb|^{-\frac{3}{4}}\sgn(\det((p_{Jjv})^{-})\det((p_{Jjv})^{-})(p^{-1}_{Iiv})^{-}\hspace{5cm}\nonumber\\
&&\underbrace{\Big(
\int\limits_0^1 dt\int\limits_{S_{e^{+}_{I}}} d^2u\dot{e}^{\sn}_{\Kn}(t) n^{S_{e^{+}_{I}}}_a(u_1,u_2)\delta(x(u),z)\delta^{k}_{\nn}\delta^{i}_{k}\delta^{b}_{a}\delta^{a}_{b}}_{\displaystyle \delta_{(\Kn,\sn,\nn,v)(I,+,i,v)}}
-\underbrace{\int\limits_0^1 dt\int\limits_{S_{e^{-}_{I}}} d^2u\dot{e}^{\sn}_{\Kn}(t) n^{S_{e^{-}_{I}}}_a(u_1,u_2)\delta(x(u),z)\delta^{k}_{\nn}\delta^{i}_{k}\delta^{b}_{a}\delta^{a}_{b}\Big)}_{\displaystyle \delta_{(\Kn,\sn,\nn,v)(I,-,i,v)}}\nonumber\\
&=&\frac{i\kappa}{8a^2}\cholloop{\In}{\Jn}{\sn}{\mn}{v}\lb|\det((p_{Jjv})^{-})\rb|^{\frac{1}{4}}\sgn(\sn)(p^{-}_{\Kn\nn v})^{-}
\ea
Consequently, we have
\ba
\label{Classical}
\lefteqn{\lb(a^{\frac{3}{2}}\cholloop{\In}{\Jn}{\snt}{\mn}{v}h_{\Kn\nn\sn v}\{h^{-1}_{\Kn\nn\sn v},V^{\half}_{\alpha,v}\}\rb)^*a^{\frac{3}{2}}\cholloop{\In}{\Jn}{\snt}{\mn}{v}h_{\Kn\nn\sn v}\{h^{-1}_{\Kn\nn\sn v},V^{\half}_{\alpha,v}\}}\nonumber\\
&=&
\lb(\frac{\kappa a^{\frac{3}{2}}}{8a^2}\rb)^2\lb(\cholloop{\Int}{\Jnt}{\snt}{\mnt}{v}\rb)^{-1}\holloop{\In}{\Jn}{\sn}{\mn}{v}
\lb(\lb|\det((p_{Jjv})^-)\rb|^{\frac{1}{4}}\rb)^2\sgn(\sn)(p^{-1}_{\Kn\nn v})^-\sgn(\snt)(p^{-1}_{\Knt\nnt v})^-\nonumber\\
\ea
From our semiclassical calculation we get
\ba
\lefteqn{\MeO}\nonumber\\
&=&
e^{+i\sum\limits_{\vt\in V}\sum\limits_{(J,\sigma,j)\atop\in L}\phigen\Del}\Big(\frac{a^{\frac{3}{2}}}{\hbar}\lb|\det(\pmgen)\rb|^{\frac{1}{4}}\Big)^2\nonumber\\
&&(-sT)f^{(1)}_{\frac{1}{8}}(1)\sgn(\sn)(\qmm)_{\Kn\nn}
(-sT)f^{(1)}_{\frac{1}{8}}(1)\sgn(\snt)(\qmm)_{\Knt\nnt}\hspace{4cm}
\ea
Therefore, we have to show that the result above agrees with the expression in eqn (\ref{Classical}) in order to show the correctness of the leading order.\\
Using
\be
(q^{-1}_{\Kn\nn v})^-=p^{-1}_{\Kn\nn v}t^{\alpha}
\ee
we obtain
\ba
\lefteqn{\MeO}\nonumber\\&=&
\lb(\frac{a^{\frac{3}{2}}}{\hbar}\rb)^2e^{+i\sum\limits_{\vt\in V}\sum\limits_{(J,\sigma,j)\atop\in L}\phigen\Del}\lb|\det(\pmgen)\rb|^{\frac{1}{2}}t^{2\alpha}\nonumber\\
&&(-sT)f^{(1)}_{\frac{1}{8}}\sgn(\sn)(p^{-1}_{\Kn\nn v})^-
(-sT)f^{(1)}_{\frac{1}{8}}\sgn(\snt)(p^{-1}_{\Knt\nnt v})^-\nonumber\\
&=&
\lb(\frac{t^{2\alpha}a^{\frac{3}{2}}}{\hbar}\rb)^2e^{+i\sum\limits_{\vt\in V}\sum\limits_{(J,\sigma,j)\atop\in L}\phigen\Del}\lb|\det(\pmgen)\rb|^{\frac{1}{2}}\lb(t^{\half-\alpha}t^{\half}\rb)^2\nonumber\\
&&\frac{1}{8}\sgn(\sn)\lb((p_{\Kn\nn v})^-\rb)^{-1}
\frac{1}{8}\sgn(\snt)\lb((p_{\Knt\nnt v})^-\rb)^{-1}\nonumber\\
&=&\lb(\frac{\lp^2a^{\frac{3}{2}}}{8\hbar a^2}\rb)^2
e^{+i\sum\limits_{\vt\in V}\sum\limits_{(J,\sigma,j)\atop\in L}\phigen\Del}\lb|\det(\pmgen)\rb|^{\frac{1}{2}}\nonumber\\
&&\sgn(\sn)(p^{-1}_{\Kn\nn v})^-\sgn(\sn)(p^{-1}_{\Knt\nnt v})^-\nonumber\\
&=&
\lb(\frac{\kappa a^{\frac{3}{2}}}{8a^2}\rb)^2e^{+i\sum\limits_{\vt\in V}\sum\limits_{(J,\sigma,j)\atop\in L}\phigen\Del}\lb|\det(\pmgen)\rb|^{\frac{1}{2}}\nonumber\\
&&\sgn(\sn)(p^{-1}_{\Kn\nn v})^-\sgn(\snt)(p^{-1}_{\Knt\nnt v})^-
\ea
whereby we used in the second line $s=t^{\half-\alpha},T=t^{\half}$, in the third line $ta^2=\lp^2$ and in the last step the fact that $\kappa=\lp^2/\hbar$.\\
Using the explicit definition of $\Del$, we get
\ba
e^{+i\sum\limits_{\vt\in V}\sum\limits_{(J,\sigma,j)\atop\in L}\phigen\Del}
&=&e^{+i\lb(\varphi_{\Jnt\snt\mnt v}+\varphi_{\Int\snt\mnt v+\snt\hat{\Jnt}}-\varphi_{\Int\snt\mnt v}+\varphi_{\Jnt\snt\mnt v+\snt\hat{\Int}}\rb)}\nonumber\\
&&e^{-i\lb(\varphi_{\Jn\sn\mn v}+\varphi_{\In\sn\mn v+\sn\hat{\Jn}}-\varphi_{\In\sn\mn v}+\varphi_{\Jn\sn\mn v+\sn\hat{\In}}\rb)}\nonumber\\
&&\cholloopin{\Int}{\Jnt}{\snt}{\mnt}{v}\cholloop{\In}{\Jn}{\sn}{\mn}{v}
\ea
Finally,
\ba
\lefteqn{\MeO}\nonumber\\&=&
\lb(\frac{\kappa a^{\frac{3}{2}}}{8a^2}\rb)^2\cholloopin{\Int}{\Jnt}{\snt}{\mnt}{v}\cholloop{\In}{\Jn}{\sigma}{\mn}{v}\lb|\det(\pmgen)\rb|^{\frac{1}{2}}
\sgn(\sn)(p^{-1}_{\Kn\nn v})^-\sgn(\snt)(p^{-1}_{\Knt\nnt v})^-\nonumber\\
&=&\lb(a^{\frac{3}{2}}\cholloopin{\Int}{\Jnt}{\snt}{\mnt}{v}\chol{\Knt}{\snt}{\nnt}{v}\{\cholin{\Knt}{\snt}{\nnt}{v},V^{\half}_{\alpha,v}\}\rb)^*
a^{\frac{3}{2}}\cholloop{\In}{\Jn}{\sn}{\mn}{v}\chol{\Kn}{\sn}{\nn}{v}\{\cholin{\Kn}{\sn}{\nn}{v},V^{\half}_{\alpha,v}\}\nonumber\\
\ea
Therefore the basic building blocks of the leading order reproduces the correct classical building blocks of the classical disretised Master constraint. Thus, the semiclassical limit of the algebraic Master constraint can indeed be interpreted as the discretisation of the classical Master constraint on a cubic lattice, since the discrete Master constraint differs from the expectation value of $(\Omnt)^{\dagger}\Omn$ only by an additional summation over $\In,\Jn,\Kn,\Int,\Jnt,\Knt,\sn,\snt$ and a multiplication with $(i/2)^2$. The same summation and multiplication has to be performed on the classical side as well, thus we have shown that each summand in the sum has the correct semiclassical limit
\be
\sum\limits_{v\in V(\alpha)}\frac{\langle\Psigen\,|\,\MCO_{v}\,|\,\Psigen\rangle}{||\Psigen||^2}\Big|_{LO}=\MC^{cubic}
\ee
and therefore, we are done with step two.\\
\underline{Step 3:}
\\
The last step that remains to show is that the discretised version of the classical Master constraint $\MC^{cubic}$ yields the continuum constraint when we shrink the parameter interval length $\epsilon$ to zero.  For this purpose we will expand the discretised Master constraint $\MC^{cubic}$ in terms of powers of $\epsilon$.
Recall the form of $\MC^{cubic}$
\ba
\MC^{cubic}&=&\lb(\frac{4}{\kappa}\rb)^2a^{\frac{3}{2}}\sum\limits_{v\in V(\alpha)}\sum\limits_{\In\Jn\Kn}\sum\limits_{\Int\Jnt\Knt}\sum\limits_{\sn=+,-}\sum\limits_{\snt=+,-}\epsilon^{\In\Jn\Kn}\epsilon^{\Int\Jnt\Knt}\Big\{\Big(\delta_{\mn,\nn}\delta_{\mnt,\nnt}+\sum\limits_{\elln=1}^{3}
\epsilon_{\elln\mn\nn}\epsilon_{\elln\mnt\nnt}\Big)\nonumber\\
&&
\cholloopin{\Int}{\Jnt}{\snt}{\mnt}{v}\chol{\Knt}{\snt}{\nnt}{v}\{\cholin{\Knt}{\snt}{\nnt}{v},V^{\half}_{\alpha,v}\}
\cholloop{\In}{\Jn}{\sn}{\mn}{v}\chol{\Kn}{\sn}{\nn}{v}\{\cholin{\Kn}{\sn}{\nn}{v},V^{\half}_{\alpha,v}\}\nonumber\\
&=&
\sum\limits_{v\in V(\alpha)}\sum\limits_{\In\Jn\Kn}\sum\limits_{\Int\Jnt\Knt}\sum\limits_{\sn=+,-}\sum\limits_{\snt=+,-}\epsilon^{\In\Jn\Kn}\epsilon^{\Int\Jnt\Knt}\Big\{\Big(\delta_{\mn,\nn}\delta_{\mnt,\nnt}+\sum\limits_{\elln=1}^{3}
\epsilon_{\elln\mn\nn}\epsilon_{\elln\mnt\nnt}\Big)\nonumber\\
&&
\lb(\frac{ a^{\frac{3}{2}}}{2a^2}\rb)^2\sgn(\snt)\sgn(\sn)\lb(\cholloop{\Int}{\Jnt}{\snt}{\mnt}{v}\rb)^{-1}\holloop{\In}{\Jn}{\sn}{\mn}{v}
\lb|\det((p_{Jjv})^-)\rb|^{\frac{1}{2}}(p^{-1}_{\Kn\nn v})^-(p^{-1}_{\Knt\nnt v})^-\nonumber\\
\ea
whereby we used the result derived in step 2 in the last line. 
The first thing we do is reexpressing $(p_{\Kn\nn v})^-$ in terms of $p_{\Kn +\nn v}$ and $p_{\Kn -\nn v}$. The relation is given in eqn (\ref{Trafonp}). Inverting this equation then yields
\ba
p_{\Kn +\nn v}&=&(p_{\Kn\nn v})^+ + (p_{\Kn\nn v})^-\nonumber\\
p_{\Kn -\nn v}&=&(p_{\Kn\nn v})^+ - (p_{\Kn\nn v})^-
\ea
So, for a general $\sn\in\{+,-\}$, we have
\be
p_{\Kn\sn\nn v}=(p_{\Kn\nn v})^+ +\sgn(\sn)(p_{\Kn\nn v})^-
\quad\Leftrightarrow\quad
\sgn(\sn)(p^{-1}_{\Kn\nn v})^-=p^{-1}_{\Kn\sn\nn v}-(p^{-1}_{\Kn\nn v})^+
\ee
Now, we have to express $p^{-1}_{\Kn\sn\nn v}$ in terms of $p_{\Kn\sn\nn v}$. The relation is given by
\ba
p^{-1}_{\Kn\sn\nn v}&=&\frac{1}{2}\epsilon_{\Kn MN}\epsilon_{\nn mn}\frac{p_{M\sn mv}p_{N\sn nv}}{\det(p_{J\sn jv})}\nonumber\\
&=&\frac{a^2}{2}\epsilon_{\Kn MN}\epsilon_{\nn mn}\frac{E^{M\sn}_{mv}E^{N\sn}_{nv}}{\det(E^{J\sn}_{jv})}
\ea
where we introduced $E^{J\sigma}_{jv}:=\int\limits_{S_{e_J^{\sigma}(v)}}n^{S_{e_J^{\sigma}(v)}}_a E^a_j$. Analogously, we get
\ba
\label{pplus}
(p^{-1}_{\Kn\nn v})^+&=&\frac{a^2}{2}\epsilon_{\Kn MN}\epsilon_{\nn mn}\lb(\frac{E^{M+}_{mv}E^{N+}_{nv}}{\det(E^{J+}_{jv})}+\frac{E^{M-}_{mv}E^{N-}_{nv}}{\det(E^{J-}_{jv})}\rb)\nonumber\\
\ea
Now, we expand the loops and the fluxes in powers of $\epsilon$. With the orientation of the edges we have chosen and explained at the beginning we have
\ba
\cholloop{\In}{\Jn}{+}{\mn}{v}=\chol{\In}{+}{\mn}{v}\circ\chol{\Jn}{+}{\mn}{v+\hat{\In}}\circ
\cholin{\In}{+}{\mn}{v+\hat{\Jn}}\circ\cholin{\Jn}{+}{\mn}{v}\nonumber\\
\cholloop{\In}{\Jn}{-}{\mn}{v}=\chol{\In}{-}{\mn}{v}\circ\chol{\Jn}{-}{\mn}{v-\hat{\In}}\circ
\cholin{\In}{-}{\mn}{v-\hat{\Jn}}\circ\cholin{\Jn}{-}{\mn}{v}
\ea
Thus, the expansion yields
\ba
\cholloop{\In}{\Jn}{\sn}{\mn}{v}&\approx&1+\epsilon^2\sgn(\sn) F^{\mn}_{\In\Jn}(v)+O(\epsilon^3)\nonumber\\
\cholloopin{\Int}{\Jnt}{\snt}{\mnt}{v}&\approx&1-\sgn(\snt)\epsilon^2 F^{\mnt}_{\Int\Int}(v)+O(\epsilon^3)=1+\epsilon^2 \sgn(\snt)F^{\mnt}_{\Jnt\Int}(v)+O(\epsilon^3)
\ea
and for the flux 
\be
E^{J\sigma}_{jv}
=
\int\limits_{S_{e_{\Kn}^{\sn}}}n^{\Kn\sn}_a E^a_{\nn}\approx E^a_{\nn}(v)\epsilon^2n^{\Kn\sn}_a(v)+O(\epsilon^3)
\ee
whereby we introduced the shorthand $n^{S_{e_{\Kn}^{\sn}}}_a(v)=n^{\Kn\sn}(v)$. The determinant of the fluxes is therefore approximated by
\be
\det(E^{J\sn}_{jv})\approx\det(E^a_{j}(v)\epsilon^2n^{J\sn}_a(v)+O(\epsilon^3))=\epsilon^6\det(E^a_j(v))\det(n^{J\sn}_a(v))+O(\epsilon^8)\ee
Due to the fact the $\det(n^{J+}_a(v))=-\det(n^{J-}_a(v))$, we conclude that $(p^{-1}_{\Kn\nn v})^+$ vanishes in the leading order, because the two terms cancel each other exactly. Therefore, we get
\ba
\sgn(\sn)(p^{-1}_{\Kn\nn v})^-&\approx&\frac{a^2}{2}\epsilon_{\Kn MN}\epsilon^{\nn mn}
\frac{E^{a}_m(v)n^{M\sn}_a(v)E^b_n(v)n^{N\sn}_b(v)}{\det(E^a_j(v))\det(n^{J\sn}_{a}(v))}\lb(\frac{1+O(\epsilon^2)}{\epsilon^2+O(\epsilon^4)}\rb)\nonumber\\
&=&
\frac{a^2}{2}\epsilon_{\Kn MN}\epsilon^{\nn mn}\sgn(\sn)
\frac{E^{a}_m(v)n^{M\sn}_a(v)E^b_n(v)n^{N\sn}_b(v)}{\det(E^a_j(v))|\det(n^{J\sn}_{a}(v))|}\lb(\frac{1+O(\epsilon^2)}{\epsilon^2+O(\epsilon^4)}\rb)
\ea
Reinserting the expanded terms  into $\MC^{cubic}$ we end up with
\ba
\MC^{cubic}
&\approx&
\sum\limits_{v\in V(\alpha)}\sum\limits_{\In\Jn\Kn}\sum\limits_{\Int\Jnt\Knt}\sum\limits_{\sn=+,-}\sum\limits_{\snt=+,-}\epsilon^{\In\Jn\Kn}\epsilon^{\Int\Jnt\Knt}\Big\{\Big(\delta_{\mn,\nn}\delta_{\mnt,\nnt}+\sum\limits_{\elln=1}^{3}
\epsilon_{\elln\mn\nn}\epsilon_{\elln\mnt\nnt}\Big)\lb(\frac{a^{\frac{3}{2}}}{2a^2}\rb)^2\nonumber\\
&&
\lb(1+\epsilon^2\sgn(\snt)F^{\mnt}_{\Jnt\Int}(v)+O(\epsilon^3)\rb)\lb(1+\sgn(\sn)\epsilon^2F^{\mn}_{\In\Jn}(v)+O(\epsilon^3)\rb)
\nonumber\\
&&\Big(\frac{\epsilon^{\frac{3}{2}}}{a^{\frac{3}{2}}}\Big|\det(E^a_j)\det(n^{J\sn}_a)\Big|^{\frac{1}{4}}+O(\epsilon^4)\Big)
\Big(\frac{\epsilon^{\frac{3}{2}}}{a^{\frac{3}{2}}}\Big|\det(E^a_j)\det(n^{J\snt}_a)\Big|^{\frac{1}{4}}+O(\epsilon^2)\Big)
\nonumber\\
&&
\lb(\frac{a^2}{2}\epsilon_{\Kn MN}\epsilon^{\nn mn}\sgn(\sn)
\frac{E^{a}_m(v)n^{M\sn}_a(v)E^b_n(v)n^{N\sn}_b(v)}{\det(E^a_j(v))|\det(n^{J\sn}_{a}(v))|}\lb(\frac{1+O(\epsilon^2)}{\epsilon^2+O(\epsilon^4)}\rb)\rb)\nonumber\\
&&
\lb(\frac{a^2}{2}\epsilon_{\Kn \wt{M}\wt{N}}\epsilon^{\nnt \wt{m}\wt{n}}\sgn(\snt)
\frac{E^{a}_{\wt{m}}(v)n^{\wt{M}\snt}_a(v)E^b_{\wt{n}}(v)n^{\wt{N}\snt}_b(v)}{\det(E^a_j(v))|\det(n^{J\snt}_{a}(v))|}\lb(\frac{1+O(\epsilon^2)}{\epsilon^2+O(\epsilon^4)}\rb)\rb)\nonumber\\
\ea
Let us consider the summation over $\In,\Jn,\Kn$ separately
\ba
\lefteqn{\sum\limits_{\In\Jn\Kn}\epsilon^{\In\Jn\Kn}\lb(1+\sgn(\sn)\epsilon^2F^{\mn}_{\In\Jn}(v)\rb)\epsilon_{\Kn MN}\epsilon^{\nn mn}\sgn(\sn)
\frac{E^{a}_m(v)n^{M\sn}_a(v)E^b_n(v)n^{N\sn}_b(v)}{\det(E^a_j(v))|\det(n^{J\sn}_{a}(v))|}}\nonumber\\
&=&2\sgn(\sn)\epsilon^2F^{\mn}_{\In\Jn}(v)\epsilon^{\nn mn}\sgn(\sn)
\frac{E^{a}_m(v)n^{\In\sn}_a(v)E^b_n(v)n^{\Jn\sn}_b(v)}{\det(E^a_j(v))|\det(n^{J\sn}_{a}(v))|}\hspace{3cm}\nonumber\\
&=&2\epsilon^2F^{\mn}_{\In\Jn}(v)\epsilon^{\nn mn}
\frac{E^{a}_m(v)n^{\In\sn}_a(v)E^b_n(v)n^{\Jn\sn}_b(v)}{\det(E^a_j(v))|\det(n^{J\sn}_{a}(v))|}
\ea
In the last step $\sgn^2(\sn)=1$ was used. Using
$
F^{\mn}_{\In\Jn}(v)=F^{\mn}_{ab}(v)\frac{\partial X^a(v)}{\partial t^{\In}}\frac{\partial X^b(v)}{\partial t^{\Jn}}
$
we obtain
\ba
\lefteqn{\sum\limits_{\In\Jn\Kn}\epsilon^{\In\Jn\Kn}\lb(1+\sgn(\sn)\epsilon^2F^{\mn}_{\In\Jn}(v)\rb)\epsilon_{\Kn MN}\epsilon^{\nn mn}\sgn(\sn)
\frac{E^{a}_m(v)n^{M\sn}_a(v)E^b_n(v)n^{N\sn}_b(v)}{\det(E^a_j(v))|\det(n^{J\sn}_{a}(v))|}}\nonumber\\
&=&2\epsilon^2F^{\mn}_{ab}(v)\epsilon^{\nn mn}
\frac{E^{c}_m(v)E^d_n(v)}{\det(E^f_j(v))|\det(n^{J\sn}_{f}(v))|}\lb(n^{\In\sn}_c(v)\frac{\partial X^a(v)}{\partial t^{\In}}\rb)\lb(n^{\Jn\sn}_d(v)\frac{\partial X^b(v)}{\partial t^{\Jn}}\rb)\nonumber\\
\ea
Taking advantage of the identity
\be
n^{I\sn}_a(v)\frac{\partial X^b(v)}{\partial t^{I}}=\delta^a_b\sgn(\sn)\Big|\det\lb(\frac{\partial X^a(v)}{\partial t^I}\rb)\Big|
\ee
we immediately see
\be
\det(n^{I\sn}_a(v))=\Big|\det\lb(\frac{\partial X^a(v)}{\partial t^I}\rb)\Big|^2
\ee
Consequently, the two last terms cancel the $|\det(n^{J\sn}_{f}(v))|$ in the denominator and we have
\ba
\lefteqn{\sum\limits_{\In\Jn\Kn}\epsilon^{\In\Jn\Kn}\lb(1+\sgn(\sn)\epsilon^2F^{\mn}_{\In\Jn}(v)\rb)\epsilon_{\Kn MN}\epsilon^{\nn mn}\sgn(\sn)
\frac{E^{a}_m(v)n^{M\sn}_a(v)E^b_n(v)n^{N\sn}_b(v)}{\det(E^a_j(v))|\det(n^{J\sn}_{a}(v))|}}\nonumber\\
&=&
\frac{2\epsilon^2F^{\mn}_{ab}(v)\epsilon^{\nn mn}E^{a}_m(v)E^b_n(v)}{\det(E^c_j(v))}\hspace{9cm}
\ea
Reinserting the result of the summation over $\In,\Jn,\Kn$ and $\Int,\Jnt,\Knt$ respectively into the expectation value of $\MC^{cubic}$ results in
\ba
\MC^{cubic}
&\approx&
\sum\limits_{v\in V(\alpha)}\sum\limits_{\sn=+,-}\sum\limits_{\snt=+,-}\Big\{\Big(\delta_{\mn,\nn}\delta_{\mnt,\nnt}+\sum\limits_{\elln=1}^{3}
\epsilon_{\elln\mn\nn}\epsilon_{\elln\mnt\nnt}\Big)\lb(\frac{a^{\frac{3}{2}}a^2}{a^{\frac{3}{2}}a^2}\rb)^2\nonumber\\
&&\lb(\frac{\epsilon^3}{4}\rb)
\lb(\frac{F^{\mn}_{ab}(v)\epsilon^{\nn mn}E^{a}_m(v)E^b_n(v)F^{\mnt}_{\wt{a}\wt{b}}(v)\epsilon^{\nnt \wt{m}\wt{n}}E^{\wt{a}}_{\wt{m}}(v)E^{\wt{b}}_{\wt{n}}(v)}{|\det(E^c_j(v))|^{\frac{3}{2}}}\rb)\lb|\det\lb(\frac{\partial X^a}{\partial t^{I}}\rb)\rb|\lb(\frac{1+O(\epsilon)}{1+O(\epsilon^4)}\rb)\nonumber\\
&{=\atop \lim\limits_{\epsilon\to 0}}&
\int\limits_{\sigma} d^3x \Big\{
\lb(\frac{\lb[\epsilon^{\nn mn}F^{\nn}_{ab}(x)E^{a}_m(x)E^b_n(x)\rb]^2}{(\sqrt{\det(q(x))})^3}\rb)
+\sum\limits_{\elln=1}^{3}\lb(\frac{\lb[\epsilon_{\elln\mn\nn}\epsilon^{\nn mn}F^{\nn}_{ab}(x)E^{a}_m(x)E^b_n(x)\rb]^2}{(\sqrt{\det(q(x))})^3}\rb)\nonumber\\
&=&\Big\{\int_\sigma \; d^3x\; 
\frac{C^2+q^{ab} C_a C_b}{(\sqrt{\det(q)})^3}(x)\Big\}\nonumber\\
&=&\MC
\ea
Here, as a first step we performed the sum over $\sn,\snt$ which leads to a factor of $4$ and cancels the $\frac{1}{4}$. Secondly, we used $|\det(E^c_j(v))|=|\det(q(v))|$ and finally in the limit $\epsilon\to 0$ we replaced
\be
\sum\limits_{v\in V(\alpha)}\epsilon^3\lb|\det\lb(\frac{\partial X^a}{\partial t^{I}}\rb)\rb|\longrightarrow \int\limits_{\sigma}d^3x
\ee
and realised that all terms above the leading order in $\epsilon$ vanish in the limit $\epsilon\to 0$.\\
Finally, summarising step 1,step2 and step 3 we have proved the following identity
\be
\fbox{$\vspace{0.2cm}\frac{\langle\Psi^t_{\alpha,m}\,|\,\MCO\,|\,\Psi^t_{\alpha,m}\rangle}{||\Psi^t_{\alpha,m}||^2}=\sum\limits_{v\in V(\alpha)}\frac{\langle\Psigen\,|\,\MCO_{v}\,|\,\Psigen\rangle}{||\Psigen||^2}{=\atop\lim\limits_{t\to 0}}\MC^{cubic}[m]{=\atop\lim\limits_{\epsilon\to 0}}\MC[m]\vspace{0.2cm}$}
\ee
This equation has to be understood in the following way. We have calculated the expectation value of the algebraic Master constraint with respect to coherent states. These states carry a classicality label $t\propto \hbar$. When taking the limit $\lim\limits_{t\to 0}$ that corresponds to $\lim\limits_{\hbar\to 0}$ we obtain an expression that can be identified with a discretisation $\MC^{cubic}$ of the classical Master constraint $\MC$ on a cubic lattice. The parameter of this discretisation is $\epsilon$, the so called parameter interval length. Considering the limit $\lim\limits_{\epsilon\to 0}$, we showed in step 3 that $\MC^{cubic}$ coincides with the classical Master constraint $\MC$.  
\section{The Next-to-Leading Order Contribution to Expectation Value of the Algebraic Master Constraint}
In this section we will discuss the next-to-leading order term of the expectation value of the algebraic Master constraint operator $\MCO_v$. As before, our first task to do is deriving the next-to-leading order terms of the $\Lambda^{\half}$-functions. These are the terms denoted by $(sT/t)s\xmm$ and  $(sT/t)s^2(\xmm)^2$ in the power counting eqn (\ref{powerc}).  The derivation for the expanded $\Lambda^{\half}$-functions up to $O(s^2(sT/t)^2)$ can be found in section \ref{NLO} in appendix. The product of these two $\Lambda^{\half}$-functions that is entering the expectation value of $(\Omnt)^{\dagger}\Omn$ is then given by
\ba
\label{LambdaNLO}
\lefteqn{\Lambda^{\half}\lb(\lb\{\xmgenn+\frac{1}{T}\lb(\pmgen\rb)+\frac{T}{4}\lb[\sgn(\snt)\Dtmt-\sgn(\sn)\Dtm\rb]\rb\},\half\sgn(\sn)\e{}{\Kn}{v},\nn\rb)}\nonumber\\
&&\hspace{-0.5cm}
\Lambda^{\half}\lb(\lb\{\xmgenn+\frac{1}{T}\lb(\pmgen\rb)-\frac{T}{4}\lb[\sgn(\snt)\Dtmt-\sgn(\sn)\Dtm\rb]\rb\},\half\sgn(\snt)\e{}{\Knt}{v},\nnt\rb)\nonumber\\
&=&\lb(\frac{a^{\frac{3}{2}}|\det(\pmm)|^{\frac{1}{4}}}{\hbar}\rb)^2\lb(sT\rb)^2\nonumber\\
&&\Big(C^{\Kn\sn\nn}+s\xmm_{Mm}C^{Mm,\Kn\sn\nn}+s^2\xmm_{Mm}\xmm_{Nn}\nonumber\\
\nonumber\\
&&\hspace{1cm}
\lb[\frac{1}{2}C^{Mm,\Kn\sn\nn}C^{Nn}+\frac{1}{3!}\epsilon_{ijk}\lb[\epsilon_{\nn mn}C^{\Kn \sn i}C^{Mj}C^{Nk}
+\epsilon_{\ell\nn n}C^{\Kn \sn j}C^{Li}C^{Nk}
+\epsilon_{\ell m\nn}C^{\Kn \sn k}C^{Li}C^{Mj}\rb]\Big)\rb]\nonumber\\
&&\Big(C^{\Knt\snt\nnt}+s\xmm_{Mm}C^{Mm,\Knt\snt\nnt}+s^2\xmm_{Mm}\xmm_{Nn}\nonumber\\
&&\hspace{1cm}
\lb[\frac{1}{2}C^{Mm,\Knt\snt\nnt}C^{Nn}+\frac{1}{3!}\epsilon_{ijk}\lb[\epsilon_{\nnt mn}C^{\Knt \snt i}C^{Mj}C^{Nk}
+\epsilon_{\ell\nnt n}C^{\Knt \snt j}C^{Li}C^{Nk}
+\epsilon_{\ell m\nnt}C^{\Knt \snt k}C^{Li}C^{Mj}\rb]\Big)\rb]\nonumber\\
&&\lb[\lb(f^{(1)}_{\frac{1}{8}}(1)\rb)^2+ s\xmm_{Mm}C^{Mm}\lb(8f^{(1)}_{\frac{1}{8}}(1)f^{(2)}_{\frac{1}{8}}(1)\rb)\rb.
\nonumber\\
&&\lb.
+s^2\xmm_{Mm}\xmm_{Nn}\lb[C^{Mm,Nn}\lb(4f^{(1)}_{\frac{1}{8}}(1)f^{(2)}_{\frac{1}{8}}(1)\rb)
+C^{Mm}C^{Nn}\lb(40f^{(1)}_{\frac{1}{8}}(1)f^{(3)}_{\frac{1}{8}}(1)+32\lb(f^{(2)}_{\frac{1}{8}}(1)\rb)^2\rb)\rb]\rb]\Big|_{s^2(sT/t)^2)}\nonumber\\
&&+O(s^2(sT/t)^2)
\ea
where we introduced the shorthands
\ba
C^{Mm}&:=&\qmm_{Mm}\nonumber\\
C^{Mm,Nn}&:=&2\qmm_{Mm}\qmm_{Nn}-\qmm_{Mn}\qmm_{Nm}\nonumber\\
C^{\Kn\sn\nn}&:=&\sgn(\sn)\qmm_{\Kn\nn}\nonumber\\
C^{Mm,\Kn\sn\nn}&:=&\sgn(\sn)\lb(2\qmm_{Mm}\qmm_{\Kn\nn}-\qmm_{\Kn n}\qmm_{N\nn}\rb)
\ea
and $\Big|_{s^2(sT/t)^2)}$ denotes that only terms up to order $s^2(sT/t)^2$ are considered.
The expansion has the following structure
\ba
\lefteqn{\lb(\frac{sT}{t}\rb)^2\lb(\alpha_0+\alpha_1 s\xmm +\alpha_2 s^2(\xmm)^2\rb)\lb(\beta_0+\beta_1 s\xmm +\beta_2 s^2(\xmm)^2\rb)\lb(\gamma_0+\gamma_1 s\xmm +\gamma_2 s^2(\xmm)^2\rb)}\nonumber\\
&=&\alpha_0\beta_0\gamma_0 + s^2(\xmm)^2\lb[\alpha_2\lb(\beta_0+\gamma_0\rb)+\beta_2\lb(\gamma_2+\alpha_2\rb)+\gamma_2\lb(\alpha_0+\beta_0\rb)+\alpha_1\beta_1\gamma_0+\alpha_1\beta_0\gamma_1+\alpha_0\beta_1\gamma_1\rb]\nonumber\\
&&+\mathrm{lin}(\xmm)+O(s^2(sT/t)^2)
\ea
whereby $\mathrm{lin}(\xmm)$ denotes all terms linear in $\xmm$ which we do not show in detail as they will not contribute to the final result, because they vanish when integrated against the even function $\exp(-2(\xmm)^2)$.
The integration of $\Lambda^{\half}$ multiplied with the Gaussian $\exp(-2(\xmm_{Mm})^2)$, which is contained in the expression of the expectation value of $(\Omnt)^{\dagger}\Omn$ yields $\sqrt{\pi/2}^9$ for the zeroth power and $(9/4)\sqrt{\pi/2}^9$ for the second power in $\xmm_{Mm}$ respectively. Note that we have a factor $e^{-\frac{t}{4}\sum\limits_{\vt\in V}\sum\limits_{(J,\sigma,j)\atop\in L}\lb(\Del\rb)^2}$ in the expression for the expectation value. Therefore we have to expand this function in powers of $t$. The linear term in $t$ leads to a term having a minimal order of $(sT)^2/t$. This order is already smaller than terms of the order $s^2(sT/t)^2$, because
$
 \lb[s^2(sT/t)^2\rb]\lb[t/(sT)^2\rb]=s^2/t=1/t^{2\alpha}\gg 1
$.
Fortunately, we can neglect the linear term in $t$ in the expansion of the exp-function. We refrain from listing the explicit form of the expectation value  of $(\Omnt)^{\dagger}\Omn$ here which can be found in section \ref{NLO} in eqn (\ref{OmnNLO}) and discuss directly the final expression of the expectation value of $\MCO$ given by
\ba
\label{MNLO}
\lefteqn{\frac{\langle\Psigen\,|\,\MCO\,|\,\Psigen\rangle}{||\Psigen||^2}}\nonumber\\
&=&\MC
+\frac{9}{4}s^2\sum\limits_{v\in V(\alpha)}\lb[\sum\limits_{\In\Jn\Kn}\sum\limits_{\Int\Jnt\Knt}\sum\limits_{\sn=+,-}\sum\limits_{\snt=+,-}\rb.\nonumber\\
&&
\Big\{
\epsilon^{\In\Jn\Kn}\epsilon^{\Int\Jnt\Knt}\lb(\delta_{\mn,\nn}\delta_{\mnt,\nnt}
+\sum\limits_{\elln=1}^3 \epsilon_{\elln \mn\nn}\epsilon_{\elln \mnt\nnt}\rb)\nonumber\\
&&\lb(\frac{4a^{\frac{3}{2}}|\det(\pmm)|^{\frac{1}{4}}}{\kappa\hbar}\rb)^2\lb(sT\rb)^2e^{+i\sum\limits_{(J,\sigma,j)}\phigen\Del}\nonumber\\
&&\Big\{
\lb(C^{\Knt\snt\nnt}+\lb(f^{(1)}_{\frac{1}{8}}(1)\rb)^2\rb)\nonumber\\
&&\hspace{0.3cm}
\lb[\frac{1}{2}C^{Mm,\Kn\sn\nn}C^{Nn}+\frac{1}{3!}\epsilon_{ijk}\lb[\epsilon_{\nn mn}C^{\Kn \sn i}C^{Mj}C^{Nk}
+\epsilon_{\ell\nn n}C^{\Kn \sn j}C^{Li}C^{Nk}
+\epsilon_{\ell m\nn}C^{\Kn \sn k}C^{Li}C^{Mj}\rb]\Big)\rb]\nonumber\\
&&+\lb(C^{\Kn\sn\nn}+\lb(f^{(1)}_{\frac{1}{8}}(1)\rb)^2\rb)\nonumber\\
&&\hspace{0.3cm}
\lb[\frac{1}{2}C^{Mm,\Knt\snt\nnt}C^{Nn}+\frac{1}{3!}\epsilon_{ijk}\lb[\epsilon_{\nn mn}C^{\Knt \snt i}C^{Mj}C^{Nk}
+\epsilon_{\ell\nnt n}C^{\Knt \snt j}C^{Li}C^{Nk}
+\epsilon_{\ell m\nnt}C^{\Knt \snt k}C^{Li}C^{Mj}\rb]\Big)\rb]\nonumber\\
&&+\lb(C^{\Kn\sn\nn}+C^{\Knt\snt\nnt}\rb)\nonumber\\
&&\hspace{0.3cm}
\lb[C^{Mm,Nn}\lb(4f^{(1)}_{\frac{1}{8}}(1)f^{(2)}_{\frac{1}{8}}(1)\rb)
+C^{Mm}C^{Nn}\lb(40f^{(1)}_{\frac{1}{8}}(1)f^{(3)}_{\frac{1}{8}}(1)+32\lb(f^{(2)}_{\frac{1}{8}}(1)\rb)^2\rb)\rb]\nonumber\\
&&\lb.\hspace{-0.5cm}
+\lb(C^{\Kn\sn\nn}C^{Mm,\Knt\snt\nnt}+C^{\Knt\snt\nnt}C^{Mm,\Kn\sn\nn}\rb)C^{Nn}\lb(8f^{(1)}_{\frac{1}{8}}(1)f^{(2)}_{\frac{1}{8}}(1)\rb)
+C^{Mm,\Kn\sn\nn}C^{Mm,\Knt\snt\nnt}\lb(f^{(1)}_{\frac{1}{8}}(1)\rb)^2\Big\}\rb]\nonumber\\
&&+O(s^2(sT/t)^2)
\ea
From the above result one can conclude that the magnitude of the quantum fluctuations (NLO) compared to the leading order  (LO) adopts an additional $s^2$-factor ,because $NLO/LO\propto s^2$. Recalling $s=t^{\half-\alpha}$, we can conclude that as long as $0<\alpha<\half$ and $t$ is a tiny small number as assumed through out all the calculations the quantum fluctuations are finite and small compared to the LO-term.\\
With this in mind, we could  proceed similar to the discussion of the LO-term and rewrite the quantum fluctuations as $s^2/2$ times a discretised integral over certain powers of the fluxes and the field strengths. In this work we are not interested in the precise value of this Riemann sum, rather in the question whether it is finite.Hence, as long as we choose $\alpha<\frac{1}{2}$,  this is indeed the case, because the $C^{\Kn\sn},C^{Mm},C^{Mm,Nn}$ and $C^{Mm,\Kn\sn}$ are all or order unity since $\qmm$ is of order unity by consruction.
\section{Conclusion}
In this paper we investigated the semiclassical limit of the (extended) algebraic Master Constraint operator $\MCO$ associated with an algebraic graph of cubic symmetry. We showed in detail that the leading order of the expectation value of $\MCO$ with respect to coherent states can be interpreted as the discretised version of the (extended) Master constraint operator on a cubic lattice, denoted by $\MC^{cubic}$. In a further analysis, we proved that $\MC^{cubic}$ agrees with the classical (extended) Master constraint $\MC$ in the limit where the lattice parameter interval length is send to zero. Hence, we have the following identity
\be
\fbox{$\vspace{0.2cm}\frac{\langle\Psi^t_{\alpha,m}\,|\,\MCO\,|\,\Psi^t_{\alpha,m}\rangle}{||\Psi^t_{\alpha,m}||^2}=\sum\limits_{v\in V(\alpha)}\frac{\langle\Psigen\,|\,\MCO_{v}\,|\,\Psigen\rangle}{||\Psigen||^2}{=\atop\lim\limits_{t\to 0}}\MC^{cubic}[m]{=\atop\lim\limits_{\epsilon\to 0}}\MC[m]\vspace{0.2cm}$}
\ee
whereby $t$ is the so called classicality parameter and the limit $t\to 0$ corresponds to extracting the leading order out of the semiclassical expectation value. The second limit $\epsilon\to 0$ denotes the transition from a discretised into a continuum theory.\\
Consequently, we have shown that the dynamics of AQG which are encoded in $\MCO$ reproduce the correct infinitesimal generators of General Relativity. \\
Furthermore, we discussed the next-to-leading order contribution of the expectation value of $\MCO$ and could show that these quantum fluctuations are finite. A more detailed analysis of the quantum fluctuations will be postponed to  future research.
\\ 
\\ \\
{\large Acknowledgments}\\
K.G. thanks the Heinrich B\"oll Stiftung for financial support. 
This research project was supported in part by a grant from NSERC of 
Canada to the Perimeter Institute for Theoretical Physics.
\begin{appendix}
\section{Application of the Poisson Resummation formula}
 The aim of this work is to discuss the semiclassical behaviour of the algebraic Master constraint, thus we are mainly interested in the properties of the expectation value in eqn (\ref{vorPR}) for tiny values of the classicality parameter $t$. Looking at eqn (\ref{vorPR}), tiny values of $t$ will correspond to a slow convergence behaviour when considering the sum over $\ngen$. Therefore we will perform a Poisson resummation in which $t$ gets replaced by $1/t$. Then the series converges rapidly when considering small, tiny values of the classicality parameter.
 Let us introduce the following quantities
\be
T:=\sqrt{t}\quad \xgen:=T\ngen\quad \xgenv:=T\ngenv
\ee
with the help of whose all quantities can be expressed in terms of $\xgenv$.
\be
\lambda^{\half}(\{\ngenv\})=T^{-\frac{3}{4}}\lambda^{\half}(\{T\ngenv\})=T^{-\frac{3}{4}}\lambda^{\half}(\{\xgenv\})
\ee
and the expectation value can be rewritten in terms of $\xgen$ as
\ba
\lefteqn{\MeO}\nonumber\\
&=&\frac{T^{-\frac{3}{2}}}{||\Psigen||^2}\nonumber\\
&&\left(e^{-\sum\limits_{\vt\neq v\atop\in V}\sum\limits_{(J,\sigma,j)\atop\in L\setminus L_v}\pgen\Del}
e^{+i\sum\limits_{\vt\neq v\atop\in V}\sum\limits_{((J,\sigma,j)\atop\in L\setminus L_v}\phigen\Del}\right.\nonumber\\
&&
e^{-\frac{t}{2}\sum\limits_{\vt\neq v\atop\in V}\sum\limits_{ (J,\sigma,j)\atop\in L\setminus L_v}\left(\Del\right)^2}\nonumber\\
&&\sum\limits_{\xgen\in\Zl T}
e^{-\sum\limits_{\vt\neq v\atop\in V}\sum\limits_{(J,\sigma,j)\atop\in L\setminus L_v}\left(\xgen\right)^2-\xgen\lb(\frac{2}{T}\pgen+T\Del\rb)}\nonumber\\
&&\hspace{1cm}
\left(e^{-\sum\limits_{(J,\sigma,j)\atop\in L_v}\pgenv\Delv}e^{+i\sum\limits_{((J,\sigma,j)\atop\in L_v}\phigenv\Delv}\right.\nonumber\\
&&
e^{-\frac{t}{2}\sum\limits_{ (J,\sigma,j)\atop\in L_v}\left(\Delv\right)^2}\nonumber\\
&&\sum\limits_{\xgenv\in\Zl T}
e^{-\sum\limits_{(J,\sigma,j)\atop\in L_v}\left(\xgenv\right)^2-\xgenv\lb(\frac{2}{T}\pgenv+T\Delv\rb)}\nonumber\\
&&\hspace{2cm}
\left[\lambda^{\frac{1}{2}}\lb(\{\xgenv\}\rb)-\lambda^{\frac{1}{2}}\lb(\{\xgenv+T\kdel{J}{\sigma}{j}{v}{\Kn}{\sn}{\nn}{v}\}\rb)\right]\nonumber\\
&&\hspace{2cm}
\left[+\lambda^{\frac{1}{2}}\lb(\{\xgenv-T\Delv\}\rb)\right.\\
&&\hspace{2.2cm}\left.
\left.-\lambda^{\frac{1}{2}}\lb(\{\xgenv-T(\Delv+\kdel{J}{\sigma}{j}{v}{\Kn}{\sn}{\nnt}{v})\}\rb)\right]\right)\nonumber
\ea
Since the first three exp-functions are not involved in the summations, we can rearrange the terms considering $\vt\neq v$ and $\vt=v$ together again
\ba
\lefteqn{\MeO}\nonumber\\
&=&\frac{T^{-\frac{3}{2}}}{||\Psigen||^2}\nonumber\\
&&e^{-\sum\limits_{\vt\in V}\sum\limits_{(J,\sigma,j)\atop\in L}\pgen\Del}
e^{+i\sum\limits_{\vt\in V}\sum\limits_{((J,\sigma,j)\atop\in L}\phigen\Del}\nonumber\\
&&
e^{-\frac{t}{2}\sum\limits_{\vt\in V}\sum\limits_{ (J,\sigma,j)\atop\in L}\left(\Del\right)^2}\nonumber\\
&&\left(\sum\limits_{\xgen\in\Zl T}
e^{-\sum\limits_{\vt\neq v\atop\in V}\sum\limits_{(J,\sigma,j)\atop\in L\setminus L_v}\left(\xgen\right)^2-\xgen\lb(\frac{2}{T}\pgen+T\Del\rb)}\right)\nonumber\\
&&
\left(\sum\limits_{\xgenv\in\Zl T}
e^{-\sum\limits_{(J,\sigma,j)\atop\in L_v}\left(\xgenv\right)^2-\xgenv\lb(\frac{2}{T}\pgenv+T\Delv\rb)}\right.\nonumber\\
&&\left.
\Lambda^{\half}\lb(\{\xgenv\},\e{\sn}{\Kn}{v},\nn\rb)
\Lambda^{\half}\lb(\{\xgenv-T\Delv\},\e{\snt}{\Knt}{v},\nnt\rb)\right)\nonumber\\
\ea
where we introduced  
\be
\Lambda^{\half}\lb(\{\xgenv\},\e{\sn}{\Kn}{v},\nn\rb):=T^{-\frac{3}{4}}\lb[\lambda^{\frac{1}{2}}\lb(\{\xgenv\}\rb)-\lambda^{\frac{1}{2}}\lb(\{\xgenv+T\kdel{J}{\sigma}{j}{v}{\Kn}{\sn}{\nn}{v}\}\rb)\rb]
\ee
in order to keep the expression as short as possible.Moreover, the denominator can be reexpressed as
\be
||\Psigen||^2=\prod\limits_{(J,\sigma,j)}\sqrt{\frac{\pi}{t}}e^{-\frac{1}{t}\lb(\pgen\rb)^2}\lb[1+K_t(p)\rb]
=\lb(\sqrt{\frac{\pi}{t}}\rb)^{30}e^{+\frac{1}{t}\sum\limits_{\vt\in V}\sum\limits_{(J,\sigma,j)\atop\in L}\lb(\pgen\rb)^2}\lb[1+K_t(p)\rb]^{30}
\ee
Here we get a power of $30$ since we have ten edges involved and each edge has 3 labels.\\
Inserting the expression for the norm above, we get
\ba
\lefteqn{\MeO}\nonumber\\
&=&\frac{1}
{(\sqrt{\frac{\pi}{t}})^{30}[1+K_t(p)]^{30}}e^{-\frac{1}{t}\sum\limits_{\vt\in V}\sum\limits_{(J,\sigma,j)\atop\in L}\lb(\pgen\rb)^2}\nonumber\\
&&e^{-\sum\limits_{\vt\in V}\sum\limits_{(J,\sigma,j)\atop\in L}\pgen\Del}
e^{+i\sum\limits_{\vt\in V}\sum\limits_{((J,\sigma,j)\atop\in L}\phigen\Del}\nonumber\\
&&
e^{-\frac{t}{2}\sum\limits_{\vt\in V}\sum\limits_{ (J,\sigma,j)\atop\in L}\left(\Del\right)^2}\nonumber\\
&&\left(\sum\limits_{\xgen\in\Zl T}
e^{-\sum\limits_{\vt\neq v\atop\in V}\sum\limits_{(J,\sigma,j)\atop\in L\setminus L_v}\left(\xgen\right)^2-\xgen\lb(\frac{2}{T}\pgen+T\Del\rb)}\right)\nonumber\\
&&
\left(\sum\limits_{\xgenv\in\Zl T}
e^{-\sum\limits_{(J,\sigma,j)\atop\in L_v}\left(\xgenv\right)^2-\xgenv\lb(\frac{2}{T}\pgenv+T\Delv\rb)}\right.\nonumber\\
&&\left.
\Lambda^{\half}\lb(\{\xgenv\},\e{\sn}{\Kn}{v},\nn\rb)
\Lambda^{\half}\lb(\{\xgenv-T\Delv\},\e{\snt}{\Knt}{v},\nnt\rb)\right)\nonumber\\
\ea
The Poisson resummation formula reads
\be
\sum\limits_{\ngen}f(T\ngen)=\frac{1}{T}\sum\limits_{\ngen}\ti{f}\lb(\frac{2\pi\ngen}{T}\rb)
\ee
In order to apply these formula we need the Fourier transformation of the following functions
\ba
f(\xgenv)&:=&
e^{-\sum\limits_{(J,\sigma,j)\atop\in L_v}\left(\xgenv\right)^2-\xgenv\lb(\frac{2}{T}\pgenv+T\Delv\rb)}\nonumber\\
&&\Lambda^{\half}\lb(\{\xgenv\},\e{\sn}{\Kn}{v},\nn\rb)
\Lambda^{\half}\lb(\{\xgenv-T\Delv\},\e{\snt}{\Knt}{v},\nnt\rb)\nonumber\\
g(\xgen)&:=&
e^{-\sum\limits_{\vt\neq v\atop\in V}\sum\limits_{(J,\sigma,j)\atop\in L\setminus L_v}\left(\xgen\right)^2-\xgen\lb(\frac{2}{T}\pgen+T\Del\rb)}
\ea
It is simply given by
\ba
\ti{f}(\kgenv)&:=&
\frac{1}{(2\pi)^{18}}\int\limits_{\Rl^{18}}\,d^{18}\xgenv e^{-i\sum\limits_{(J,\sigma,j)\atop\in L_v}\kgenv\xgenv}\nonumber\\ &&e^{-\sum\limits_{(J,\sigma,j)\atop\in L_v}\left(\xgenv\right)^2-\xgenv\lb(\frac{2}{T}\pgenv+T\Delv\rb)}
\\
&&\Lambda^{\half}\lb(\{\xgenv\},\e{\sn}{\Kn}{v},\nn\rb)
\Lambda^{\half}\lb(\{\xgenv-T\Delv\},\e{\snt}{\Knt}{v},\nnt\rb)\nonumber
\ea
and
\ba
\ti{g}(\kgen)&:=&
\frac{1}{(2\pi)^{12}}\int\limits_{\Rl^{12}}\,d^{12}\xgen e^{-i\sum\limits_{\vt\neq v\in V}\sum\limits_{(J,\sigma,j)\atop\in L\setminus L_v}\kgen\xgen}\nonumber\\ 
&&e^{-\sum\limits_{\vt\neq v\atop\in V}\sum\limits_{(J,\sigma,j)\atop\in L\setminus L_v}\left(\xgen\right)^2-\xgen\lb(\frac{2}{T}\pgen+T\Del\rb)}\nonumber\\
&=&
e^{+\sum\limits_{\vt\neq v\atop\in V}\sum\limits_{(J,\sigma,j)\atop\in L\setminus L_v}\pgen\Del}e^{+\frac{1}{t}\sum\limits_{\vt\neq v\atop\in V}\sum\limits_{(J,\sigma,j)\atop\in L\setminus L_v}(\pgen)^2}
e^{+\sum\limits_{\vt\neq v\in V}\sum\limits_{(J,\sigma,j)\atop\in L\setminus L_v}(\kgen)^2}\nonumber\\
&&
e^{+\frac{t}{4}\sum\limits_{\vt\neq v\atop\in V}\sum\limits_{(J,\sigma,j)\atop\in L\setminus L_v}(\Del)^2}
e^{-\sum\limits_{\vt\neq v\in V}\sum\limits_{(J,\sigma,j)\atop\in L\setminus L_v}\lb(\pgenv+\frac{T^2}{2}\Del\rb)\kgen}
\nonumber\\
&&\frac{1}{(2\pi)^{12}}\int\limits_{\Rl^{12}}\,d^{12}\xgen 
e^{-\sum\limits_{\vt\neq v\in V}\sum\limits_{(J,\sigma,j)\atop\in L\setminus L_v}\left(\xgen-\frac{1}{T}\lb(\pgen+\frac{T^2}{2}\Del-T\kgen\rb)\right)^2}\nonumber\\
&=&\frac{(\sqrt{\pi})^{12}}{(2\pi)^{12}}
e^{+\sum\limits_{\vt\neq v\atop\in V}\sum\limits_{(J,\sigma,j)\atop\in L\setminus L_v}\pgen\Del}e^{+\frac{1}{t}\sum\limits_{\vt\neq v\atop\in V}\sum\limits_{(J,\sigma,j)\atop\in L\setminus L_v}(\pgen)^2}
e^{+\sum\limits_{\vt\neq v\in V}\sum\limits_{(J,\sigma,j)\atop\in L\setminus L_v}(\kgen)^2}\nonumber\\
&&
e^{+\frac{t}{4}\sum\limits_{\vt\neq v\atop\in V}\sum\limits_{(J,\sigma,j)\atop\in L\setminus L_v}(\Del)^2}
e^{-\sum\limits_{\vt\neq v\in V}\sum\limits_{(J,\sigma,j)\atop\in L\setminus L_v}\lb(\pgenv+\frac{T^2}{2}\Del\rb)\kgen}
\nonumber\\
\ea
The last line follows from the fact that the integral is a usual (complex) Gaussian integral that can easily be performed.
Thus,
\ba
\lefteqn{\ti{f}\lb(\frac{2\pi\ngenv}{T}\rb)}\nonumber\\
&=&
\frac{1}{(2\pi)^{18}}\int\limits_{\Rl^{18}}\,d^{18}\xgenv e^{-i\frac{2\pi}{T}\sum\limits_{(J,\sigma,j)\atop\in L_v}\xgenv\ngenv}\nonumber\\ 
&&e^{-\sum\limits_{(J,\sigma,j)\atop\in L_v}\left(\xgenv\right)^2-\xgenv\lb(\frac{2}{T}\pgenv+T\Delv\rb)}
\nonumber\\
&&\Lambda^{\half}\lb(\{\xgenv\},\e{\sn}{\Kn}{v},\nn\rb)
\Lambda^{\half}\lb(\{\xgenv-T\Delv\},\e{\sn}{\Kn}{v},\nnt\rb)\nonumber\\
&=&
\frac{1}{(2\pi)^{18}}\int\limits_{\Rl^{18}}\,d^{18}\xgenv 
e^{-\sum\limits_{(J,\sigma,j)\atop\in L_v}\left(\xgenv\right)^2-\frac{2}{T}\xgenv\lb(\pgenv+\frac{T^2}{2}\Delv-i\pi\ngenv\rb)}\nonumber
\\
&&\Lambda^{\half}\lb(\{\xgenv\},\e{\sn}{\Kn}{v},\nn\rb)
\Lambda^{\half}\lb(\{\xgenv-T\Delv\},\e{\snt}{\Knt}{v},\nnt\rb)\nonumber\\
\ea
and
\ba
\lefteqn{\ti{g}\lb(\frac{2\pi\ngen}{T}\rb)}\nonumber\\
&=&\frac{(\sqrt{\pi})^{12}}{(2\pi)^{12}}
e^{+\sum\limits_{\vt\neq v\atop\in V}\sum\limits_{(J,\sigma,j)\atop\in L\setminus L_v}\pgen\Del}e^{+\frac{1}{t}\sum\limits_{\vt\neq v\atop\in V}\sum\limits_{(J,\sigma,j)\atop\in L\setminus L_v}(\pgen)^2}\nonumber\\
&&e^{+\frac{t}{4}\sum\limits_{\vt\neq v\atop\in V}\sum\limits_{(J,\sigma,j)\atop\in L\setminus L_v}(\Del)^2}\nonumber\\
&&
e^{-\frac{\pi^2}{t}\sum\limits_{\vt\neq v\atop\in V}\sum\limits_{(J,\sigma,j)\atop\in L\setminus L_v}\lb(\ngen\rb)^2}
e^{-2i\frac{\pi}{t}\sum\limits_{\vt\neq v\atop\in V}\sum\limits_{(J,\sigma,j)\atop\in L\setminus L_v}\ngen\lb(\pgen+\frac{T^2}{2}\Del\rb)}
\ea
Additionally, we have to Fourier transform the expression in the denominator as well, thus the application of
the Poisson resummation formula leads therefore to the following expectation value
\ba
\lefteqn{\MeO}\nonumber\\
&=&\frac{1}
{(\sqrt{\frac{\pi}{t}})^{30}[1+K_t(p)]^{30}}\frac{(\sqrt{\pi})^{12}}{T^{30}}\nonumber\\
&&
\left(
e^{+i\sum\limits_{\vt\neq v\atop\in V}\sum\limits_{(J,\sigma,j)\atop\in L\setminus L_v}\phigen\Del}e^{-\frac{t}{4}\sum\limits_{\vt\neq v\atop\in V}\sum\limits_{(J,\sigma,j)\atop\in L\setminus L_v}\lb(\Del\rb)^2}\right.\nonumber\\
&&\left.
\sum\limits_{\ngen\in\Zl}e^{-\frac{\pi^2}{t}\sum\limits_{\vt\neq v\atop\in V}\sum\limits_{(J,\sigma,j)\atop\in L\setminus L_v}\lb(\ngen\rb)^2}e^{-2i\frac{\pi}{t}\sum\limits_{\vt\neq v\atop\in V}\sum\limits_{(J,\sigma,j)\atop\in L\setminus L_v}\ngen\lb(\pgen+\frac{T^2}{2}\Del\rb)}\right)\nonumber\\
&&\left(e^{-\sum\limits_{(J,\sigma,j)\atop\in L_v}\pgenv\Delv}
e^{+i\sum\limits_{(J,\sigma,j)\atop\in L_v}\phigenv\Delv}\right.\nonumber\\
&&e^{-\frac{t}{2}\sum\limits_{(J,\sigma,j)\atop\in L_v}\lb(\Delv\rb)^2}
e^{-\frac{1}{t}\sum\limits_{(J,\sigma,j)\atop\in L_v}\lb(\pgenv\rb)^2}\nonumber\\
&&\sum\limits_{\ngenv\in\Zl}e^{-\frac{\pi^2}{t}\sum\limits_{(J,\sigma,j)\atop\in L_v}\lb(\ngenv\rb)^2}e^{-2i\frac{\pi}{t}\sum\limits_{(J,\sigma,j)\atop\in L_v}\ngenv\lb(\pgenv+\frac{T^2}{2}\Del\rb)}\nonumber\\
&&\int\limits_{\Rl^{18}}\,d^{18}\xgenv e^{-\sum\limits_{(J,\sigma,j)\atop\in L_v}\left(\xgenv\right)^2-\frac{2}{T}\xgenv\lb(\pgenv+\frac{T^2}{2}\Delv-i\pi\ngenv\rb)}\nonumber
\\
&&\left.
\Lambda^{\half}\lb(\{\xgenv\},\e{\sn}{\Kn}{v},\nn\rb)
\Lambda^{\half}\lb(\{\xgenv-T\Delv\},\e{\snt}{\Knt}{v},\nnt\rb)\right)\nonumber\\
\ea
\section{$\xmgen$-Transformation}
\label{Trafo}
Furthermore, similar to \cite{Hanno} we introduce another transformation of variables that reduces the 18-dimensional integral down to a 9-dimensional integral times another 9-dimensional integral which contains no $\Lambda^{\half}$-functions.\\
This transformation leads to 18 new variables called $\xmgen$ and $\xpgen$ which are defined by
\ba
\xmgen:=\frac{x_{J+jv}-x_{J-jv}}{2}&\, ,\,&\xpgen:=\frac{x_{J+jv}+x_{J-jv}}{2}
\ea
while the other quantities as $\ngen$ and $\pgen$ undergo an analogous transformation given by
\ba
\label{Trafonp}
\nmgen:=\frac{n_{J+jv}-n_{J-jv}}{2}&\, ,\,&\npgen:=\frac{n_{J+jv}+n_{J-jv}}{2}\nonumber\\
\pmgen:=\frac{p_{J+jv}-p_{J-jv}}{2}&\, ,\,&\ppgen:=\frac{p_{J+jv}+p_{J-jv}}{2}
\ea
The corresponding transformation for the Kronecker-$\delta$-function reads
\ba
\frac{1}{2}\sgn(\sn)\dmgen&:=&\frac{\kdel{J}{+}{j}{v}{\Kn}{\sn}{\nn}{v}-\kdel{J}{-}{j}{v}{\Kn}{\sn}{\nn}{v}}{2}\nonumber\\
\half\dpgen&:=&\frac{\kdel{J}{+}{j}{v}{K}{\sn}{\nn}{v}+\kdel{J}{-}{j}{v}{\Kn}{\sn}{\nn}{v}}{2}\nonumber\\
\frac{1}{2}\sgn(\snt)\dmgent&:=&\frac{\kdel{J}{+}{j}{v}{\Knt}{\snt}{\nnt}{v}-\kdel{J}{-}{j}{v}{\Knt}{\snt}{\nnt}{v}}{2}\nonumber\\
\half\dpgent&:=&\frac{\kdel{J}{+}{j}{v}{\Knt}{\snt}{\nnt}{v}+\kdel{J}{-}{j}{v}{\Knt}{\snt}{\nnt}{v}}{2}\nonumber\\
\ea
The Jacobean of this transformation is simply
\be
\lb|\det\lb(\frac{\partial\xgenv}{\partial(\xmgen,\xpgen)}\rb)\rb|=2^{\half dim(\xgenv)}=2^9
\ee
The corresponding transformation for the $\Delta$-functions can only be given when decomposing \\$\Delv$ again into $\Bdelv$ and $\Bdeltv$ since we cannot factor out a global
$\sgn(\sigma)$ factor due to the fact that $\Delv$ contains $\sn$ and $\snt$. 
\be
\Delv:=\Bdeltv-\Bdelv
\ee
We then define
\ba
\half\sgn(\sn)\Bdelvm&:=&\frac{\Delta(\In,\Jn,\sn,\mn,v,J,+,j,v)-\Delta(\In,\Jn,\sn,\mn,v,J,-,j,v)}{2}\nonumber\\
\half\Bdelvp&:=&\frac{\Delta(\In,\Jn,\sn,\mn,v,J,+,j,v)+\Delta(\In ,\Jn,\sn,\mn,v,J,-,j,v)}{2}\nonumber\\
\half\sgn(\snt)\Bdelvtm&:=&\frac{\Delta(\Int,\Jnt,\snt,\mnt,v,J,+,j,v)-\Delta(\Int,\Jnt,\snt,\mnt,v,J,-,j,v)}{2}\nonumber\\
\half\Bdelvtp&:=&\frac{\Delta(\Int,\Jnt,\snt,\mnt,v,J,+,j,v)+\Delta(\Int,\Jnt,\snt,\mnt,v,J,-,j,v)}{2}
\ea
When now expressing $x_{J+jv}$ by the sum of $\xpgen$ and $\xmgen$ and $x_{J-jv}$ by $\xpgen-\xmgen$ (the same for all the other occurring terms) and performing the square in the exponential, we realise that all terms that involve mixed $()^+$ and $()^-$-terms as for instance $\xpgen\xmgen$ will drop out, while the terms involving only $()^+$ or $()^-$ respectively occur twice. Furthermore, the $\Lambda^{\half}$-functions do only depend on the variable $\xmgen$, consequently, we can 
can rewrite the integral as
\be
\int\limits_{\Rl^{18}}d^{18}\xgenv F(\xgenv,\ngenv)=\int\limits_{\Rl^9}d^9\xpgen G_1(\xpgen,\npgen)\int\limits_{\Rl^9}d^9 \xmgen G_2(\xmgen,\nmgen)
\ee
where
\ba 
G_1(\xpgen)&:=&e^{-2\sum\limits_{(J,j,v)\atop\in L_v}\left((\xpgen)^2-\frac{2}{T}\xpgen\lb(\ppgen+\frac{T^2}{4}\lb[\Bdelvtp-\Bdelvp\rb]-i\pi\npgen\rb)\right)}\nonumber\\
G_2(\xmgen)&:=&e^{-2\sum\limits_{(J,j,v)\atop\in L_v}\left(\xmgen)^2-\frac{2}{T}\xmgen\lb(\pmgen+\frac{T^2}{4}\lb[\sgn(\snt)\Bdelvtm-\sgn(\sn)\Bdelvm\rb]-i\pi \nmgen\rb)\right)}\nonumber\\
&&\Lambda^{\half}\lb(\{\xmgen-\frac{T}{2}\lb[\sgn(\snt)\Bdelvtm-\sgn(\sn)\Bdelvm\rb]\},\frac{1}{2}\sgn(\snt)\e{-}{\Knt}{v},\nnt\rb)\nonumber\\
&&\Lambda^{\half}\lb(\{\xmgen,\frac{1}{2}\sgn(\sn)\e{-}{\Kn}{v},\nn\rb)
\ea
 Thus we get 
\ba
2^9\int\limits_{\Rl^9}d^9\xpgen&&\nonumber\\
&&\hspace{-3.2cm}
e^{-2\sum\limits_{(J,j)\atop\in L_v}\left((\xpgen)^2-\frac{2}{T}\xpgen\lb(\ppgen+\frac{T^2}{4}\lb[\Bdelvtp-\Bdelvp\rb]-i\pi\npgen\rb)\right)}\nonumber\\ 
2^9\int\limits_{\Rl^9}d^9\xmgen&&\nonumber\\
&&\hspace{-3.2cm}
e^{-2\sum\limits_{(J,j)\atop\in L_v}\left((\xmgen)^2-\frac{2}{T}\xmgen\lb(\pmgen+\frac{T^2}{4}\lb[\sgn(\snt)\Bdelvtm-\sgn(\sn)\Bdelvm\rb]-i\pi\nmgen\rb)\right)}\nonumber\\ 
&&\hspace{-3.2cm}
\Lambda^{\half}\lb(\xmgen-\frac{T}{2}\lb[\sgn(\snt)\Bdelvtm-\sgn(\sn)\Bdelvm\rb]\},\half\sgn(\snt)\e{}{\Knt}{v},\nnt\rb)\nonumber\\
&&\hspace{-3.2cm}
\Lambda^{\half}\lb(\{\xmgen\},\half\sgn(\sn)\e{}{\Kn}{v},\nn\rb)\nonumber\\
&&\hspace{-3.2cm}
=2^9\lb(\sqrt{\frac{\pi}{2}}\rb)^9
e^{-\half\sum\limits_{(J,j)\atop\in L_v}\lb(\ppgen(\Bdelvtp-\Bdelvp)s\rb)}e^{+\frac{2}{t}\sum\limits_{(J,j,v)\atop\in L_v}(\ppgen)^2}\nonumber\\
&&\hspace{-3.2cm}
e^{+\frac{t}{8}\sum\limits_{(J,j,v)\atop\in L_v}(\Bdelvtp-\Bdelvp)^2}\nonumber\\
&&\hspace{-3.2cm}
e^{-2\frac{\pi^2}{t}\sum\limits_{(J,j)\atop\in L_v}\lb(\npgen\rb)^2}
e^{-4i\frac{\pi}{t}\sum\limits_{(J,j)\atop\in L_v}\npgen\lb(\ppgen+\frac{T^2}{4}(\Bdelvtp-\Bdelvp)\rb)}\nonumber\\
&&\hspace{-3.2cm}
2^9\int\limits_{\Rl^9}d^9\xmgen\nonumber\\
&&\hspace{-3.2cm}
e^{-2\sum\limits_{(J,j)\atop\in L_v}\left((\xmgen)^2-\frac{2}{T}\xmgen\lb(\pmgen+\frac{T^2}{4}\lb[\half\sgn(\snt)\Bdelvtm-\half\sgn(\sn)\Bdelvm\rb]-i\pi\nmgen\rb)\right)}\nonumber\\ 
&&\hspace{-3.2cm}
\Lambda^{\half}\lb(\xmgen-T\lb[\Bdelvtm-\Bdelvm\rb]\},\half\sgn(\snt)\e{}{\Knt}{v},\nnt\rb)\nonumber\\
&&\hspace{-3.2cm}
\Lambda^{\half}\lb(\{\xmgen\},\half\sgn(\sn)\e{}{\Kn}{v},\nn\rb)\nonumber\\
\ea
with 
\ba
\Lambda^{\half}\lb(\{\xmgen\},\half\sgn(\sn)\e{}{\Kn}{v},\nn\rb)
&:=&T^{-\frac{3}{4}}\lb[\lambda^{\frac{1}{2}}\lb(\{\xmgen\}\rb)-\lambda^{\frac{1}{2}}\lb(\{\xmgen+\frac{T}{2}\sgn(\sn)\dmgen\}\rb)\rb]\nonumber\\
\lambda^{\frac{1}{2}}\lb(\{\xmgen\}\rb)&=&t^{\frac{3}{4}}\lb(\sqrt{\lb|\det(\xmgen)\rb|}\rb)^{\half}
\ea
The integral over $\xpgen$ is a usual (complex) Gaussian integral that can
easily be performed and yields a factor $(\sqrt{\pi/2})^9$. Thus, the expectation simplifies to
 \ba
\lefteqn{\MeO}\nonumber\\
&=&
\frac{e^{+i\sum\limits_{\vt\neq v\atop\in V}\sum\limits_{(J,\sigma,j)\atop\in L\setminus L_v}\phigen\Del}e^{-\frac{t}{4}\sum\limits_{\vt\neq v\atop\in V}\sum\limits_{(J,\sigma,j)\atop\in L\setminus L_v}\lb(\Del\rb)^2}}
{(\sqrt{\frac{\pi}{t}})^{30}[1+K_t(p)]^{30}}\nonumber\\
&&\frac{(\sqrt{\pi})^{12}}{T^{30}}2^9\lb(\sqrt{\frac{\pi}{2}}\rb)^9\nonumber\\
&&\left(
\sum\limits_{\ngen\in\Zl}e^{-\frac{\pi^2}{t}\sum\limits_{\vt\neq v\atop\in V}\sum\limits_{(J,\sigma,j)\atop\in L\setminus L_v}\lb(\ngen\rb)^2}e^{-2i\frac{\pi}{t}\sum\limits_{\vt\neq v\atop\in V}\sum\limits_{(J,\sigma,j)\atop\in L\setminus L_v}\ngen\lb(\pgen+\frac{T^2}{2}\Del\rb)}\right)\nonumber\\
&&\left(
e^{-\frac{t}{8}\sum\limits_{(J,j)\atop\in L_v}\lb(\lb(\Bdelvtp-\Bdelvp\rb)^2\rb)}
e^{+\frac{i}{2}\sum\limits_{(J,j)\atop\in L_v}\lb(\phipgen(\Bdelvtp-\Bdelvp)\rb)}\rb.\nonumber\\
&&\sum\limits_{\npgen\in\Zl}\lb.
e^{-2\frac{\pi^2}{t}\sum\limits_{(J,j)\atop\in L_v}\lb(\npgen\rb)^2}
e^{-4i\frac{\pi}{t}\sum\limits_{(J,j)\atop\in L_v}\npgen\lb(\ppgen+\frac{T^2}{4}(\Bdelvtp-\Bdelvp)\rb)}\rb)\nonumber\\
&&\lb(
e^{-\half\sum\limits_{(J,j)\atop\in L_v}\lb(\pmgen(\sgn(\snt)\Bdelvtm-\sgn(\sn)\Bdelvm)\rb)}\rb.\nonumber\\
&&e^{+\frac{i}{2}\sum\limits_{(J,j)\atop\in L_v}\lb(\phimgen(\sgn(\snt)\Bdelvtm-\sgn(\sn)\Bdelvm)\rb)}\nonumber\\
&&
e^{-\frac{2}{t}\sum\limits_{(J,j)\atop\in L_v}\lb(\lb(\pmgen\rb)^2\rb)}
e^{-\frac{t}{4}\sum\limits_{(J,j)\atop\in L_v}\lb(\lb(\sgn(\snt)\Bdelvtm-\sgn(\snt)\Bdelvm\rb)^2\rb)}\nonumber\\
&&\sum\limits_{\nmgen\in\Zl}
\int\limits_{\Rl^9}d^9\xmgen\nonumber\\
&&
e^{-2\sum\limits_{(J,j,v)\atop\in L_v}\left((\xmgen)^2-\frac{2}{T}\lb(\pmgen+\frac{T^2}{2}\lb[\half\sgn(\snt)\Bdelvtm-\half\sgn(\sn)\Bdelvm\rb]-i\pi\nmgen\rb)\right)}\nonumber\\ 
&&
\Lambda^{\half}\lb(\lb\{\xmgen-\frac{T}{2}\lb[\sgn(\snt)\Bdelvtm-\sgn(\sn)\Bdelvm\rb]\rb\},\half\sgn(\snt)\e{}{\Knt}{v},\nnt\rb)\nonumber\\
&&\lb.
\Lambda^{\half}\lb(\lb\{\xmgen\rb\},\half\sgn(\sn)\e{}{\Kn}{v},\nn\rb)\rb)
\ea
where we used
\ba
\lefteqn{e^{-\sum\limits_{(J,\sigma,j)\atop\in L_v}\pgenv\Delv}}\nonumber\\
&=&
e^{-\half\sum\limits_{(J,j)\atop\in L_v}\lb(\ppgen(\Bdelvtp-\Bdelvp)+\pmgen(\sgn(\snt)\Bdelvtm-\sgn(\sn)\Bdelvm)\rb)}\nonumber\\
\lefteqn{e^{+i\sum\limits_{(J,\sigma,j)\atop\in L_v}\phigenv\Delv}}\nonumber\\
&=&e^{+\frac{i}{2}\sum\limits_{(J,j)\atop\in L_v}\lb(\phipgen(\Bdelvtp-\Bdelvp)+\phimgen(\sgn(\snt)\Bdelvtm-\sgn(\sn)\Bdelvm)\rb)}\nonumber\\
\lefteqn{e^{-\frac{1}{t}\sum\limits_{(J,\sigma,j)\atop\in L_v}\lb(\pgenv\rb)^2}}\nonumber\\
&=&e^{-\frac{2}{t}\sum\limits_{(J,j)\atop\in L_v}\lb(\lb(\ppgen\rb)^2+\lb(\pmgen\rb)^2\rb)}\nonumber\\
\lefteqn{e^{-\frac{t}{2}\sum\limits_{(J,\sigma,j)\atop\in L_v}\lb(\Delv\rb)^2}}\nonumber\\
&=&e^{-\frac{t}{4}\sum\limits_{(J,j)\atop\in L_v}\lb(\lb(\Bdelvtp-\Bdelvp\rb)^2+\lb(\sgn(\snt)\Bdelvtm-\sgn(\snt)\Bdelvm\rb)^2\rb)}\nonumber\\
\ea
\section{Only the Term with $\ngenv=0$ matters}
\label{nNull}
In the following we will show that only the term with $\ngenv=0$ contributes and all other terms are of order $O(t^{\infty})$.
Thus we consider the following estimation
\ba
\lefteqn{\lb|\frac{\langle .\, ,\, . \rangle}{||\, .\, ||^2}-\frac{\langle .\, ,\,  .\rangle}{||\, .\,|
|^2}\lb|_{\ngen=0}\rb|\right.}\\ \nonumber
&=&\Big|\frac{\langle .\, ,\, . \rangle}{||\, .\, ||^2}-\frac{e^{+i\sum\limits_{\vt\in V}\sum\limits_{(J,\sigma,j)\atop\in L}\phigen\Del}e^{-\frac{t}{4}\sum\limits_{\vt\in V}\sum\limits_{(J,\sigma,j)\atop\in L}\lb(\Del\rb)^2}}
{(\sqrt{\frac{\pi}{t}})^{30}[1+K_t(p)]^{30}}\nonumber\\
&&\frac{(\sqrt{\pi})^{12}}{T^{30}}2^9\lb(\sqrt{\frac{\pi}{2}}\rb)^9
\int\limits_{\Rl^9}d^9\xmgenn
e^{-2\sum\limits_{(J,j)\atop\in L_v}\left(\xmgen\right)^2}\nonumber\\ 
&&\hspace{-0.5cm}
\Lambda^{\half}\lb(\lb\{\xmgenn+\frac{1}{T}\lb(\pmgen\rb)-\frac{T}{4}\lb[\sgn(\snt)\Dtmt-\sgn(\sn)\Dtm\rb]\rb\},\half\sgn(\snt)\e{}{\Knt}{v},\nnt\rb)\nonumber\\
&&\hspace{-0.5cm}\left.
\Lambda^{\half}\lb(\lb\{\xmgenn+\frac{1}{T}\lb(\pmgen\rb)+\frac{T}{4}\lb[\sgn(\snt)\Dtmt-\sgn(\sn)\Dtm\rb]\rb\},\half\sgn(\sn)\e{}{\Kn}{v},\nn\rb)\rb|\nonumber\\
\ea
Let us neglect the explicit $J,j,v$ and introduce the following abbreviations in order to make the expressions more convenient
\be
\xmgen:=\xmm,\quad\pmgen:=\pmm,\quad\ngen:=\nmm
\ee
and  use the expression of the $\Lambda^{\half}$-functions in terms of the determinants in eqn (\ref{lambdadet}) yields 
\ba
\lefteqn{\lb|\frac{\langle .\, ,\, . \rangle}{||\, .\, ||^2}-\frac{\langle .\, ,\,  .\rangle}{||\, .\,|
|^2}\lb|_{\ngen=0}\rb|\right.}\\ \nonumber
&\leq&
\lb|\frac{(\sqrt{\pi})^{12}}{T^{30}}2^9\lb(\sqrt{\frac{\pi}{2}}\rb)^9\rb.\nonumber\\
&&\frac{e^{+i\sum\limits_{\vt\in V}\sum\limits_{(J,\sigma,j)\atop\in L}\phigen\Del}e^{-\frac{t}{4}\sum\limits_{\vt\in V}\sum\limits_{(J,\sigma,j)}\lb(\Del\rb)^2}}
{(\sqrt{\frac{\pi}{t}})^{30}[1+K_t(p)]^{30}}\nonumber\\
&&\left(
\sum\limits_{\ngen\neq 0\atop\in\Zl}e^{-\frac{\pi^2}{t}\sum\limits_{\vt\neq v\atop\in V}\sum\limits_{(J,\sigma,j)\atop\in L\setminus L_v}\lb(\ngen\rb)^2}e^{-2i\frac{\pi}{t}\sum\limits_{\vt\neq v\atop\in V}\sum\limits_{(J,\sigma,j)\atop\in L\setminus L_v}\ngen\lb(\pgen+\frac{T^2}{2}\Del\rb)}\right)\nonumber\\
&&\left(\sum\limits_{\npgen\neq 0\atop\in\Zl}
e^{-2\frac{\pi^2}{t}\sum\limits_{(J,j)\atop\in L_v}\lb(\npgen\rb)^2}
e^{-4i\frac{\pi}{t}\sum\limits_{(J,j)\atop\in L_v}\npgen\lb(\ppgen+\frac{T^2}{4}(\Bdelvtp-\Bdelvp)\rb)}\rb)\nonumber\\
&&\left(\sum\limits_{\nmgen\neq 0\atop\in\Zl}
e^{-2\frac{\pi^2}{t}\sum\limits_{(J,j)\atop\in L_v}\lb(\nmgen\rb)^2}
e^{-4i\frac{\pi}{t}\sum\limits_{(J,j)\atop\in L_v}\nmgen\lb(\pmgen+\frac{T^2}{4}(\sgn(\snt)\Bdelvtm-\sgn(\sn)\Bdelvm)\rb)}\rb.\nonumber\\
&&\int\limits_{\Rl^9}d^9\xmgenn
e^{-2\sum\limits_{(J,j)}\left(\xmgenn\right)^2}\nonumber\\ 
&&\hspace{-0.75cm}
\lb[\exp\lb(\ln\lb(\frac{1}{4}\det\lb(\lb\{\xmm+\frac{1}{T}\lb(\pmm-i\pi\nmm\rb)+\frac{T}{4}\lb[\sgn(\snt)\Dtmt-\sgn(\sn)\Dtm\rb]\rb\}\rb)\rb)\rb)\right.\nonumber\\
&&\hspace{-0.95cm}\left.
+\exp\lb(\ln\lb(\frac{1}{4}\det\lb(\lb\{\xmm+\frac{1}{T}\lb(\pmm-i\pi\nmm\rb)+\frac{T}{4}\lb[\sgn(\snt)\Dtmt-\sgn(\sn)\Dtm\rb]-\frac{T}{2}\sgn(\sn)\dmgen\rb\}\rb)\rb)\rb)\rb]\nonumber\\
&&\hspace{-0.75cm}
\lb[\exp\lb(\ln\lb(\frac{1}{4}\det\lb(\lb\{\xmm+\frac{1}{T}\lb(\pmm-i\pi\nmm\rb)-\frac{T}{4}\lb[\sgn(\snt)\Dtmt-\sgn(\sn)\Dtm\rb]\rb\}\rb)\rb)\rb)\right.\nonumber\\
&&\hspace{-1.4cm}\left.\left.
+\exp\lb(\ln\lb(\frac{1}{4}\det\lb(\lb\{\xmm+\frac{1}{T}\lb(\pmm-i\pi\nmm\rb)-\frac{T}{4}\lb[\sgn(\snt)\Dtmt-\sgn(\sn)\Dtm\rb]-\frac{T}{2}\sgn(\snt)\dmgent\rb\}\rb)\rb)\rb)\rb]\Big)\rb|\nonumber\\
\ea,
where we have used $|a-b|\leq|a+b|$ for the exp-functions\\
Let $(\omega_{\{J,j\}})$ be a matrix of complex numbers and define the norm to be $||\omega||^2=\sum\limits_{J,j}|\omega_{\{J,j\}}|^2.$ Then, certainly, $||\omega_1+\omega_2||\leq||\omega_1||+||\omega_2||$ and $||\omega||\geq|\omega_{\{J,j\}}|$ for all $J,j$. In particular $\det(\{\omega_{\{J,j\}}\})\leq 6||\omega||^3$.
Consequently, we get
\ba
\lefteqn{\det\lb(\lb\{\xmm+\frac{1}{T}\lb(\pmm-i\pi\nmm\rb)+\frac{T}{4}\lb[\sgn(\snt)\Dtmt-\sgn(\sn)\Dtm\rb]\rb\}\rb)}\nonumber\\
&\leq&6\lb(T||\xmm||+||\pmm||+\pi||\nmm||+\frac{T^2}{4}||\sgn(\snt)\Dtmt-\sgn(\sn)\Dtm||\rb)^3
\nonumber\\
&\leq&6\lb(T||\xmm||+||\pmm||+\pi||\nmm||+\frac{T^2}{4}||\sgn(\snt)\Dtmt||+\frac{T}{4}||\sgn(\sn)\Dtm||\rb)^3
\nonumber\\
&\leq&6\lb(T||\xmm||+||\pmm||+\pi||\nmm||+\frac{T^2}{4}+\frac{T^2}{4}\rb)^3\nonumber\\
&\leq&6\lb(T||\xmm||+||\pmm||+\pi||\nmm||+T^2\rb)^3\nonumber\\
\ea
Analogously,
\ba
\lefteqn{\det\lb(\lb\{\xmm+\frac{1}{T}\lb(\pmm-i\pi\nmm\rb)+\frac{T}{4}\lb[\sgn(\snt)\Dtmt-\sgn(\sn)\Dtm\rb]-\frac{T}{2}\sgn(\sn)\dmgen\rb\}\rb)}\nonumber\\
&\leq&6\lb(T||\xmm||+||\pmm||+\pi||\nmm||+\frac{T^2}{4}||\sgn(\snt)\Dtmt-\sgn(\sn)\Dtm||-\frac{T^2}{2}||\sgn(\sn)\dmgen||\rb)^3
\nonumber\\
&\leq&6\lb(T||\xmm||+||\pmm||+\pi||\nmm||+\frac{T^2}{4}||\sgn(\snt)\Dtmt||+\frac{T}{4}||\sgn(\sn)\Dtm||+\frac{T^2}{2}||\sgn(\sn)\dmgen||\rb)^3
\nonumber\\
&\leq&6\lb(T||\xmm||+||\pmm||+\pi||\nmm||+\frac{T^2}{4}+\frac{T^2}{4}+\frac{T^2}{2}\rb)^3\nonumber\\
&=&6\lb(T||\xmm||+||\pmm||+\pi||\nmm||+T^2\rb)^3\nonumber\\
\ea
The same is true for the second term involving the determinants, thus we obtain
\ba
\lefteqn{\lb|\frac{\langle .\, ,\, . \rangle}{||\, .\, ||^2}-\frac{\langle .\, ,\,  .\rangle}{||\, .\,|
|^2}\lb|_{\ngen=0}\rb|\right.}\\ \nonumber
&\leq&
\lb|\frac{(\sqrt{\pi})^{12}}{T^{30}}2^9\lb(\sqrt{\frac{\pi}{2}}\rb)^9\rb.\nonumber\\
&&\frac{e^{+i\sum\limits_{\vt\in V}\sum\limits_{(J,\sigma,j)\atop\in L}\phigen\Del}e^{-\frac{t}{4}\sum\limits_{\vt\in V}\sum\limits_{(J,\sigma,j)}\lb(\Del\rb)^2}}
{(\sqrt{\frac{\pi}{t}})^{30}[1+K_t(p)]^{30}}\nonumber\\
&&\left(
\sum\limits_{\ngen\neq 0\atop\in\Zl}e^{-\frac{\pi^2}{t}\sum\limits_{\vt\neq v\atop\in V}\sum\limits_{(J,\sigma,j)\atop\in L\setminus L_v}\lb(\ngen\rb)^2}e^{-2i\frac{\pi}{t}\sum\limits_{\vt\neq v\atop\in V}\sum\limits_{(J,\sigma,j)\atop\in L\setminus L_v}\ngen\lb(\pgen+\frac{T^2}{2}\Del\rb)}\right)\nonumber\\
&&\left(\sum\limits_{\npgen\neq 0\atop\in\Zl}
e^{-2\frac{\pi^2}{t}\sum\limits_{(J,j)\atop\in L_v}\lb(\npgen\rb)^2}
e^{-4i\frac{\pi}{t}\sum\limits_{(J,j)\atop\in L_v}\npgen\lb(\ppgen+\frac{T^2}{4}(\Bdelvtp-\Bdelvp)\rb)}\rb)\nonumber\\
&&\left(\sum\limits_{\nmgen\neq 0\atop\in\Zl}
e^{-2\frac{\pi^2}{t}\sum\limits_{(J,j)\atop\in L_v}\lb(\nmgen\rb)^2}
e^{-4i\frac{\pi}{t}\sum\limits_{(J,j)\atop\in L_v}\nmgen\lb(\pmgen+\frac{T^2}{4}(\sgn(\snt)\Bdelvtm-\sgn(\sn)\Bdelvm)\rb)}\rb.\nonumber\\
&&\int\limits_{\Rl^9}d^9\xmgenn
e^{-2\sum\limits_{(J,j)}\left(\xmgenn\right)^2}\nonumber\\ 
&&\hspace{-0.75cm}
\lb[\exp\lb(\ln\lb(\frac{1}{4}6\lb(T||\xmm||+||\pmm||+\pi||\nmm||+T^2\rb)^3\rb)\rb)
+\exp\lb(\ln\lb(\frac{1}{4}6\lb(T||\xmm||+||\pmm||+\pi||\nmm||+T^2\rb)^3\rb)\rb)\rb]\nonumber\\
&&\hspace{-0.75cm}
\lb[\exp\lb(\ln\lb(\frac{1}{4}6\lb(T||\xmm||+||\pmm||+\pi||\nmm||+T^2\rb)^3\rb)\rb)
+\exp\lb(\ln\lb(\frac{1}{4}6\lb(T||\xmm||+||\pmm||+\pi||\nmm||+T^2\rb)^3\rb)\rb)\rb]\Big)\Big|\nonumber\\
\ea
and
\ba
\lefteqn{\lb|\frac{\langle .\, ,\, . \rangle}{||\, .\, ||^2}-\frac{\langle .\, ,\,  .\rangle}{||\, .\,|
|^2}\lb|_{\ngen=0}\rb|\right.}\\ \nonumber
&\leq&
\lb|\frac{(\sqrt{\pi})^{12}}{T^{30}}2^9\lb(\sqrt{\frac{\pi}{2}}\rb)^9\rb.\nonumber\\
&&\frac{e^{+i\sum\limits_{\vt\in V}\sum\limits_{(J,\sigma,j)\atop\in L}\phigen\Del}e^{-\frac{t}{4}\sum\limits_{\vt\in V}\sum\limits_{(J,\sigma,j)}\lb(\Del\rb)^2}}
{(\sqrt{\frac{\pi}{t}})^{30}[1+K_t(p)]^{30}}\nonumber\\
&&\left(
\sum\limits_{\ngen\neq 0\atop\in\Zl}e^{-\frac{\pi^2}{t}\sum\limits_{\vt\neq v\atop\in V}\sum\limits_{(J,\sigma,j)\atop\in L\setminus L_v}\lb(\ngen\rb)^2}e^{-2i\frac{\pi}{t}\sum\limits_{\vt\neq v\atop\in V}\sum\limits_{(J,\sigma,j)\atop\in L\setminus L_v}\ngen\lb(\pgen+\frac{T^2}{2}\Del\rb)}\right)\nonumber\\
&&\left(\sum\limits_{\npgen\neq 0\atop\in\Zl}
e^{-2\frac{\pi^2}{t}\sum\limits_{(J,j)\atop\in L_v}\lb(\npgen\rb)^2}
e^{-4i\frac{\pi}{t}\sum\limits_{(J,j)\atop\in L_v}\npgen\lb(\ppgen+\frac{T^2}{4}(\Bdelvtp-\Bdelvp)\rb)}\rb)\nonumber\\
&&\left(\sum\limits_{\nmgen\neq 0\atop\in\Zl}
e^{-2\frac{\pi^2}{t}\sum\limits_{(J,j)\atop\in L_v}\lb(\nmgen\rb)^2}
e^{-4i\frac{\pi}{t}\sum\limits_{(J,j)\atop\in L_v}\nmgen\lb(\pmgen+\frac{T^2}{4}(\sgn(\snt)\Bdelvtm-\sgn(\sn)\Bdelvm)\rb)}\rb.\nonumber\\
&&\left.
\int\limits_{\Rl^9}d^9\xmm
e^{-2||\xmm||^2}
\lb[\frac{1}{4}+T^2||\xmm||^2+||\pmm||+\pi||\nmm||+T^2\rb]^{\lb[\frac{3}{2}\rb]+1}\right|
\ea
where we used $x\leq x^2+\frac{1}{4}$ in the last step and $[\frac{3}{2}]$ is the Gau\ss{} bracket, i.e. the smallest integer equal or lower than $\frac{3}{2}$, hence $[\frac{3}{2}]=1$.\\
Estimation of the integral:
\be
I_k:=\lb(\sqrt{\frac{2}{\pi}}\rb)^9\int\limits_{\Rl^9}d^9x e^{-2||x||^2}||x||^{2k},\quad\quad
I_k=\frac{9+2(k-1)}{4}I_{k-1},\quad  I_0=1
\ee
In our case $k=2$ therefore we get $I_1=\frac{9}{4},I_2=\frac{99}{16}$.
Using 
\ba
\lefteqn{\lb(\frac{1}{4}+T^2||\xmm||^2+||\pmm||+\pi||\nmm||+T^2\rb)^2}\nonumber\\ &=&T^4||\xmm||^4+2T^2||\xmm||^2(\frac{1}{4}+T^2||\xmm||^2+||\pmm||+\pi||\nmm||+T^2)\hspace{4cm}\nonumber\\
&&+(\frac{1}{4}+T^2||\xmm||^2+||\pmm||+\pi||\nmm||+T^2)^2
\ea
we get
\ba
\lefteqn{\int\limits_{\Rl^9}d^9\xmm
e^{-2||\xmm||^2}
\lb[\frac{1}{4}+T^2||\xmm||^2+||\pmm||+\pi||\nmm||+T^2\rb]^{2}}\nonumber\\
&=&\lb(\sqrt{\frac{\pi}{2}}\rb)^9\lb(T^4 I_2+2 T^2(\frac{1}{4}+T^2||\xmm||^2+||\pmm||+\pi||\nmm||+T^2)I_1\rb.\space{4cm}\nonumber\\
&&\lb.
+(\frac{1}{4}+T^2||\xmm||^2+||\pmm||+\pi||\nmm||+T^2)^2 I_0\rb)\nonumber\\
&=&\lb(\sqrt{\frac{\pi}{2}}\rb)^9\lb(\lb(\frac{9}{4}T^2+\frac{1}{4}+T^2||\xmm||^2+||\pmm||+\pi||\nmm||+T^2)\rb)^2+\frac{9}{4}T^4\rb)
\ea
Hence, we have
\ba
\lefteqn{\lb|\frac{\langle .\, ,\, . \rangle}{||\, .\, ||^2}-\frac{\langle .\, ,\,  .\rangle}{||\, .\,|
|^2}\lb|_{\ngen=0}\rb|\right.}\\ \nonumber
&\leq&
\lb|\frac{(\sqrt{\pi})^{12}}{T^{30}}2^9\lb(\sqrt{\frac{\pi}{2}}\rb)^9\rb.\nonumber\\
&&\frac{e^{+i\sum\limits_{\vt\in V}\sum\limits_{(J,\sigma,j)\atop\in L}\phigen\Del}e^{-\frac{t}{4}\sum\limits_{\vt\in V}\sum\limits_{(J,\sigma,j)}\lb(\Del\rb)^2}}
{(\sqrt{\frac{\pi}{t}})^{30}[1+K_t(p)]^{30}}\nonumber\\
&&\left(
\sum\limits_{\ngen\neq 0\atop\in\Zl}e^{-\frac{\pi^2}{t}\sum\limits_{\vt\neq v\atop\in V}\sum\limits_{(J,\sigma,j)\atop\in L\setminus L_v}\lb(\ngen\rb)^2}e^{-2i\frac{\pi}{t}\sum\limits_{\vt\neq v\atop\in V}\sum\limits_{(J,\sigma,j)\atop\in L\setminus L_v}\ngen\lb(\pgen+\frac{T^2}{2}\Del\rb)}\right)\nonumber\\
&&\left(\sum\limits_{\npgen\neq 0\atop\in\Zl}
e^{-2\frac{\pi^2}{t}\sum\limits_{(J,j)\atop\in L_v}\lb(\npgen\rb)^2}
e^{-4i\frac{\pi}{t}\sum\limits_{(J,j)\atop\in L_v}\npgen\lb(\ppgen+\frac{T^2}{4}(\Bdelvtp-\Bdelvp)\rb)}\rb)\nonumber\\
&&\left(\sum\limits_{\nmgen\neq 0\atop\in\Zl}
e^{-2\frac{\pi^2}{t}\sum\limits_{(J,j)\atop\in L_v}\lb(\nmgen\rb)^2}
e^{-4i\frac{\pi}{t}\sum\limits_{(J,j)\atop\in L_v}\nmgen\lb(\pmgen+\frac{T^2}{4}(\sgn(\snt)\Bdelvtm-\sgn(\sn)\Bdelvm)\rb)}\rb.\nonumber\\
&&\left.
\lb(\sqrt{\frac{\pi}{2}}\rb)^9\lb(\lb(\frac{9}{4}T^2+\frac{1}{4}+T^2||\xmm||^2+||\pmm||+\pi||\nmm||+T^2)\rb)^2+\frac{9}{4}T^4\rb)\right|
\ea
This is obviously of the order $O(t^{\infty})$ for $\ngen\neq 0$. Consequently, up to order $O(t^{\infty})$ we have
\ba
\lefteqn{\MeO}\nonumber\\
&=&
\frac{e^{+i\sum\limits_{\vt\in V}\sum\limits_{(J,\sigma,j)\atop\in L}\phigen\Del}e^{-\frac{t}{4}\sum\limits_{\vt\in V}\sum\limits_{(J,\sigma,j)\atop\in L}\lb(\Del\rb)^2}}
{(\sqrt{\frac{\pi}{t}})^{30}[1+K_t(p)]^{30}}\nonumber\\
&&\frac{(\sqrt{\pi})^{12}}{T^{30}}2^9\lb(\sqrt{\frac{\pi}{2}}\rb)^9
\int\limits_{\Rl^9}d^9\xmgenn
e^{-2\sum\limits_{(J,j)\atop\in L_v}\left(\xmgenn\right)^2}\nonumber\\ 
&&\hspace{-0.75cm}
\Lambda^{\half}\lb(\lb\{\xmgenn+\frac{1}{T}\lb(\pmgen\rb)-\frac{T}{4}\lb[\sgn(\snt)\Dtmt-\sgn(\sn)\Dtm\rb]\rb\},\half\sgn(\snt)\e{}{\Knt}{v},\nnt\rb)\nonumber\\
&&\hspace{-0.75cm}
\Lambda^{\half}\lb(\lb\{\xmgenn+\frac{1}{T}\lb(\pmgen\rb)+\frac{T}{4}\lb[\sgn(\snt)\Dtmt-\sgn(\sn)\Dtm\rb]\rb\},\half\sgn(\sn)\e{}{\Kn}{v},\nn\rb)\nonumber\\
\ea
\section{The Expansion of the $\Lambda^{\half}$-Functions}
\subsection{Calculation of the $z_A$-terms}
\label{zA}
 As mentioned in the main text we have to calculate the necessary $z^{\prime}_A$ up to order $O(sT)$ and $O(s^3)$ respectively. For this purpose we use the following index notation $A_{ik}=s\sum_M\qmm_{Mi}\xmm_{Mk}$
\ba
\lefteqn{\tr\lb(s\qmm\xmm\pm\frac{T}{4}s\qmm\lb[\sgn(\snt)\Dtmt-\sgn(\sn)\Dtm\rb]\rb)}\nonumber\\
&=&
s\qmm_{Mm}\xmm_{Mm}\pm\frac{sT}{4}\qmm_{Mm}\lb(\sgn(\snt)\Dtmt_{Mm}-\sgn(\sn)\Dtm_{Mm}\rb)
\ea
Thus
\ba
\lefteqn{\lb[\tr\lb(s\qmm\xmm\pm\frac{T}{4}s\qmm\lb[\sgn(\snt)\Dtmt-\sgn(\sn)\Dtm\rb]\rb)\rb]^2}\nonumber\\
&=&
s^2\qmm_{Mm}\qmm_{Nn}\xmm_{Mm}\xmm_{Nn}\pm\frac{s^2T}{2}\qmm_{Mm}\qmm_{Nn}\xmm_{Mm}\lb(\sgn(\snt)\Dtmt_{Nn}-\sgn(\sn)\Dtm_{Nn}\rb)\nonumber\\
&&+\frac{(sT)^2}{16}\qmm_{Mm}\qmm_{Nn}\lb(\sgn(\snt)\Dtmt_{Mm}-\sgn(\sn)\Dtm_{Mm}\rb)
\lb(\sgn(\snt)\Dtmt_{Nn}-\sgn(\sn)\Dtm_{Nn}\rb)
\nonumber\\
\ea
Moreover
\ba
\lefteqn{\tr\lb(\lb[s\qmm\xmm\pm\frac{T}{4}s\qmm\lb[\sgn(\snt)\Dtmt-\sgn(\sn)\Dtm\rb]\rb]^2\rb)}\nonumber\\
&=&
s^2\qmm_{Mn}\qmm_{Nm}\xmm_{Mm}\xmm_{Nn}\pm\frac{s^2T}{2}\qmm_{Mn}\qmm_{Nm}\xmm_{Mm}\lb(\sgn(\snt)\Dtmt_{Nn}-\sgn(\sn)\Dtm_{Nn}\rb)\nonumber\\
&&+\frac{(sT)^2}{16}\qmm_{Mn}\qmm_{Nm}\lb(\sgn(\snt)\Dtmt_{Mm}-\sgn(\sn)\Dtm_{Mm}\rb)
\lb(\sgn(\snt)\Dtmt_{Nn}-\sgn(\sn)\Dtm_{Nn}\rb)\nonumber\\
\ea
\ba
\lefteqn{\det\lb(s\qmm\xmm\pm\frac{T}{4}s\qmm\lb[\sgn(\snt)\Dtmt-\sgn(\sn)\Dtm\rb]\rb)}\\
&=&\frac{1}{3!}\epsilon_{ijk}\epsilon_{\ell mn}
\lb((s\qmm\xmm)_{i\ell}\pm\frac{T}{4}s(\qmm\lb[\sgn(\snt)\Dtmt-\sgn(\sn)\Dtm\rb])_{i\ell}\rb)\nonumber\\
&&\hspace{2.2cm}
\lb((s\qmm\xmm)_{jm}\pm\frac{T}{4}s(\qmm\lb[\sgn(\snt)\Dtmt-\sgn(\sn)\Dtm\rb])_{jm}\rb)\nonumber\\
&&\hspace{2.2cm}
\lb((s\qmm\xmm)_{kn}\pm\frac{T}{4}s(\qmm\lb[\sgn(\snt)\Dtmt-\sgn(\sn)\Dtm\rb])_{kn}\rb)\nonumber\\
&=&s^3\det(\qmm)\det(\xmm)\pm
\frac{s^3}{3!}\epsilon_{ijk}\epsilon_{\ell mn}\nonumber\\
&&\lb[
+\lb(\frac{T}{4}\qmm_{Li}\lb[\sgn(\snt)\Dtmt_{L\ell}-\sgn(\sn)\Dtm_{L\ell}\rb]\rb)
\lb(\qmm_{Mj}\xmm_{Mm}\rb)\lb(\qmm_{Nk}\xmm_{Nn}\rb)\rb.\nonumber\\
&&+\lb(\frac{T}{4}\qmm_{Mj}\lb[\sgn(\snt)\Dtmt_{Mm}-\sgn(\sn)\Dtm_{Mm}\rb]\rb)
\lb(\qmm_{Li}\xmm_{L\ell}\rb)\lb(\qmm_{Nk}\xmm_{Nn}\rb)\nonumber\\
&&\lb.
+\lb(\frac{T}{4}\qmm_{Nk}\lb[\sgn(\snt)\Dtmt_{Nn}-\sgn(\sn)\Dtm_{Nn}\rb]\rb)
\lb(\qmm_{Li}\xmm_{L\ell}\rb)\lb(\qmm_{Mj}\xmm_{Mm}\rb)\rb]+O((sT)^2)\nonumber\\
\ea
Hence, we obtain
\ba
z^{\prime}_{sq^{-1}x\pm\Delta}&=&
s\qmm_{Mm}\xmm_{Mm}\pm\frac{sT}{4}\lb(\sgn(\snt)\qmm_{Mm}\Dtmt_{Mm}-\sgn(\sn)\qmm_{Mm}\Dtm_{Mm}\rb)\nonumber\\
&&
+\frac{s^2}{2}\lb(\qmm_{Mm}\qmm_{Nn}-\qmm_{Mn}\qmm_{Nm}\rb)\xmm_{Mm}\xmm_{Nn}\nonumber\\
&&
\pm\frac{s^2T}{4}\lb(\qmm_{Mm}\qmm_{Nn}-\qmm_{Mn}\qmm_{Nm}\rb)\xmm_{Mm}\lb(\sgn(\snt)\Dtmt_{Nn}-\sgn(\sn)\Dtm_{Nn}\rb)\nonumber\\
&&\lb(\sgn(\snt)\Dtmt_{Nn}-\sgn(\sn)\Dtm_{Nn}\rb)\nonumber\\
&&+s^3\det(\qmm)\det(\xmm)\pm
\frac{s^3}{3!}\epsilon_{ijk}\epsilon_{\ell mn}\nonumber\\
&&\lb[
+\lb(\frac{T}{4}\qmm_{Li}\lb[\sgn(\snt)\Dtmt_{L\ell}-\sgn(\sn)\Dtm_{L\ell}\rb]\rb)
\lb(\qmm_{Mj}\xmm_{Mm}\rb)\lb(\qmm_{Nk}\xmm_{Nn}\rb)\rb.\nonumber\\
&&+\lb(\frac{T}{4}\qmm_{Mj}\lb[\sgn(\snt)\Dtmt_{Mm}-\sgn(\sn)\Dtm_{Mm}\rb]\rb)
\lb(\qmm_{Li}\xmm_{L\ell}\rb)\lb(\qmm_{Nk}\xmm_{Nn}\rb)\nonumber\\
&&\lb.
+\lb(\frac{T}{4}\qmm_{Nk}\lb[\sgn(\snt)\Dtmt_{Nn}-\sgn(\sn)\Dtm_{Nn}\rb]\rb)
\lb(\qmm_{Li}\xmm_{L\ell}\rb)\lb(\qmm_{Mj}\xmm_{Mm}\rb)\rb]+O((sT)^2)\nonumber\\
\ea
and
\ba
\lefteqn{(z^{\prime}_{sq^{-1}x\pm\Delta})^2}\nonumber\\
&=&
s^2\qmm_{Mn}\qmm_{Nm}\xmm_{Mm}\xmm_{Nn}\pm\frac{s^2T}{4}\qmm_{Mm}\qmm_{Nn}\xmm_{Mm}\lb(\sgn(\snt)\Dtmt_{Nn}-\sgn(\sn)\Dtm_{Nn}\rb)\nonumber\\
&&+\frac{s^3}{2}\qmm_{L\ell}\lb(\qmm_{Mm}\qmm_{Nn}-\qmm_{Mn}\qmm_{Nm}\rb)\xmm_{Mm}\xmm_{Nn}\xmm_{L\ell}\nonumber\\
&&
\pm\frac{s^3T}{4}\qmm_{Nn}\lb(\qmm_{Mm}\qmm_{L\ell}-\qmm_{M\ell}\qmm_{Lm}\rb)\lb(\sgn(\snt)\Dtmt_{L\ell}-\sgn(\sn)\Dtm_{L\ell}\rb)\xmm_{Mm}\xmm_{Nn}\nonumber\\
&&+s^4\qmm_{Mm}\xmm_{Mm}\det(\qmm)\det(\xmm)+o((sT)^2)
\ea
Therefore we get
\ba
\lefteqn{\lb[\det\lb(1+s\qmm\xmm\pm\frac{T}{4}s\qmm\lb[\sgn(\snt)\Dtmt-\sgn(\sn)\Dtm\rb]\rb)\rb]^2}\nonumber\\
&=&
1+2z^{\prime}_{sq^{-1}x\pm\Delta}+\lb(z^{\prime}_{q^{-1}xs\pm\Delta}\rb)^2\nonumber\\
&=&1+2s\qmm_{Mm}\xmm_{Mm}\pm\frac{sT}{2}\lb(\sgn(\snt)\qmm_{Mm}\Dtmt_{Mm}-\sgn(\sn)\qmm_{Mm}\Dtm_{Mm}\rb)\nonumber\\
&&
+s^2\lb(2\qmm_{Mm}\qmm_{Nn}-\qmm_{Mn}\qmm_{Nm}\rb)\xmm_{Mm}\xmm_{Nn}+2s^3\det(\qmm)\det(\xmm)\nonumber\\
&&
\pm\frac{s^2T}{2}\lb(2\qmm_{Mm}\qmm_{Nn}-\qmm_{Mn}\qmm_{Nm}\rb)\xmm_{Mm}\lb(\sgn(\snt)\Dtmt_{Nn}-\sgn(\sn)\Dtm_{Nn}\rb)\nonumber\\
&&+\frac{s^3}{2}\qmm_{L\ell}\lb(\qmm_{Mm}\qmm_{Nn}-\qmm_{Mn}\qmm_{Nm}\rb)\xmm_{Mm}\xmm_{Nn}\xmm_{L\ell}\nonumber\\
&&\pm 2
\frac{s^3}{3!}\epsilon_{ijk}\epsilon_{\ell mn}\nonumber\\
&&\lb[
+\lb(\frac{T}{4}\qmm_{Li}\lb[\sgn(\snt)\Dtmt_{L\ell}-\sgn(\sn)\Dtm_{L\ell}\rb]\rb)
\lb(\qmm_{Mj}\xmm_{Mm}\rb)\lb(\qmm_{Nk}\xmm_{Nn}\rb)\rb.\nonumber\\
&&+\lb(\frac{T}{4}\qmm_{Mj}\lb[\sgn(\snt)\Dtmt_{Mm}-\sgn(\sn)\Dtm_{Mm}\rb]\rb)
\lb(\qmm_{Li}\xmm_{L\ell}\rb)\lb(\qmm_{Nk}\xmm_{Nn}\rb)\nonumber\\
&&\lb.
+\lb(\frac{T}{4}\qmm_{Nk}\lb[\sgn(\snt)\Dtmt_{Nn}-\sgn(\sn)\Dtm_{Nn}\rb]\rb)
\lb(\qmm_{Li}\xmm_{L\ell}\rb)\lb(\qmm_{Mj}\xmm_{Mm}\rb)\rb]\nonumber\\
&&+s^4\qmm_{Mm}\xmm_{Mm}\det(\qmm)\det(\xmm)\nonumber\\
&&
\pm\frac{s^3T}{4}\qmm_{Nn}\lb(\qmm_{Mm}\qmm_{L\ell}-\qmm_{M\ell}\qmm_{Lm}\rb)\lb(\sgn(\snt)\Dtmt_{L\ell}-\sgn(\sn)\Dtm_{L\ell}\rb)\xmm_{Mm}\xmm_{Nn}+O((sT)^2)\nonumber\\
\ea 
Now we have to repeat this calculation for the second term involving the Kronecker-deltas
\ba
\lefteqn{\tr\lb(s\qmm\xmm\pm\frac{T}{4}s\qmm\lb[\sgn(\snt)\Dtmt-\sgn(\sn)\Dtm\rb]\rb)+\frac{T}{2}s\qmm\sgn(\sn)\dmgen}\nonumber\\
&=&
s\qmm_{mM}\xmm_{Mm}\pm\frac{sT}{4}\lb(\sgn(\snt)\qmm_{Mm}\Dtmt_{Mm}-\sgn(\sn)\qmm_{Mm}\Dtm_{Mm}\rb)\nonumber\\
&&
+\frac{sT}{2}\qmm_{Mm}\sgn(\sn)(\dmgen)_{Mm}
\ea
Thus
\ba
\lefteqn{\lb[\tr\lb(s\qmm\xmm\pm\frac{T}{4}s\qmm\lb[\sgn(\snt)\Dtmt-\sgn(\sn)\Dtm\rb]+\frac{T}{2}s\qmm\sgn(\sn)\dmgen\rb)\rb]^2}\nonumber\\
&=&
s^2\qmm_{Mm}\qmm_{Nn}\xmm_{Mm}\xmm_{Nn}\pm\frac{s^2T}{2}\qmm_{Mm}\qmm_{Nn}\xmm_{Mm}\lb(\sgn(\snt)\Dtmt_{Nn}-\sgn(\sn)\Dtm_{Nn}\rb)\nonumber\\
&&\pm s^2T\qmm_{Mm}\qmm_{Nn}\xmm_{Mm}\sgn(\sn)(\dmgen)_{Nn}
+o((sT)^2)\nonumber\\
\ea
\ba
\lefteqn{\tr\lb(\lb[s\qmm\xmm\pm\frac{T}{4}s\qmm\lb[\sgn(\snt)\Dtmt-\sgn(\sn)\Dtm\rb]+\frac{T}{2}s\qmm\sgn(\sn)\dmgen\rb]^2\rb)}\nonumber\\
&=&
s^2\qmm_{Mn}\qmm_{Nm}\xmm_{Mm}\xmm_{Nn}\pm\frac{s^2T}{2}\qmm_{Mn}\qmm_{Nm}\xmm_{Mm}\lb(\sgn(\snt)\Dtmt_{Nn}-\sgn(\sn)\Dtm_{Nn}\rb)\nonumber\\
&&\pm s^2T\qmm_{Mn}\qmm_{Nm}\xmm_{Mm}\sgn(\sn)(\dmgen)_{Nn}+o((sT)^2)
\ea
\ba
\lefteqn{\det\lb(s\qmm\xmm\pm\frac{T}{4}s\qmm\lb[\sgn(\snt)\Dtmt-\sgn(\sn)\Dtm\rb]+\frac{T}{2}s\qmm\sgn(\sn)\dmgen\rb)}\nonumber\\
&=&s^3\det(\qmm)\det(\xmm)\nonumber\\
&&
\pm\frac{s^3}{3!}\epsilon_{ijk}\epsilon_{\ell mn}\nonumber\\
&&\lb[
+\lb(\frac{T}{4}\qmm_{Li}\lb[\sgn(\snt)\Dtmt_{L\ell}-\sgn(\sn)\Dtm_{L\ell}\rb]\rb)
\lb(\qmm_{Jm}\xmm_{Mm}\rb)\lb(\qmm_{Kn}\xmm_{Nn}\rb)\rb.\nonumber\\
&&+\lb(\frac{T}{4}\qmm_{Mj}\lb[\sgn(\snt)\Dtmt_{Mm}-\sgn(\sn)\Dtm_{Mm}\rb]\rb)
\lb(\qmm_{Li}\xmm_{L\ell}\rb)\lb(\qmm_{Nk}\xmm_{Nn}\rb)\nonumber\\
&&\lb.
+\lb(\frac{T}{4}\qmm_{Nk}\lb[\sgn(\snt)\Dtmt_{Nn}-\sgn(\sn)\Dtm_{Nn}\rb]\rb)
\lb(\qmm_{Li}\xmm_{L\ell}\rb)\lb(\qmm_{Mj}\xmm_{Mm}\rb)\rb]\nonumber\\
&&+\frac{s^3}{3!}\epsilon_{ijk}\epsilon_{\ell mn}\nonumber\\
&&\lb[
+\lb(\frac{T}{2}\qmm_{Li}\lb(\sgn(\sn)\dmgen\rb)_{L\ell}\rb)
\lb(\qmm_{Mj}\xmm_{Mm}\rb)\lb(\qmm_{Nk}\xmm_{Nn}\rb)\rb.\nonumber\\
&&+\lb(\frac{T}{2}\qmm_{Mj}\lb(\sgn(\sn)\dmgen\rb)_{Mm}\rb)
\lb(\qmm_{Li}\xmm_{L\ell}\rb)\lb(\qmm_{Nk}\xmm_{Nn}\rb)\nonumber\\
&&\lb.
+\lb(\frac{T}{2}\qmm_{Nk}\lb(\sgn(\sn)\dmgen\rb)_{Nn}\rb)
\lb(\qmm_{Li}\xmm_{L\ell}\rb)\lb(\qmm_{Mj}\xmm_{Mm}\rb)\rb]+O((sT)^2)\nonumber\\
\ea
Hence, we obtain
\ba
z^{\prime}_{sq^{-1}x\pm\Delta+\delta}&=&
s\qmm_{Mm}\xmm_{Mm}\pm\frac{sT}{4}\lb(\sgn(\snt)\qmm_{Mm}\Dtmt_{Mm}-\sgn(\sn)\qmm_{Mm}\Dtm_{Mm}\rb)\nonumber\\
&&+\frac{sT}{2}\qmm_{Mm}\sgn(\sn)(\dmgen)_{Mm}\nonumber\\
&&
+\frac{s^2}{2}\lb(\qmm_{Mm}\qmm_{Nn}-\qmm_{Mn}\qmm_{Nm}\rb)\xmm_{Mm}\xmm_{Nn}\nonumber\\
&&
\pm\frac{s^2T}{4}\lb(\qmm_{Mm}\qmm_{Nn}-\qmm_{Mm}\qmm_{Nm}\rb)\xmm_{Mm}\lb(\sgn(\snt)\Dtmt_{Nn}-\sgn(\sn)\Dtm_{Nn}\rb)\nonumber\\
&& +\frac{s^2T}{2}\lb(\qmm_{Mm}\qmm_{Nn}-\qmm_{Mn}\qmm_{Nm}\rb)\xmm_{Mm}\sgn(\sn)(\dmgen)_{Nn}\nonumber\\
&&+s^3\det(\qmm)\det(\xmm)\nonumber\\
&&
\pm\frac{s^3}{3!}\epsilon_{ijk}\epsilon_{\ell mn}\nonumber\\
&&\lb[
+\lb(\frac{T}{4}\qmm_{Li}\lb[\sgn(\snt)\Dtmt_{L\ell}-\sgn(\sn)\Dtm_{L\ell}\rb]\rb)
\lb(\qmm_{Mj}\xmm_{Mm}\rb)\lb(\qmm_{Nk}\xmm_{Nn}\rb)\rb.\nonumber\\
&&+\lb(\frac{T}{4}\qmm_{Mj}\lb[\sgn(\snt)\Dtmt_{Mm}-\sgn(\sn)\Dtm_{Mm}\rb]\rb)
\lb(\qmm_{Li}\xmm_{L\ell}\rb)\lb(\qmm_{Nk}\xmm_{Nn}\rb)\nonumber\\
&&\lb.
+\lb(\frac{T}{4}\qmm_{Nk}\lb[\sgn(\snt)\Dtmt_{Nn}-\sgn(\sn)\Dtm_{Nn}\rb]\rb)
\lb(\qmm_{Li}\xmm_{L\ell}\rb)\lb(\qmm_{Mj}\xmm_{Mm}\rb)\rb]\nonumber\\
&&+\frac{s^3}{3!}\epsilon_{ijk}\epsilon_{\ell mn}\nonumber\\
&&\lb[
+\lb(\frac{T}{2}\qmm_{Li}\lb(\sgn(\sn)\dmgen\rb)_{L\ell}\rb)
\lb(\qmm_{Mj}\xmm_{Mm}\rb)\lb(\qmm_{Nk}\xmm_{Nn}\rb)\rb.\nonumber\\
&&+\lb(\frac{T}{2}\qmm_{Mj}\lb(\sgn(\sn)\dmgen\rb)_{Mm}\rb)
\lb(\qmm_{Li}\xmm_{L\ell}\rb)\lb(\qmm_{Nk}\xmm_{Nn}\rb)\nonumber\\
&&+\lb.
\lb(\frac{T}{2}\qmm_{KN}\lb(\sgn(\sn)\dmgen\rb)_{Nn}\rb)
\lb(\qmm_{IL}\xmm_{L\ell}\rb)\lb(\qmm_{JM}\xmm_{Mm}\rb)\rb]+O((sT)^2)\nonumber\\
\ea
\ba
\lefteqn{(z^{\prime}_{sq^{-1}x\pm\Delta+\delta})^2}\nonumber\\
&=&
s^2\qmm_{Mn}\qmm_{Nm}\xmm_{Mm}\xmm_{Nn}\pm\frac{s^2T}{4}\qmm_{Mm}\qmm_{Nn}\xmm_{Mm}\lb(\sgn(\snt)\Dtmt_{Nn}-\sgn(\sn)\Dtm_{Nn}\rb)\nonumber\\
&&+\frac{s^2T}{2}\qmm_{Mm}\qmm_{Nn}\xmm_{Mm}\sgn(\sn)(\dmgen)_{Nn}\nonumber\\
&&
+\frac{s^3}{2}\qmm_{L\ell}\lb(\qmm_{Mm}\qmm_{Nn}-\qmm_{Mn}\qmm_{Nm}\rb)\xmm_{Mm}\xmm_{Nn}\xmm_{L\ell}\nonumber\\
&&
\pm\frac{s^3T}{4}\qmm_{Nn}\lb(\qmm_{Mm}\qmm_{L\ell}-\qmm_{M\ell}\qmm_{Lm}\rb)\lb(\sgn(\snt)\Dtmt_{L\ell}-\sgn(\sn)\Dtm_{L\ell}\rb)\xmm_{Mm}\xmm_{Nn}\nonumber\\
&& +\frac{s^3T}{2}\qmm_{Nn}\lb(\qmm_{Mm}\qmm_{L\ell}-\qmm_{M\ell}\qmm_{Lm}\rb)\xmm_{Mm}\sgn(\sn)(\dmgen)_{L\ell}\xmm_{Nn}\nonumber\\
&&+s^4\qmm_{Mm}\xmm_{Mm}\det(\qmm)\det(\xmm)+O((sT)^2)
\ea
Therefore we obtain
\ba
\lefteqn{\lb[\det\lb(1+s\qmm\xmm\pm\frac{T}{4}s\qmm\lb[\sgn(\snt)\Dtmt-\sgn(\sn)\Dtm\rb]+\frac{T}{2}s\qmm\sgn(\sn)\dmgen\rb)\rb]^2}\nonumber\\
&=&
1+2z^{\prime}_{sq^{-1}x\pm\Delta+\delta}+\lb(z^{\prime}_{q^{-1}xs\pm\Delta+\delta}\rb)^2\nonumber\\
&=&1+2s\qmm_{Mm}\xmm_{Mm}\pm\frac{sT}{2}\lb(\sgn(\snt)\qmm_{Mm}\Dtmt_{Mm}-\sgn(\sn)\qmm_{Mm}\Dtm_{Mm}\rb)\nonumber\\
&&+sT\qmm_{Mm}\sgn(\sn)(\dmgen)_{Mm}\nonumber\\
&&
+s^2\lb(2\qmm_{Mm}\qmm_{Nn}-\qmm_{Mn}\qmm_{Nm}\rb)\xmm_{Mm}\xmm_{Nn}\nonumber\\
&&
\pm\frac{s^2T}{2}\lb(2\qmm_{Mm}\qmm_{Nn}-\qmm_{Mn}\qmm_{Nm}\rb)\xmm_{Mm}\lb(\sgn(\snt)\Dtmt_{Nn}-\sgn(\sn)\Dtm_{Nn}\rb)\nonumber\\
&&+s^2T\lb(2\qmm_{Mm}\qmm_{Nn}-\qmm_{Mn}\qmm_{Nm}\rb)\xmm_{Mm}\sgn(\sn)(\dmgen)_{Nn}\nonumber\\
&&+2s^3\det(\qmm)\det(\xmm)\nonumber\\
&&
\pm 2\frac{s^3}{3!}\epsilon_{ijk}\epsilon_{\ell mn}\nonumber\\
&&\lb[
+\lb(\frac{T}{4}\qmm_{Li}\lb[\sgn(\snt)\Dtmt_{L\ell}-\sgn(\sn)\Dtm_{L\ell}\rb]\rb)
\lb(\qmm_{Mj}\xmm_{Mm}\rb)\lb(\qmm_{Nk}\xmm_{Nn}\rb)\rb.\nonumber\\
&&+\lb(\frac{T}{4}\qmm_{Mj}\lb[\sgn(\snt)\Dtmt_{Mm}-\sgn(\sn)\Dtm_{Mm}\rb]\rb)
\lb(\qmm_{Li}\xmm_{L\ell}\rb)\lb(\qmm_{Nk}\xmm_{Nn}\rb)\nonumber\\
&&\lb.
+\lb(\frac{T}{4}\qmm_{Nk}\lb[\sgn(\snt)\Dtmt_{Nn}-\sgn(\sn)\Dtm_{Nn}\rb]\rb)
\lb(\qmm_{Li}\xmm_{L\ell}\rb)\lb(\qmm_{Mj}\xmm_{Mm}\rb)\rb]\nonumber\\
&&+2\frac{s^3}{3!}\epsilon_{ijk}\epsilon_{\ell mn}\nonumber\\
&&\lb[
+\lb(\frac{T}{2}\qmm_{Li}\lb(\sgn(\sn)\dmgen\rb)_{L\ell}\rb)
\lb(\qmm_{Mj}\xmm_{Mm}\rb)\lb(\qmm_{Nk}\xmm_{Nn}\rb)\rb.\nonumber\\
&&+\lb(\frac{T}{2}\qmm_{Mj}\lb(\sgn(\sn)\dmgen\rb)_{Mm}\rb)
\lb(\qmm_{Li}\xmm_{L\ell}\rb)\lb(\qmm_{Nk}\xmm_{Nn}\rb)\nonumber\\
&&\lb.
+\lb(\frac{T}{2}\qmm_{Nk}\lb(\sgn(\sn)\dmgen\rb)_{Nn}\rb)
\lb(\qmm_{Li}\xmm_{L\ell}\rb)\lb(\qmm_{Mj}\xmm_{Mm}\rb)\rb]\nonumber\\
&&
+\frac{s^3}{2}\qmm_{L\ell}\lb(\qmm_{Mm}\qmm_{Nn}-\qmm_{Mn}\qmm_{Nm}\rb)\xmm_{Mm}\xmm_{Nn}\xmm_{L\ell}\nonumber\\
&&
\pm\frac{s^3T}{4}\qmm_{Nn}\lb(\qmm_{Mm}\qmm_{L\ell}-\qmm_{M\ell}\qmm_{Lm}\rb)\lb(\sgn(\snt)\Dtmt_{L\ell}-\sgn(\sn)\Dtm_{L\ell}\rb)\xmm_{Mm}\xmm_{Nn}\nonumber\\
&& +\frac{s^3T}{2}\qmm_{Nn}\lb(\qmm_{Mm}\qmm_{L\ell}-\qmm_{M\ell}\qmm_{Lm}\rb)\xmm_{Mm}\sgn(\sn)(\dmgen)_{L\ell}\xmm_{Nn}\nonumber\\
&&+s^4\qmm_{Mm}\xmm_{Mm}\det(\qmm)\det(\xmm)+O((sT)^2)
\ea 
We easily see that we have the following equality:
\ba
\lefteqn{\lb[\det\lb(1+s\qmm\xmm\pm\frac{T}{4}s\qmm\lb[\sgn(\snt)\Dtmt-\sgn(\sn)\Dtm\rb]+\frac{T}{2}s\qmm\sgn(\sn)\dmgen\rb)\rb]^2}\nonumber\\
&=&
\lb[\det\lb(1+s\qmm\xmm\pm\frac{T}{4}s\qmm\lb[\sgn(\snt)\Dtmt-\sgn(\sn)\Dtm\rb]\rb)\rb]^2\hspace{3.7cm}\nonumber\\
&&+sT\qmm_{Mm}\sgn(\sn)(\dmgen)_{Mm}\nonumber\\
&&+s^2T\lb(2\qmm_{Mm}\qmm_{Nn}-\qmm_{Mn}\qmm_{Nm}\rb)\xmm_{Mm}\sgn(\sn)(\dmgen)_{Nn}\nonumber\\
&&+2\frac{s^3}{3!}\epsilon_{ijk}\epsilon_{\ell mn}\nonumber\\
&&\lb[
+\lb(\frac{T}{2}\qmm_{Li}\lb(\sgn(\sn)\dmgen\rb)_{L\ell}\rb)
\lb(\qmm_{Mj}\xmm_{Mm}\rb)\lb(\qmm_{Nk}\xmm_{Nn}\rb)\rb.\nonumber\\
&&+\lb(\frac{T}{2}\qmm_{Mj}\lb(\sgn(\sn)\dmgen\rb)_{Mm}\rb)
\lb(\qmm_{Li}\xmm_{L\ell}\rb)\lb(\qmm_{Nk}\xmm_{Nn}\rb)\nonumber\\
&&\lb.
+\lb(\frac{T}{2}\qmm_{Nk}\lb(\sgn(\sn)\dmgen\rb)_{Nn}\rb)
\lb(\qmm_{Li}\xmm_{L\ell}\rb)\lb(\qmm_{Mj}\xmm_{Mm}\rb)\rb]\nonumber\\
&& +\frac{s^3T}{2}\qmm_{Nn}\lb(\qmm_{Mm}\qmm_{L\ell}-\qmm_{M\ell}\qmm_{Lm}\rb)\xmm_{Mm}\sgn(\sn)(\dmgen)_{L\ell}\xmm_{Nn}+O((sT)^2)\nonumber\\
\ea 
Considering the terms that contain Kronecker-deltas, we get
\ba
\qmm_{Mm}(\dmgen)_{Mm}&=&\qmm_{Mm}\delta^-_{(M,m,v),(\Kn,\sn,v)}=\qmm_{\Kn,\nn}\nonumber\\
\qmm_{Mm}(\Dtmt)_{Mm}&=&\qmm_{\Jnt\mnt}-\qmm_{\Int\mnt}\nonumber\\
\qmm_{Mm}(\Dtm)_{Mm}&=&\qmm_{\Jn\mn}-\qmm_{\In\mn}
\ea
The only difference that occurs when considering the term $z^{\prime}_{sq^{-1}x-\Delta+\delta}$ is the $\Kn,\sn,\nn$ get replaced by $\Knt,\snt,\nnt$. Consequently, by
Reinserting the above results into the expression for $[\det]^2$, we obtain the following four expressions
\ba
\lefteqn{\lb[\det\lb(1+s\qmm\xmm+\frac{T}{4}s\qmm\lb[\sgn(\snt)\Dtmt-\sgn(\sn)\Dtm\rb]\rb)\rb]^2}\nonumber\\
&=&1+2s\qmm_{Mm}\xmm_{Mm}+\frac{sT}{2}\lb(\sgn(\snt)\lb(\qmm_{\Jnt\mnt}-\qmm_{\Int\mnt}\rb)-\sgn(\sn)\lb(\qmm_{\Jn\mn}-\qmm_{\In\mn}\rb)\rb)\nonumber\\
&&
+s^2\lb(2\qmm_{Mm}\qmm_{Nn}-\qmm_{Mn}\qmm_{Nm}\rb)\xmm_{Mm}\xmm_{Nn}\nonumber\\
&&
+\frac{s^2T}{2}\xmm_{Mm}\lb[2\lb(\sgn(\snt)\lb(\qmm_{\Jnt\mnt}-\qmm{\Int\mnt}\rb)-\sgn(\sn)\lb(\qmm_{\Jn\mn}-\qmm_{\In\mn}\rb)
\rb)\qmm_{Mm}\rb.\nonumber\\
&&\lb.
-\lb(\qmm_{M\mnt}\sgn(\snt)\lb(\qmm_{\Jnt m}-\qmm_{\Int m}\rb)-\qmm_{M\mn}\sgn(\sn)\lb(\qmm_{\Jn m}-\qmm_{\In m}\rb)\rb)\rb]
\nonumber\\
&&+2s^3\det(\qmm)\det(\xmm)\nonumber\\
&&
+ 2\frac{s^3}{3!}\epsilon_{ijk}\nonumber\\
&&\lb[
+\lb(\frac{T}{4}\lb[\epsilon_{\mnt mn}\sgn(\snt)\lb(\qmm_{\Jnt i}-\qmm_{\Int i}\rb)-\epsilon_{\mn mn}\sgn(\sn)\lb(\qmm_{\Jn i}-\qmm_{\In i}\rb)\rb]\rb)\rb.\nonumber\\
&&\hspace{1.5cm}
\lb(\qmm_{Mj}\xmm_{Mm}\rb)\lb(\qmm_{Nk}\xmm_{Nn}\rb)\nonumber\\
&&+\lb(\frac{T}{4}\lb[\epsilon_{\ell\mnt n}\sgn(\snt)\lb(\qmm_{\Jnt j}-\qmm_{\Int j}\rb)-\epsilon_{\ell\mn n}\sgn(\sn)\lb(\qmm_{\Jn j}-\qmm_{\In j}\rb)\rb]\rb)\nonumber\\
&&\hspace{1.5cm}
\lb(\qmm_{Li}\xmm_{L\ell}\rb)\lb(\qmm_{Nk}\xmm_{Nn}\rb)\nonumber\\
&&
+\lb(\frac{T}{4}\lb[\epsilon_{\ell m\mnt}\sgn(\snt)\lb(\qmm_{\Jnt k}-\qmm_{\Int k}\rb)-\epsilon_{\ell m\mn}\sgn(\sn)\lb(\qmm_{\Jn k}-\qmm_{\In k}\rb)\rb]\rb)\nonumber\\
&&\hspace{1.5cm}\lb.
\lb(\qmm_{Li}\xmm_{L\ell}\rb)\lb(\qmm_{Mj}\xmm_{Mm}\rb)\rb]\nonumber\\
&&
+\frac{s^3}{2}\qmm_{L\ell}\lb(\qmm_{Mm}\qmm_{Nn}-\qmm_{Mn}\qmm_{Nm}\rb)\xmm_{Mm}\xmm_{Nn}\xmm_{L\ell}\nonumber\\
&&
+\frac{s^3T}{4}
\lb[\qmm_{Nn}\lb[\qmm_{Mm}\lb(\sgn(\snt)(\qmm_{\Jnt\mnt}-\qmm_{\Int\mnt})-\sgn(\sn)(\qmm_{\Jn\mn}-\qmm_{\In\mn})\rb)\rb)\rb.\nonumber\\
&&\hspace{1cm}\lb.\lb.
-\lb(\qmm_{M\mnt}\sgn(\snt)(\qmm_{\Jnt m}-\qmm_{\Int m})-\qmm_{M\mn}\sgn(\sn)(\qmm_{\Jn m}-\qmm_{\In m})\rb)\rb]\xmm_{Mm}\xmm_{Nn}\rb]\nonumber\\
&&+s^4\qmm_{Mm}\xmm_{Mm}\det(\qmm)\det(\xmm)+O((sT)^2)
\ea 
\ba
\lefteqn{\lb[\det\lb(1+s\qmm\xmm+\frac{T}{4}s\qmm\lb[\sgn(\snt)\Dtmt-\sgn(\sn)\Dtm\rb]+\frac{T}{2}s\qmm\sgn(\sn)\dmgen\rb)\rb]^2}\nonumber\\
&=&
\lb[\det\lb(1+s\qmm\xmm+\frac{T}{4}s\qmm\lb[\sgn(\snt)\Dtmt-\sgn(\sn)\Dtm\rb]\rb)\rb]^2\nonumber\\
&&+sT\sgn(\sn)\qmm_{\Kn\nn}+s^2T\sgn(\sn)\lb(2\qmm_{\Kn\nn}\qmm_{Mm}-\qmm_{\Kn m}\qmm_{M\nn}\rb)\xmm_{Mm}\nonumber\\
&&+2\frac{s^3}{3!}\epsilon_{ijk}\nonumber\\
&&\lb[
+\lb(\frac{T}{2}\epsilon_{\nn mn}\qmm_{\Kn i}\sgn(\sn)\rb)
\lb(\qmm_{Mj}\xmm_{Mm}\rb)\lb(\qmm_{Nk}\xmm_{Nn}\rb)\rb.\nonumber\\
&&+\lb(\frac{T}{2}\epsilon_{\ell\nn n}\qmm_{\Kn j}\sgn(\sn)\rb)
\lb(\qmm_{Li}\xmm_{L\ell}\rb)\lb(\qmm_{Nk}\xmm_{Nn}\rb)\nonumber\\
&&\lb.
+\lb(\frac{T}{2}\epsilon_{\ell m\nn}\qmm_{\Kn k}\sgn(\sn)\rb)
\lb(\qmm_{Li}\xmm_{L\ell}\rb)\lb(\qmm_{Mj}\xmm_{Mm}\rb)\rb]\nonumber\\
&& +\frac{s^3T}{2}\sgn(\sn)\qmm_{Nn}\lb(\qmm_{Mm}\qmm_{\Kn\nn}-\qmm_{M\nn}\qmm_{\Kn m}\rb)\xmm_{Mm}\xmm_{Nn}+O((sT)^2)\nonumber\\
\ea 
\ba
\lefteqn{\lb[\det\lb(1+s\qmm\xmm-\frac{T}{4}s\qmm\lb[\sgn(\snt)\Dtmt-\sgn(\sn)\Dtm\rb]\rb)\rb]^2}\nonumber\\
&=&1+2s\qmm_{Mm}\xmm_{Mm}-\frac{sT}{2}\lb(\sgn(\snt)(\qmm_{\Jnt\mnt}-\qmm_{\Int\mnt})-\sgn(\sn)(\qmm_{\Jn\mn}-\qmm_{\In\mn})\rb)\nonumber\\
&&
+s^2\lb(2\qmm_{Mm}\qmm_{Nn}-\qmm_{Mn}\qmm_{Nm}\rb)\xmm_{Mm}\xmm_{Nn}\nonumber\\
&&
-\frac{s^2T}{2}\xmm_{Mm}\lb[2\lb(\sgn(\snt)\lb(\qmm_{\Jnt\mnt}-\qmm_{\Int\mnt}\rb)-\sgn(\sn)\lb(\qmm_{\Jn\mn}-\qmm_{\In\mn}\rb)\rb)\qmm_{Mm}\rb.\nonumber\\
&&\lb.
-\lb(\qmm_{M\mnt}\sgn(\snt)\lb(\qmm_{\Jnt m}-\qmm_{\Int m}\rb)-\qmm_{M\mn}\sgn(\sn)\lb(\qmm_{\Jn m}-\qmm_{\In m}\rb)\rb)\rb]
\nonumber\\
&&
- 2\frac{s^3}{3!}\epsilon_{ijk}\nonumber\\
&&\lb[
+\lb(\frac{T}{4}\lb[\epsilon_{\mnt mn}\sgn(\snt)\lb(\qmm_{\Jnt i}-\qmm_{\Int i}\rb)-\epsilon_{\mn mn}\sgn(\sn)\lb(\qmm_{\Jn i}-\qmm_{\In i}\rb)\rb]\rb)\rb.\nonumber\\
&&\hspace{1.5cm}
\lb(\qmm_{Mj}\xmm_{Mm}\rb)\lb(\qmm_{Nk}\xmm_{Nn}\rb)\nonumber\\
&&+\lb(\frac{T}{4}\lb[\epsilon_{\ell\mnt n}\sgn(\snt)\lb(\qmm_{\Jnt j}-\qmm_{\Int j}\rb)-\epsilon_{\ell\mn n}\sgn(\sn)\lb(\qmm_{\Jn j}-\qmm_{\In j}\rb)\rb]\rb)\nonumber\\
&&\hspace{1.5cm}
\lb(\qmm_{Li}\xmm_{L\ell}\rb)\lb(\qmm_{Nk}\xmm_{Nn}\rb)\nonumber\\
&&
+\lb(\frac{T}{4}\lb[\epsilon_{\ell m\mnt}\sgn(\snt)\lb(\qmm_{\Jnt k}-\qmm_{\Int k}\rb)-\epsilon_{\ell m\mn}\sgn(\sn)\lb(\qmm_{\Jn k}-\qmm_{\In k}\rb)\rb]\rb)\nonumber\\
&&\hspace{1.5cm}\lb.
\lb(\qmm_{Li}\xmm_{L\ell}\rb)\lb(\qmm_{Mj}\xmm_{Mm}\rb)\rb]\nonumber\\
&&
+\frac{s^3}{2}\qmm_{L\ell}\lb(\qmm_{Mm}\qmm_{Nn}-\qmm_{Mn}\qmm_{Nm}\rb)\xmm_{Mm}\xmm_{Nn}\xmm_{L\ell}\nonumber\\
&&
-\frac{s^3T}{4}
\lb[\qmm_{Nn}\lb[\qmm_{Mm}\lb(\sgn(\snt)(\qmm_{\Jnt\mnt}-\qmm_{\Int\mnt})-\sgn(\sn)(\qmm_{\Jn\mn}-\qmm_{\In\mn})\rb)\rb)\rb.\nonumber\\
&&\hspace{1cm}\lb.\lb.
-\lb(\qmm_{M\mnt}\sgn(\snt)(\qmm_{\Jnt m}-\qmm_{\Int m})-\qmm_{M\mn}\sgn(\sn)(\qmm_{\Jn m}-\qmm_{\In m})\rb)\rb]\xmm_{Mm}\xmm_{Nn}\rb]\nonumber\\
&&+s^4\qmm_{Mm}\xmm_{Mm}\det(\qmm)\det(\xmm)+O((sT)^2)
\ea 
\ba
\lefteqn{\lb[\det\lb(1+s\qmm\xmm-\frac{T}{4}s\qmm\lb[\sgn(\snt)\Dtmt-\sgn(\sn)\Dtm\rb]+\frac{T}{2}s\qmm\sgn(\snt)\dmgent\rb)\rb]^2}\nonumber\\
&=&
\lb[\det\lb(1+s\qmm\xmm-\frac{T}{4}s\qmm\lb[\sgn(\snt)\Dtmt-\sgn(\sn)\Dtm\rb]\rb)\rb]^2\nonumber\\
&&+sT\sgn(\snt)\qmm_{\Knt\nnt}+s^2T\sgn(\snt)\lb(2\qmm_{\Knt\nnt}\qmm_{Mm}-\qmm_{\Knt m}\qmm_{M\nnt}\rb)\xmm_{Mm}\nonumber\\
&&+2\frac{s^3}{3!}\epsilon_{ijk}\nonumber\\
&&\lb[
+\lb(\frac{T}{2}\epsilon_{\nnt mn}\qmm_{\Knt i}\sgn(\snt)\rb)
\lb(\qmm_{Mj}\xmm_{Mm}\rb)\lb(\qmm_{Nk}\xmm_{Nn}\rb)\rb.\nonumber\\
&&+\lb(\frac{T}{2}\epsilon_{\ell\nnt n}\qmm_{\Knt j}\sgn(\snt)\rb)
\lb(\qmm_{Li}\xmm_{L\ell}\rb)\lb(\qmm_{Nk}\xmm_{Nn}\rb)\nonumber\\
&&\lb.
+\lb(\frac{T}{2}\epsilon_{\ell m\nnt}\qmm_{\Knt k}\sgn(\snt)\rb)
\lb(\qmm_{Li}\xmm_{L\ell}\rb)\lb(\qmm_{Mj}\xmm_{Mm}\rb)\rb]\nonumber\\
&& +\frac{s^3T}{2}\sgn(\snt)\qmm_{Nn}\lb(\qmm_{Mm}\qmm_{\Kn\nnt}-\qmm_{M\nnt}\qmm_{\Knt m}\rb)\xmm_{Mm}\xmm_{Nn}+O((sT)^2)\nonumber\\
\ea 
\subsection{Expansion of the $y$- and $y_1$-terms: Estimation of the Remainder in the Expansion}
\label{yExp}
In this section we will estimate the remainder in the expansion of $y^{\frac{1}{8}},y_1^{\frac{1}{8}},\wt{y}^{\frac{1}{8}},\wt{y}_1^{\frac{1}{8}}$ around $y=y_1=\wt{y}=\wt{y}_1=1$. For this we use the tools developed in \cite{Hanno}. Here we have the case where $L=1$ and $M=8$. The expansion of $\Lambda^{\half}$ \cite{Hanno} reads in this case
\ba
\lefteqn{\Lambda^{\half}\lb(\lb\{\xmgenn+\frac{1}{T}\lb(\pmgen\rb)+\frac{T}{4}\lb[\sgn(\snt)\Dtmt-\sgn(\sn)\Dtm\rb]\rb\},\half\sgn(\sn)\e{}{\Kn}{v},\nn\rb)}\nonumber\\
&=&\frac{a^{\frac{3}{2}}\lb|\det(\pmm)\rb|^{\frac{1}{4}}}{i\hbar}\nonumber\\
&&\Big\{(y-y_1)\lb(\sum\limits_{k=1}^n f^{(k)}_{\frac{1}{8}}(1)
\sum\limits_{l=0}^{k-1}(y-1)^l(y_1-1)^{k-1-l}\rb)
+\underbrace{\Big[f^{(n+1)}_{\frac{1}{8}}(y)(y-1)^{n+1} - f^{(n+1)}_{\frac{1}{8}}(y_1)(y_1-1)^{(n+1)}\Big]}_{\displaystyle\mathrm{remainder}}\Big\}\nonumber\\
\ea
  By using the explicit expression for $\Lambda^{\half}$ in terms of $y^{\frac{1}{8}},y_1^{\frac{1}{8}}$ and $\wt{y}^{\frac{1}{8}},\wt{y}_1^{\frac{1}{8}}$ respectively, the remainder in lemma 4.1 of \cite{Hanno} will lead to a Gaussian integral of the form
\be
\int\limits_{\Rl^9}d^9\xmgen e^{-2||\xmm||^2}\lb(f^{(n+1)}_{\frac{1}{8}}(y)(y-1)^{(n+1)}-f^{(n+1)}_{\frac{1}{8}}(y_1)(y_1-1)^{(n+1)}\rb)
\ee
for the one $\Lambda^{\half}-$function and
\be
\int\limits_{\Rl^9}d^9\xmgen e^{-2||\xmm||^2}\lb(f^{(n+1)}_{\frac{1}{8}}(\wt{y})(\wt{y}-1)^{(n+1)}-f^{(n+1)}_{\frac{1}{8}}(\wt{y}_1)(\wt{y}_1-1)^{(n+1)}\rb)
\ee
for the other one respectively.\\
We come to the estimation of the two remainders now:\\
First of all we need an estimation for $z_{sq^{-1}x+\Delta},z_{sq^{-1}x-\Delta},z_{sq^{-1}x+\Delta+\delta},z_{sq^{-1}x-\Delta+\tilde{\delta}}$.
\ba
\lefteqn{z_{sq^{-1}x\pm\Delta}}\nonumber\\
&=&\tr(s\qmm\xmm\pm\Delta)+\frac{1}{2}\lb[\lb(\tr(s\qmm\xmm\pm\Delta)\rb)^2-\tr([s\qmm\xmm\pm\Delta)^2)\rb]+\det(s\qmm\xmm\pm\Delta)\nonumber\\
\ea
We obtain
\ba
\lb|\tr(s\qmm\xmm\pm\Delta)\rb|
&\leq&s||\qmm|| ||\xmm||+\frac{sT}{4}\lb(|\qmm_{\Jnt\mnt}|+|\qmm_{\Int\mnt}|+|\qmm_{\Jn\mn}|+|\qmm_{\In\mn}|\rb)\nonumber\\
&\leq&s||\qmm|| ||\xmm||+s T||\qmm||
\ea
\ba
\lb|[\tr(s\qmm\xmm\pm\Delta)]^2\rb|
&\leq&|\tr(s\qmm\xmm\pm\Delta)|^2\nonumber\\
&\leq&s^2||\qmm||^2||\xmm||^2+\frac{s^2T^2}{16}\lb(|\qmm_{\Jnt\mnt}|+|\qmm_{\Int\mnt}|+|\qmm_{\Jn\mn}|+|\qmm_{\In\mn}|\rb)^2
\nonumber\\
&&+\frac{sT}{2}||\qmm||||\xmm||\lb(|\qmm_{\Jnt\mnt}|+|\qmm_{\Int\mnt}|+|\qmm_{\Jn\mn}|+|\qmm_{\In\mn}|\rb)\nonumber\\
&\leq&s^2||\qmm||^2||\xmm||^2+2 s T||\qmm||^2||\xmm||+s^2T^2||\qmm||^2
\ea
\ba
\lb|\tr([s\qmm\xmm\pm\Delta]^2)\rb|
&\leq&s^2||\qmm||^2||\xmm||^2+\frac{s^2T^2}{16}\lb(|\qmm_{\Jnt\mnt}|+|\qmm_{\Int\mnt}|+|\qmm_{\Jn\mn}|+|\qmm_{\In\mn}|\rb)^2
\nonumber\\
&&+\frac{sT}{2}\lb(|(\qmm\xmm\qmm)_{\Jnt\mnt}|+|(\qmm\xmm\qmm)_{\Int\mnt}|\rb.\nonumber\\
&&\hspace{1cm}\lb.
+|(\qmm\xmm\qmm)_{\Jn\mn}|+|(\qmm\xmm\qmm)_{\In\mn}|\rb.\Big)\nonumber\\
&\leq&s^2||\qmm||^2||\xmm||^2+2 s T||\qmm||^2||\xmm||+s^2T^2||\qmm||^2
\ea
\ba
\lb|\det(s\qmm\xmm\pm\Delta)\rb|
&\leq&
6s^3\lb(||\qmm|| ||\xmm|| +\frac{T}{4}\lb(|\qmm_{\Jnt\mnt}|+|\qmm_{\Int\mnt}|+|\qmm_{\Jn\mn}|+|\qmm_{\In\mn}|\rb)\rb)^3\nonumber\\
&\leq&
6s^3\lb(||\qmm||^3 ||\xmm||^3+3 T||\qmm||^3 ||\xmm||^2+3T||\xmm||||\qmm||^3+T^3||\qmm||^3\rb)\nonumber\\
\ea
With these intermediate estimations, we can estimate $z_{sq^{-1}x+\Delta}$ and $z_{sq^{-1}x-\Delta}$ now.
\ba
\lb|z_{sq^{-1}x\pm\Delta}\rb|
&\leq&s||\qmm|| ||\xmm||+s T||\qmm||\nonumber\\
&&+\frac{1}{2}\lb[+s^2||\qmm||^2||\xmm||^2+2 s T||\qmm||^2||\xmm||+s^2T^2||\qmm||^2\rb.\nonumber\\
&&\lb.\hspace{0.75cm}
-s^2||\qmm||^2||\xmm||^2+2 s T||\qmm||^2||\xmm||+s^2T^2||\qmm||^2\rb]\nonumber\\
&&
+6s^3\lb(||\qmm||^3 ||\xmm||^3+3 T||\qmm||^3 ||\xmm||^2+3T||\xmm||||\qmm||^3+T^3||\qmm||^3\rb)\nonumber\\
&=&s||\qmm|| ||\xmm||+18s^3T||\qmm||^3||\xmm||+18s^3T||\qmm||^3 ||\xmm||^2+6s^3||\qmm||^3 ||\xmm||^3\nonumber\\
&&+s T||\qmm||+6s^3T^3||\qmm||^3\nonumber\\
&=:&u(||\xmm||)
\ea
then
\be
|y-1|\leq 2u +u^2=:P(||x||)\quad\mathrm{and}\quad|\wt{y}-1|\leq 2u +u^2=:P(||x||)
\ee
In an analogous way we obtain
\ba
\lb|z_{sq^{-1}x+\Delta+\delta}\rb|
&\leq&\lb|z_{sq^{-1}x\pm\Delta}\rb|+sT||\qmm||+s^2T||\qmm||^2||\xmm||=:u_1(||x||)
\ea
and
\ba
\lb|z_{sq^{-1}x-\Delta+\tilde{\delta}}\rb|
&\leq&\lb|z_{sq^{-1}x\pm\Delta}\rb|+sT||\qmm||+s^2T||\qmm||^2||\xmm||=:u_1(||x||)
\ea
Thus we have
\be
|y_1-1|\leq 2u_1 +u_1^2=:P_1(||x||)\quad\mathrm{and}\quad |\wt{y}_1-1|\leq 2u_1 +u_1^2=:P_1(||x||)
\ee
Since $y$ and $\wt{y}$ and $y_1$ and $\wt{y}_1$ respectively are estimated by exactly the same polynomials $P(||\xmm||)$ and $P_1(||\xmm||)$ respectively, we can estimate the two remainders by the same term, thus
\ba
\lefteqn{\lb|\int\limits_{\Rl^9}d^9\xmgen e^{-2||\xmm||^2}\lb(f^{(n+1)}_{\frac{1}{8}}(y)(y-1)^{(n+1)}-f^{(n+1)}_{\frac{1}{8}}(y_1)(y_1-1)^{(n+1)}\rb)\rb|}\nonumber\\
&\leq&
\int\limits_{\Rl^9}d^9\xmgen e^{-2||\xmm||^2}(3\cdot 8)^{(n+1)}\lb[P(||\xmm||)^{(n+2)}+2 P(||\xmm||)^{(n+1)}\rb.\hspace{3cm}\nonumber\\
&&\lb.\hspace{5.75cm}
+P_1(||\xmm||)^{(n+2)}+2 P_1(||\xmm||)^{(n+1)}\rb]
\ea
and
\ba
\lefteqn{\lb|\int\limits_{\Rl^9}d^9\xmgen e^{-2||\xmm||^2}\lb(f^{(n+1)}_{\frac{1}{8}}(\wt{y})(\wt{y}-1)^{(n+1)}-f^{(n+1)}_{\frac{1}{8}}(\wt{y}_1)(\wt{y}_1-1)^{(n+1)}\rb)\rb|}\nonumber\\
&\leq&
\int\limits_{\Rl^9}d^9\xmgen e^{-2||\xmm||^2}(3\cdot 8)^{(n+1)}\lb[P(||\xmm||)^{(n+2)}+2 P(||\xmm||)^{(n+1)}\rb.\hspace{3cm}\nonumber\\
&&\lb.\hspace{5.75cm}
+P_1(||\xmm||)^{(n+2)}+2 P_1(||\xmm||)^{(n+1)}\rb]
\ea
thus we can restrict our further estimation to one term only\\
Express $P(||\xmm||)$ as an arbitrary polynomial of the order sixth:
\be
P(||\xmm||)=\sum\limits_{k=0}^6a_k||\xmm||^k
\ee
By the multinomial theorem we obtain
\be
\lb[P(||\xmm||)\rb]^n=\sum\limits_{n_0+..+n_6=n}\frac{n!}{(n_0!)...(n_6)!}\lb[\prod\limits_{k=0}^6a_k^n\rb]||\xmm||^{\sum\limits_{k=0}^6k n_k}
\ee
Consider the Gaussian integral of the form
\be
\sqrt{\frac{2}{\pi}}^9\int\limits_{\Rl^9} d^9\xmm e^{-2||\xmm||^2}||\xmm||^n=V_{8}\sqrt{\frac{2}{\pi}}^9\int\limits_0^{\infty}dr e^{-2r^2}r^{n+8}=:V_8\sqrt{\frac{2}{\pi}}^9J_{n+8}
\ee
where $V_m=\frac{2\pi^{\frac{m}{2}}}{\Gamma(\frac{m}{2})}$ is the volume of $S^m$. 
Now,
\ba
J_n&=&\frac{\sqrt{2\pi}}{4}2^{-\frac{3n}{2}}\frac{n!}{(\frac{n}{2})!}\quad\mbox{for n even}\nonumber\\
J_n&=&\half 2^{-\frac{(n-1)}{2}}(\frac{n-1}{2})!\quad\mbox{for n odd}
\ea
Introducing again the Gau\ss{} bracket $[.]$, we can immediately check that
\be
J_n\leq\frac{\sqrt{2\pi}[\frac{n}{2}]!}{4\cdot 2^{[\frac{n}{2}]}}
\ee
Using $n!\leq e(\frac{(n+1)}{e})^{(n+1)}$ we may further estimate 
\be
J_n\leq\frac{e\sqrt{2\pi}}{4}\frac{(\frac{(n+1)}{2e}^{\frac{(n+1)}{2}}}{2^{\frac{(n-1)}{2}}}=\frac{e\sqrt{2\pi}}{4}2^{-n}\lb(\frac{n+1}{e}\rb)^{\frac{(n+1)}{2}}
\ee
where $\frac{(n-1)}{2}\leq[\frac{n}{2}]\leq\frac{n}{2}$.\\
Finally, if $n\leq n_9$, then
\ba
\sqrt{\frac{2}{\pi}}^9\int\limits_{\Rl^9} d^9 \xmm e^{-2||\xmm||^2}P(||\xmm||)^n&\leq&V_8\sqrt{\frac{2}{\pi}}^9\lb(\frac{9+6n}{4e}\rb)^{\frac{9}{2}}\lb[\sum\limits_{k=0}^{6}a_k\sqrt{\frac{9+6n}{4e}}^k\rb]^n\nonumber\\
&=:&K_{9,6}\lb(\frac{9+6n}{4e}\rb)^{\frac{9}{2}}P(\sqrt{\frac{9+6n}{4e}})
\ea
Using the above estimate, we can bound the remainder from above
\ba
\label{error}
\lefteqn{\lb|\int\limits_{\Rl^9}d^9\xmgen e^{-2||\xmm||^2}\lb(f^{(n+1)}_{\frac{1}{8}}(\wt{y})(\wt{y}-1)^{(n+1)}-f^{(n+1)}_{\frac{1}{8}}(\wt{y}_1)(\wt{y}_1-1)^{(n+1)}\rb)\rb|}\nonumber\\
&\leq&
\int\limits_{\Rl^9}d^9\xmgen e^{-2||\xmm||^2}(3\cdot 8)^{(n+1)}\lb[P(||\xmm||)^{(n+2)}+2 P(||\xmm||)^{(n+1)}\rb.\hspace{3cm}\nonumber\\
&&\lb.\hspace{5.75cm}
+P_1(||\xmm||)^{(n+2)}+2 P_1(||\xmm||)^{(n+1)}\rb]\nonumber\\
&\leq&
K_{9,6}(3\cdot 8)^{(n+1)}\lb\{\lb(\frac{9+6(n+2)}{4e}\rb)^{\frac{9}{2}}\lb[\lb(P\lb(\sqrt{\frac{9+6(n+2)}{4e}}\rb)\rb)^{(n+2)}+\lb(P_1\lb(\sqrt{\frac{9+6(n+2)}{4e}}\rb)\rb)^{(n+2)}\rb]\rb.\nonumber\\
&&\lb.
+2\lb(\frac{9+6(n+1)}{4e}\rb)^{\frac{9}{2}}\lb[\lb(P\lb(\sqrt{\frac{9+6(n+1)}{4e}}\rb)\rb)^{(n+1)}+\lb(P_1\lb(\sqrt{\frac{9+6(n+1)}{4e}}\rb)\rb)^{(n+1)}\rb]\rb\}
\ea
As pointed out in \cite{Hanno}, for small values of $n$ the error connected with the remainder is proportional to $s^{n+1}$. However, for larger values of $n$ the size of the error becomes comparable to the order of accuracy (in powers of $s$) up to which we have performed the expansion. Thus, we are interested in the value $n_0$ from where onwards the error becomes so large that it does not make sene to compute corrections. An estimate for $n_0$ can be derived from the condition
\ba
\frac{\lb|\int\limits_{\Rl^9}d^9\xmgen e^{-2||\xmm||^2}\lb(f^{(n+2)}_{\frac{1}{8}}(\wt{y})(\wt{y}-1)^{(n+2)}-f^{(n+2)}_{\frac{1}{8}}(\wt{y}_1)(\wt{y}_1-1)^{(n+2)}\rb)\rb|}{\lb|\int\limits_{\Rl^9}d^9\xmgen e^{-2||\xmm||^2}\lb(f^{(n+1)}_{\frac{1}{8}}(\wt{y})(\wt{y}-1)^{(n+1)}-f^{(n+1)}_{\frac{1}{8}}(\wt{y}_1)(\wt{y}_1-1)^{(n+1)}\rb)\rb|}&\ge&1
\ea
Since the upper bound in eqn (\ref{error}) looks rather complicated $n_0$ cannot be computed analytically. Nevertheless, the order of magnitude of $n_0$ can be obtained under the assumption that the value of $n_0$ is supposed to be quite large and therefore that the change of $P(\frac{9+6(n_0+2)}{4e})$ as we replace $n_0$ by $n_0+1$ is much smaller than the value of $n_0$ itself. Under this assumption the value of $n_0$ \cite{Hanno} is given by
\be
n_0=\frac{4e\lb(\frac{\tau_0(8)}{s||\qmgen||}\rb)^2-9}{6}-3
\ee
whereby $\tau_0(8)$ is of order unity\cite{Hanno}. Consequently, since $\qmgen$ is of order unity, we conclude as long as $s=t^{\frac{1}{2}-\alpha}$ is small, the value of $n_0\gg 1$. Hence, the precise value of $n_0$ depends on the chosen value for $\alpha$.
\section{Explicit Expressions for $y,y_1,\wt{y}$ and $\wt{y}_1$}
\label{Expy}
In this section we will derive the explicit terms for $y,y_1,\wt{y}$ and $\wt{y}_1$ that occur in the expansion of the $\Lambda^{\half}$-functions up to order $O((sT)^2)$.
\ba
\lefteqn{\Lambda^{\half}\lb(\lb\{\xmgenn+\frac{1}{T}\lb(\pmgen\rb)+\frac{T}{4}\lb[\sgn(\snt)\Dtmt-\sgn(\sn)\Dtm\rb]\rb\},\half\sgn(\sn)\e{}{\Kn}{v},\nn\rb)}\nonumber\\
&=&\frac{a^{\frac{3}{2}}}{i\hbar}|\det(\pmm)|^{\frac{1}{4}}\nonumber\\
&&\lb\{(y-y_1)\lb(f^{(1)}_{\frac{1}{8}}(1)+f^{(2)}_{\frac{1}{8}}(1)\lb(y-y_1-2\rb)+f^{(3)}_{\frac{1}{8}}(1)\lb((y_1-1)^2+(y-1)(y_1-1)+(y-1)^2\rb)\rb)\rb\}\hspace{0.5cm}
\ea
and
\ba
\lefteqn{\Lambda^{\half}\lb(\lb\{\xmgenn+\frac{1}{T}\lb(\pmgen\rb)+\frac{T}{4}\lb[\sgn(\snt)\Dtmt-\sgn(\sn)\Dtm\rb]\rb\},\half\sgn(\snt)\e{}{\Knt}{v},\nnt\rb)}\nonumber\\
&=&\frac{a^{\frac{3}{2}}}{(-i\hbar)}|\det(\pmm)|^{\frac{1}{4}}\nonumber\\
&&\lb\{(\wt{y}-\wt{y}_1)\lb(f^{(1)}_{\frac{1}{8}}(1)+f^{(2)}_{\frac{1}{8}}(1)\lb(\wt{y}-\wt{y}_1-2\rb)+f^{(3)}_{\frac{1}{8}}(1)\lb((\wt{y}_1-1)^2+(\wt{y}-1)(\wt{y}_1-1)+(\wt{y}-1)^2\rb)\rb)\rb\}\hspace{0.5cm}
\ea
 The lowerst order in the term $(y-y_1)$ and $(\wt{y}-\wt{y}_1)$ respectively is $sT$. Since we have a global term $(y-y_1)$ and $(\wt{y}-\wt{y}_1)$ respectively and the highest order we want to consider is $s^3T$, we will expand $(y-y_1)$ and $(\wt{y}-\wt{y}_1)$ respectively up to order $O((sT)^2)$ and all the others terms that are multiplied with these terms up to order $O(s^3)$. 
Using the definition of  $y,y_1,\wt{y}$ and $\wt{y}_1$ in terms of the corresponding $z_{sq^{-1}x+\Delta},z_{sq^{-1}x+\Delta+\delta},z_{sq^{-1}x-\Delta}$ and $z_{sq^{-1}x-\Delta+\delta}$
we obtain
\ba
\label{yminy1}
y-y_1&=&z_{sq^{-1}x+\Delta}-z_{sq^{-1}x+\Delta+\delta}\nonumber\\
&=&-sT\sgn(\sn)\qmm_{\Kn\nn}-s^2T\sgn(\sn)\lb(2\qmm_{\Kn\nn}\qmm_{Mm}-\qmm_{\Kn m}\qmm_{M\nn}\rb)\xmm_{Mm}\nonumber\\
&&-2\frac{s^3T}{3!}\epsilon_{ijk}\nonumber\\
&&\lb[
+\lb(\frac{1}{2}\epsilon_{\nn mn}\qmm_{\Kn i}\sgn(\sn)\rb)
\lb(\qmm_{Mj}\xmm_{Mm}\rb)\lb(\qmm_{Nk}\xmm_{Nn}\rb)\rb.\nonumber\\
&&+\lb(\frac{1}{2}\epsilon_{\ell\nn n}\qmm_{\Kn j}\sgn(\sn)\rb)
\lb(\qmm_{Li}\xmm_{L\ell}\rb)\lb(\qmm_{Nk}\xmm_{Nn}\rb)\nonumber\\
&&\lb.
-\lb(\frac{1}{2}\epsilon_{\ell m\nnt}\qmm_{\Kn k}\sgn(\sn)\rb)
\lb(\qmm_{Li}\xmm_{L\ell}\rb)\lb(\qmm_{Mj}\xmm_{Mm}\rb)\rb]\nonumber\\
&&-\frac{s^3T}{2}\sgn(\sn)\qmm_{Nn}\lb(\qmm_{Mm}\qmm_{\Kn\nn}-\qmm_{M\nnt}\qmm_{\Kn m}\rb)\xmm_{Mm}\xmm_{Nn}+O((sT)^2)\nonumber\\
\ea 
\ba
\label{ymin1}
y-1&=&z_{sq^{-1}x+\Delta}\nonumber\\
&=&2s\qmm_{Mm}\xmm_{Mm}+\frac{sT}{2}\lb(\sgn(\snt)(\qmm_{\Jnt\mnt}-\qmm_{\Int\mnt})-\sgn(\sn)(\qmm_{\Jn\mn}-\qmm_{\In\mn})\rb)\nonumber\\
&&
+s^2\lb(2\qmm_{Mm}\qmm_{Nn}-\qmm_{Mn}\qmm_{Nm}\rb)\xmm_{Mm}\xmm_{Nn}\nonumber\\
&&
+\frac{s^2T}{2}\xmm_{Mm}\lb[\lb(2\sgn(\snt)((\qmm_{\Jnt\mnt}-\qmm_{\Int\mnt})-\sgn(\sn)(\qmm_{\Jn\mn}-\qmm_{\In\mn})\rb)\qmm_{Mm}\rb.\nonumber\\
&&\lb.
-\lb(\qmm_{M\mnt}\sgn(\snt)((\qmm_{\Jnt m}-\qmm_{\Int m})-\qmm_{M\mn}\sgn(\sn)(\qmm_{\Jn m}-\qmm_{\In m})\rb)\rb]
+O(s^3)\nonumber\\
\ea 
\ba
\label{y1min1}
y_1-1&=&(y-1)+sT\sgn(\sn)\qmm_{\Kn\nn}+s^2T\lb(2\qmm_{\Kn\nn}\qmm_{Mm}-\qmm_{\Kn m}\qmm_{M\nn}\rb)\xmm_{Mm}\nonumber\\
&&+2\frac{s^3T}{3!}\epsilon_{ijk}\nonumber\\
&&\lb[
+\lb(\frac{1}{2}\epsilon_{\nn mn}\qmm_{\Kn i}\sgn(\sn)\rb)
\lb(\qmm_{Mj}\xmm_{Mm}\rb)\lb(\qmm_{Nk}\xmm_{Nn}\rb)\rb.\nonumber\\
&&+\lb(\frac{1}{2}\epsilon_{\ell\nn n}\qmm_{\Kn j}\sgn(\sn)\rb)
\lb(\qmm_{Li}\xmm_{L\ell}\rb)\lb(\qmm_{Nk}\xmm_{Nn}\rb)\nonumber\\
&&\lb.
+\lb(\frac{1}{2}\epsilon_{\ell m\nn}\qmm_{\Kn k}\sgn(\sn)\rb)
\lb(\qmm_{Li}\xmm_{L\ell}\rb)\lb(\qmm_{Mj}\xmm_{Mm}\rb)\rb]\nonumber\\
&&+\frac{s^3T}{2}\sgn(\sn)\qmm_{Nn}\lb(\qmm_{Mm}\qmm_{\Kn\nn}-\qmm_{M\nnt}\qmm_{\Kn m}\rb)\xmm_{Mm}\xmm_{Nn}+O((sT)^2)\nonumber\\
\ea 
and
\ba
\label{tyminy1}
\wt{y}-\wt{y}_1&=&z_{sq^{-1}x-\Delta}-z_{sq^{-1}x-\Delta+\tilde{\delta}}\nonumber\\
&=&-sT\qmm_{Mm}\sgn(\snt)\qmm_{\Knt\nnt}-s^2T\sgn(\snt)\lb(2\qmm_{\Knt\nnt}\qmm_{Mm}-\qmm_{\Knt m}\qmm_{M\nnt}\rb)\xmm_{Mm}\nonumber\\
&&-2\frac{s^3T}{3!}\epsilon_{ijk}\nonumber\\
&&\lb[
+\lb(\frac{1}{2}\epsilon_{\nnt mn}\qmm_{\Knt i}\sgn(\snt)\rb)
\lb(\qmm_{Mj}\xmm_{Mm}\rb)\lb(\qmm_{Nk}\xmm_{Nn}\rb)\rb.\nonumber\\
&&+\lb(\frac{1}{2}\epsilon_{\ell\nnt n}\qmm_{\Knt j}\sgn(\snt)\rb)
\lb(\qmm_{Li}\xmm_{L\ell}\rb)\lb(\qmm_{Nk}\xmm_{Nn}\rb)\nonumber\\
&&\lb.
+\lb(\frac{1}{2}\epsilon_{\ell m\nnt}\qmm_{\Knt k}\sgn(\snt)\rb)
\lb(\qmm_{Li}\xmm_{L\ell}\rb)\lb(\qmm_{Mj}\xmm_{Mm}\rb)\rb]\nonumber\\
&&-\frac{s^3T}{2}\sgn(\snt)\qmm_{Nn}\lb(\qmm_{Mm}\qmm_{\Kn\nnt}-\qmm_{M\nnt}\qmm_{\Knt m}\rb)\xmm_{Mm}\xmm_{Nn}+O((sT)^2)\nonumber\\
\ea  
\ba
\label{tymin1}
\wt{y}-1&=&z_{sq^{-1}x-\Delta}\nonumber\\
&=&2s\qmm_{Mm}\xmm_{Mm}-\frac{sT}{2}\lb(\sgn(\snt)(\qmm_{\Jnt\mnt}-\qmm_{\Int\mnt})-\sgn(\sn)(\qmm_{\Jn\mn}-\qmm_{\In\mn})\rb)\nonumber\\
&&
+s^2\lb(2\qmm_{Mm}\qmm_{Nn}-\qmm_{Mn}\qmm_{Nm}\rb)\xmm_{Mm}\xmm_{Nn}\nonumber\\
&&
-\frac{s^2T}{2}\xmm_{Mm}\lb[2\lb(\sgn(\snt)\lb(\qmm_{\Jnt\mnt}-\qmm_{\Int\mnt}\rb)-\sgn(\sn)\lb(\qmm_{\Jn\mn}-\qmm_{\In\mn}\rb)\rb)\qmm_{Mm}\rb.\nonumber\\
&&\lb.
-\lb(\qmm_{M\mnt}\sgn(\snt)\lb(\qmm_{\Jnt m}-\qmm_{\Int m}\rb)-\qmm_{M\mn}\sgn(\sn)\lb(\qmm_{\Jn m}-\qmm_{\In m}\rb)\rb)\rb]
+O(s^3)\nonumber\\
\ea 
\ba
\label{ty1min1}
\wt{y}_1-1&=&(\wt{y}-1)+sT\sgn(\snt)\qmm_{\Knt\nnt}+s^2T\lb(2\qmm_{\Knt\nnt}\qmm_{Mm}-\qmm_{\Knt m}\qmm_{M\nn}\rb)\xmm_{Mm}\nonumber\\
&&+2\frac{s^3T}{3!}\epsilon_{ijk}\nonumber\\
&&\lb[
+\lb(\frac{1}{2}\epsilon_{\nnt mn}\qmm_{\Knt i}\sgn(\snt)\rb)
\lb(\qmm_{Mj}\xmm_{Mm}\rb)\lb(\qmm_{Nk}\xmm_{Nn}\rb)\rb.\nonumber\\
&&+\lb(\frac{1}{2}\epsilon_{\ell\nnt n}\qmm_{\Knt j}\sgn(\snt)\rb)
\lb(\qmm_{Li}\xmm_{L\ell}\rb)\lb(\qmm_{Nk}\xmm_{Nn}\rb)\nonumber\\
&&\lb.
+\lb(\frac{1}{2}\epsilon_{\ell m\nnt}\qmm_{\Knt k}\sgn(\snt)\rb)
\lb(\qmm_{Li}\xmm_{L\ell}\rb)\lb(\qmm_{Mj}\xmm_{Mm}\rb)\rb]\nonumber\\
&&+\frac{s^3T}{2}\sgn(\snt)\qmm_{Nn}\lb(\qmm_{Mm}\qmm_{\Knt\nnt}-\qmm_{M\nnt}\qmm_{\Knt m}\rb)\xmm_{Mm}\xmm_{Nn}+O((sT)^2)\nonumber\\
\ea 
Since
\ba
y_1-1&=&y-1+(y_1-y)\nonumber\\
\wt{y}_1-1&=&\wt{y}-1+(\wt{y}_1-\wt{y})
\ea
we get
\ba
(y_1-1)^2&=&(y-1)^2+2(y-1)(y_1-y)+(y_1-y)^2\nonumber\\
(\wt{y}_1-1)^2&=&(\wt{y}-1)^2+2(\wt{y}-1)(\wt{y}_1-\wt{y})+(\wt{y}_1-\wt{y})^2\nonumber\\
(y-1)(y_1-1)&=&(y-1)^2+(y_1-y)(y-1)\nonumber\\
(\wt{y}-1)(\wt{y}_1-1)&=&(\wt{y}-1)^2+(\wt{y}_1-\wt{y})(\wt{y}-1)
\ea
Reinserting this into the expansion of the $\Lambda^{\half}-$functions we obtain
\ba
\label{ExpL12}
\lefteqn{\Lambda^{\half}\lb(\lb\{\xmgenn+\frac{1}{T}\lb(\pmgen\rb)+\frac{T}{4}\lb[\sgn(\snt)\Dtmt-\sgn(\sn)\Dtm\rb]\rb\},\half\sgn(\sn)\e{}{\Kn}{v},\nn\rb)}\nonumber\\
&=&\frac{a^{\frac{3}{2}}|\det(\pmm)|^{\frac{1}{4}}}{i\hbar}
(y-y_1)\lb\{\lb(f^{(1)}_{\frac{1}{8}}(1)+f^{(2)}_{\frac{1}{8}}(1)\lb(2(y-1)+(y_1-y)\rb)\rb.\rb.\hspace{2cm}\nonumber\\
&&\lb.\lb.
+f^{(3)}_{\frac{1}{8}}(1)\lb(2(y_1-1)^2+3(y-1)(y_1-1)+(y_1-y)^2\rb)\rb)\rb\}
\ea
and
\ba
\label{ExpL22}
\lefteqn{\Lambda^{\half}\lb(\lb\{\xmgenn+\frac{1}{T}\lb(\pmgen\rb)-\frac{T}{4}\lb[\sgn(\snt)\Dtmt-\sgn(\sn)\Dtm\rb]\rb\},\half\sgn(\snt)\e{}{\Knt}{v},\nnt\rb)}\nonumber\\
&=&\frac{a^{\frac{3}{2}}|\det(\pmm)|^{\frac{1}{4}}}{(-i\hbar)}
(\wt{y}-\wt{y}_1)\lb\{\lb(f^{(1)}_{\frac{1}{8}}(1)+f^{(2)}_{\frac{1}{8}}(1)\lb(2(\wt{y}-1)+(\wt{y}_1-\wt{y})\rb)\rb.\rb.\hspace{2cm}\nonumber\\
&&\lb.\lb.
+f^{(3)}_{\frac{1}{8}}(1)\lb(2(\wt{y}_1-1)^2+3(\wt{y}-1)(\wt{y}_1-1)+(\wt{y}_1-\wt{y})^2\rb)\rb)\rb\}
\ea
\subsection{The Leading Order Term of $\Lambda^{\half}$-functions}
\label{LO}
The leading order term of $\Lambda^{\half}$ is of the order $sT/t$ due to the $\hbar$ in the denominator in the equations (\ref{ExpL12}) and (\ref{ExpL22}). Hence, the leading order contribution of $\Lambda^{\half}$ is given by
\ba
\lefteqn{\Lambda^{\half}\lb(\lb\{\xmgenn+\frac{1}{T}\lb(\pmgen\rb)+\frac{T}{4}\lb[\sgn(\snt)\Dtmt-\sgn(\sn)\Dtm\rb]\rb\},\half\sgn(\sn)\e{}{\Kn}{v},\nn\rb)}\nonumber\\
&=&\frac{a^{\frac{3}{2}}|\det(\pmm)|^{\frac{1}{4}}}{i\hbar}f^{(1)}_{\frac{1}{8}}(1)(y-y_1)\Big|_{sT}\hspace{9cm}
\ea
and
\ba
\lefteqn{\Lambda^{\half}\lb(\lb\{\xmgenn+\frac{1}{T}\lb(\pmgen\rb)-\frac{T}{4}\lb[\sgn(\snt)\Dtmt-\sgn(\sn)\Dtm\rb]\rb\},\half\sgn(\snt)\e{}{\Knt}{v},\nnt\rb)}\nonumber\\
&=&\frac{a^{\frac{3}{2}}|\det(\pmm)|^{\frac{1}{4}}}{i\hbar}f^{(1)}_{\frac{1}{8}}(1)
(\wt{y}-\wt{y}_1)\Big|_{sT}\hspace{9cm}
\ea
whereby in $(y-y_1)$ and $(\wt{y}-\wt{y}_1)$ only terms of order $sT$ are considered. Recalling the equations (\ref{yminy1})) and (\ref{tyminy1}), we obtain  for the  $\Lambda^{\half}$-functions in leading order the following result
\ba
\lefteqn{\Lambda^{\half}\lb(\lb\{\xmgenn+\frac{1}{T}\lb(\pmgen\rb)+\frac{T}{4}\lb[\sgn(\snt)\Dtmt-\sgn(\sn)\Dtm\rb]\rb\},\half\sgn(\sn)\e{}{\Kn}{v},\nn\rb)}\nonumber\\
&=&\frac{a^{\frac{3}{2}}|\det(\pmm)|^{\frac{1}{4}}}{i\hbar}f^{(1)}_{\frac{1}{8}}(1)\lb(-sT\sgn(\sn)\qmm_{\Kn\nn}\rb)\hspace{9cm}
\ea
and
\ba
\lefteqn{\Lambda^{\half}\lb(\lb\{\xmgenn+\frac{1}{T}\lb(\pmgen\rb)-\frac{T}{4}\lb[\sgn(\snt)\Dtmt-\sgn(\sn)\Dtm\rb]\rb\},\half\sgn(\snt)\e{}{\Knt}{v},\nnt\rb)}\nonumber\\
&=&\frac{a^{\frac{3}{2}}|\det(\pmm)|^{\frac{1}{4}}}{(-i\hbar)}f^{(1)}_{\frac{1}{8}}(1)
\lb(-sT\sgn(\snt)\qmm_{\Knt\nnt}\rb)\hspace{9cm}
\ea
\section{The Next-to-Leading Order Contribution to the Algebraic Master Constraint Expectation Value}
\label{NLO}
Here we expand each $\Lambda^{\half}$ function up to order $O(s^3T/t)$. Afterwards we take the product of these two functions and consider all terms up to the order $O((sT/t)^2s^2)$.\\
The precise expresion for the expansion reads
\ba
\lefteqn{\Lambda^{\half}\lb(\lb\{\xmgenn+\frac{1}{T}\lb(\pmgen\rb)+\frac{T}{4}\lb[\sgn(\snt)\Dtmt-\sgn(\sn)\Dtm\rb]\rb\},\half\sgn(\sn)\e{}{\Kn}{v},\nn\rb)}\nonumber\\
&&\hspace{-0.5cm}
\Lambda^{\half}\lb(\lb\{\xmgenn+\frac{1}{T}\lb(\pmgen\rb)-\frac{T}{4}\lb[\sgn(\snt)\Dtmt-\sgn(\sn)\Dtm\rb]\rb\},\half\sgn(\snt)\e{}{\Knt}{v},\nnt\rb)\nonumber\\
&=&\lb(\frac{a^{\frac{3}{2}}|\det(\pmm)|^{\frac{1}{4}}}{\hbar}\rb)^2\lb(y-y_1\rb)\lb(\wt{y}-\wt{y}_1\rb)\nonumber\\
&&
\lb[\lb(f^{(1)}_{\frac{1}{8}}(1)\rb)^2 + f^{(1)}_{\frac{1}{8}}(1)f^{(2)}_{\frac{1}{8}}(1)\lb[2(y-1)+(y_1-1)+2(\wt{y}-1)+(\wt{y}_1-1)\rb]\rb.\nonumber\\
&&
+ f^{(1)}_{\frac{1}{8}}(1)f^{(3)}_{\frac{1}{8}}(1)\lb[2(y-1)^2+3(y-1)(y_1-1)+(y_1-y)^2 + 2(\wt{y}-1)^2+3(\wt{y}-1)(\wt{y}_1-1)+(\wt{y}_1-\wt{y})^2\rb]\nonumber\\
&&
+\lb(f^{(2)}_{\frac{1}{8}}(1)\rb)^2\lb[2(y-1)+(y_1-1)\rb]\lb[2(\wt{y}-1)+(\wt{y}_1-1)\rb]\nonumber\\
&&
+ f^{(2)}_{\frac{1}{8}}(1)f^{(3)}_{\frac{1}{8}}(1)\lb(
\lb[2(\wt{y}-1)+(\wt{y}_1-1)\rb]\lb[2(y-1)^2+3(y-1)(y_1-1)+(y_1-y)^2\rb]\rb.\nonumber\\
&&\lb.\hspace{2.5cm}
+\lb[2(y-1)+(y_1-1)\rb]\lb[2(\wt{y}-1)^2+3(\wt{y}-1)(\wt{y}_1-1)+(\wt{y}_1-1)^2\rb]\rb)\nonumber\\
&&\lb.
+\lb(f^{(3)}_{\frac{1}{8}}(1)\rb)^2\lb[2(\wt{y}-1)^2+3(\wt{y}-1)(\wt{y}_1-1)+(\wt{y}_1-1)^2\rb]
\lb[2(y_-1)^2+3(y-1)(y_1-1)+(y_1-1)^2\rb]\rb]\Big|_{s^2(sT/t)^2}\nonumber\\
&&+O(s^2(sT/t)^2)
\ea
whereby $\Big|_{s^2(sT/t)^2}$ means that only terms of maximal this order are considered although apparently higher order terms will occur due to for instance the squares and products of $y,y_1,\wt{y}$ and $\wt{y}_1$ respectively.
 As before we expand $(y-y_1)$ and $(\wt{y}-\wt{y}_1)$ up to order $O((sT)^2)$ while the other occurring terms have to be expanded up to $O(s^3)$ only.
The separated terms are given by
\ba
y-y_1
&=&-sT\sgn(\sn)\qmm_{\Kn\nn}-s^2T\sgn(\sn)\lb(2\qmm_{\Kn\nn}\qmm_{Mm}-\qmm_{\Kn m}\qmm_{M\nn}\rb)\xmm_{Mm}\nonumber\\
&&-2\frac{s^3T}{3!}\epsilon_{ijk}\nonumber\\
&&\lb[
+\lb(\frac{1}{2}\epsilon_{\nn mn}\qmm_{\Kn i}\sgn(\sn)\rb)
\lb(\qmm_{Mj}\xmm_{Mm}\rb)\lb(\qmm_{Nk}\xmm_{Nn}\rb)\rb.\nonumber\\
&&+\lb(\frac{1}{2}\epsilon_{\ell\nn n}\qmm_{\Kn j}\sgn(\sn)\rb)
\lb(\qmm_{Li}\xmm_{L\ell}\rb)\lb(\qmm_{Nk}\xmm_{Nn}\rb)\nonumber\\
&&\lb.
+\lb(\frac{1}{2}\epsilon_{\ell m\nn}\qmm_{\Kn k}\sgn(\sn)\rb)
\lb(\qmm_{Li}\xmm_{L\ell}\rb)\lb(\qmm_{Mj}\xmm_{Mm}\rb)\rb]\nonumber\\
&&-\frac{s^3T}{2}\sgn(\sn)\qmm_{Nn}\lb(\qmm_{Mm}\qmm_{\Kn\nn}-\qmm_{M\nnt}\qmm_{\Kn m}\rb)\xmm_{Mm}\xmm_{Nn}+O((sT)^2)\nonumber\\
\ea
\ba
\wt{y}-\wt{y}_1
&=&-sT\qmm_{Mm}\sgn(\snt)\qmm_{\Knt\nnt}-s^2T\sgn(\snt)\lb(2\qmm_{\Knt\nnt}\qmm_{Mm}-\qmm_{\Knt m}\qmm_{M\nnt}\rb)\xmm_{Mm}\nonumber\\
&&-2\frac{s^3T}{3!}\epsilon_{ijk}\nonumber\\
&&\lb[
+\lb(\frac{1}{2}\epsilon_{\nnt mn}\qmm_{\Knt i}\sgn(\snt)\rb)
\lb(\qmm_{Mj}\xmm_{Mm}\rb)\lb(\qmm_{Nk}\xmm_{Nn}\rb)\rb.\nonumber\\
&&+\lb(\frac{1}{2}\epsilon_{\ell\nnt n}\qmm_{\Knt j}\sgn(\snt)\rb)
\lb(\qmm_{Li}\xmm_{L\ell}\rb)\lb(\qmm_{Nk}\xmm_{Nn}\rb)\nonumber\\
&&\lb.
+\lb(\frac{1}{2}\epsilon_{\ell m\nnt}\qmm_{\Knt k}\sgn(\snt)\rb)
\lb(\qmm_{Li}\xmm_{L\ell}\rb)\lb(\qmm_{Mj}\xmm_{Mm}\rb)\rb]\nonumber\\
&&-\frac{s^3T}{2}\sgn(\snt)\qmm_{Nn}\lb(\qmm_{Mm}\qmm_{\Kn\nnt}-\qmm_{M\nnt}\qmm_{\Knt m}\rb)\xmm_{Mm}\xmm_{Nn}+O((sT)^2)\nonumber\\
\ea   
\ba
2(y-1)+(y_1-y)&=&
4s\qmm_{Mm}\xmm_{Mm}
+2s^2\lb(2\qmm_{Mm}\qmm_{Nn}-\qmm_{Mn}\qmm_{Nm}\rb)\xmm_{Mm}\xmm_{Nn}+O(s^3)\nonumber\\
2(\wt{y}-1)+(\wt{y}_1-\wt{y})&=&
4s\qmm_{Mm}\xmm_{Mm}+2s^2\lb(2\qmm_{Mm}\qmm_{Nn}-\qmm_{Mn}\qmm_{Nm}\rb)\xmm_{Mm}\xmm_{Nn}+O(s^3)\nonumber\\
(y-1)^2&=&4s^2\qmm_{Mm}\qmm_{Nn}\xmm_{Mm}\xmm_{Nn}+O(s^3)\nonumber\\
(y-1)(y_1-1)&=&4s^2\qmm_{Mm}\qmm_{Nn}\xmm_{Mm}\xmm_{Nn}+O(s^3)\nonumber\\
(\wt{y}-1)(\wt{y}_1-1)&=&4s^2\qmm_{Mm}\qmm_{Nn}\xmm_{Mm}\xmm_{Nn}+O(s^3)\nonumber\\
\ea
The term $(y_1-y)^2$ and $(\wt{y}_1-y)$ are already of order $O(s^3)$ and hence do not have to be considered.
\ba
(\wt{y}-1)^2&=&4s^2\qmm_{Mm}\qmm_{Nn}\xmm_{Mm}\xmm_{Nn}+O(s^3)
\ea
The terms $(\wt{y}_1-\wt{y})^2$ and $(\wt{y}-1)(\wt{y}_1-y)$ are of order $O(s^3)$ and will therefore be neglected in the further discussion.
\ba
(y-1)(\wt{y}-1)&=&4s^2\qmm_{Mm}\qmm_{Nn}\xmm_{Mm}\xmm_{Nn}+O(s^3)\nonumber\\
(y-1)(\wt{y}_1-1)&=&4s^2\qmm_{Mm}\qmm_{Nn}\xmm_{Mm}\xmm_{Nn}+O(s^3)\nonumber\\
(y_1-1)(\wt{y}-1)&=&4s^2\qmm_{Mm}\qmm_{Nn}\xmm_{Mm}\xmm_{Nn}+O(s^3)
\ea
Thus, when neglecting the terms of order $O(s^2(sT/t)^2)$ the expansion reduces to
\ba
\lefteqn{\Lambda^{\half}\lb(\lb\{\xmgenn+\frac{1}{T}\lb(\pmgen\rb)+\frac{T}{4}\lb[\sgn(\snt)\Dtmt-\sgn(\sn)\Dtm\rb]\rb\},\half\sgn(\sn)\e{}{\Kn}{v},\nn\rb)}\nonumber\\
&&\hspace{-0.5cm}
\Lambda^{\half}\lb(\lb\{\xmgenn+\frac{1}{T}\lb(\pmgen\rb)-\frac{T}{4}\lb[\sgn(\snt)\Dtmt-\sgn(\sn)\Dtm\rb]\rb\},\half\sgn(\snt)\e{}{\Knt}{v},\nnt\rb)\nonumber\\
&=&\lb(\frac{a^{\frac{3}{2}}|\det(\pmm)|^{\frac{1}{4}}}{\hbar}\rb)^2\lb(y-y_1\rb)\lb(\wt{y}-\wt{y}_1\rb)
\lb[\lb(f^{(1)}_{\frac{1}{8}}(1)\rb)^2 + f^{(1)}_{\frac{1}{8}}(1)f^{(2)}_{\frac{1}{8}}(1)\lb[2(y-1)+(y_1-1)+2(\wt{y}-1)+(\wt{y}_1-1)\rb]\rb.\nonumber\\
&&\hspace{3.2cm}
+ f^{(1)}_{\frac{1}{8}}(1)f^{(3)}_{\frac{1}{8}}(1)
\lb[2(y-1)^2+3(y-1)(y_1-1) + 2(\wt{y}-1)^2+3(\wt{y}-1)(\wt{y}_1-1)\rb]\nonumber\\
&&\lb.\hspace{3.2cm}
+\lb(f^{(2)}_{\frac{1}{8}}(1)\rb)^2\lb[4(y-1)(\wt{y}-1)+2(y-1)(\wt{y}_1-1)+2(y_1-1)(\wt{y}-1)\rb]\rb]+O(s^2(sT/t)^2)\nonumber\\
\ea
Reinserting these various terms into the expansion of the $\Lambda^{\half}-$functions, we obtain
\ba
\lefteqn{\Lambda^{\half}\lb(\lb\{\xmgenn+\frac{1}{T}\lb(\pmgen\rb)+\frac{T}{4}\lb[\sgn(\snt)\Dtmt-\sgn(\sn)\Dtm\rb]\rb\},\half\sgn(\sn)\e{}{\Kn}{v},\nn\rb)}\nonumber\\
&&\hspace{-0.5cm}
\Lambda^{\half}\lb(\lb\{\xmgenn+\frac{1}{T}\lb(\pmgen\rb)-\frac{T}{4}\lb[\sgn(\snt)\Dtmt-\sgn(\sn)\Dtm\rb]\rb\},\half\sgn(\snt)\e{}{\Knt}{v},\nnt\rb)\nonumber\\
&=&\lb(\frac{a^{\frac{3}{2}}|\det(\pmm)|^{\frac{1}{4}}}{\hbar}\rb)^2\lb(sT\rb)^2\nonumber\\
&&\lb(\sgn(\sn)\qmm_{\Kn\nn}+s\sgn(\sn)\lb(2\qmm_{\Kn\nn}\qmm_{Nn}-\qmm_{\Kn n}\qmm_{N\nn}\rb)\xmm_{Mm}\rb.\nonumber\\
&&+2\frac{s^2}{3!}\epsilon_{ijk}\nonumber\\
&&\lb[
+\lb(\frac{1}{2}\epsilon_{\nn mn}\qmm_{\Kn i}\sgn(\sn)\rb)
\lb(\qmm_{Mj}\xmm_{Mm}\rb)\lb(\qmm_{Nk}\xmm_{Nn}\rb)\rb.\nonumber\\
&&+\lb(\frac{1}{2}\epsilon_{\ell\nn n}\qmm_{\Kn j}\sgn(\sn)\rb)
\lb(\qmm_{Li}\xmm_{L\ell}\rb)\lb(\qmm_{Nk}\xmm_{Nn}\rb)\nonumber\\
&&\lb.
+\lb(\frac{1}{2}\epsilon_{\ell m\nn}\qmm_{\Kn k}\sgn(\sn)\rb)
\lb(\qmm_{Li}\xmm_{L\ell}\rb)\lb(\qmm_{Mj}\xmm_{Mm}\rb)\rb]\nonumber\\
&&\lb.
+\frac{s^2}{2}\sgn(\sn)\qmm_{Nn}\lb(\qmm_{Mm}\qmm_{\Kn\nn}-\qmm_{M\nnt}\qmm_{\Kn m}\rb)\xmm_{Mm}\xmm_{Nn}\rb)\nonumber\\
&&\lb(\sgn(\snt)\qmm_{\Knt\nnt}+s\sgn(\snt)\lb(2\qmm_{\Knt\nnt}\qmm_{Nn}-\qmm_{\Knt n}\qmm_{N\nnt}\rb)\xmm_{Mm}\rb.\nonumber\\
&&+2\frac{s^2}{3!}\epsilon_{ijk}\nonumber\\
&&\lb[
+\lb(\frac{1}{2}\epsilon_{\nn mn}\qmm_{\Knt i}\sgn(\snt)\rb)
\lb(\qmm_{Mj}\xmm_{Mm}\rb)\lb(\qmm_{Nk}\xmm_{Nn}\rb)\rb.\nonumber\\
&&+\lb(\frac{1}{2}\epsilon_{\ell\nnt n}\qmm_{\Knt j}\sgn(\snt)\rb)
\lb(\qmm_{Li}\xmm_{L\ell}\rb)\lb(\qmm_{Nk}\xmm_{Nn}\rb)\nonumber\\
&&\lb.
+\lb(\frac{1}{2}\epsilon_{\ell m\nnt}\qmm_{\Knt k}\sgn(\snt)\rb)
\lb(\qmm_{Li}\xmm_{L\ell}\rb)\lb(\qmm_{Mj}\xmm_{Mm}\rb)\rb]\nonumber\\
&&\lb.
+\frac{s^2}{2}\sgn(\sn)\qmm_{Nn}\lb(\qmm_{Mm}\qmm_{\Kn\nn}-\qmm_{M\nnt}\qmm_{\Kn m}\rb)\xmm_{Mm}\xmm_{Nn}\rb)\nonumber\\
&&\lb[\lb(f^{(1)}_{\frac{1}{8}}(1)\rb)^2+ s\qmm_{Mm}\xmm_{Mm}8f^{(1)}_{\frac{1}{8}}(1)f^{(2)}_{\frac{1}{8}}(1)\rb.
\nonumber\\
&&+s^2\lb(2\qmm_{Mm}\qmm_{Nn}-\qmm_{Mn}\qmm_{Nm}\rb)\xmm_{Mm}\xmm_{Nn}\lb(4f^{(1)}_{\frac{1}{8}}(1)f^{(2)}_{\frac{1}{8}}(1)\rb)\nonumber\\
&&\lb.
+s^2\qmm_{Mm}\qmm_{Nn}\xmm_{Mm}\xmm_{Nn}\lb(40f^{(1)}_{\frac{1}{8}}(1)f^{(3)}_{\frac{1}{8}}(1)+32\lb(f^{(2)}_{\frac{1}{8}}(1)\rb)^2\rb)\rb]+O(s^2(sT/t)^2)\nonumber\\
\ea
We will order the expansion in powers of $\xmm$ since this is quite useful for the later integration over $\xmm$. Moreover we will neglect the linear powers of $\xmm$, because they will cancel in the integration when integrating against the even function $\exp(-2(\xmm)^2)$. Let us introduce the following shorthands
\ba
C^{Mm}&:=&\qmm_{Mm}\nonumber\\
C^{Mm,Nn}&:=&2\qmm_{Mm}\qmm_{Nn}-\qmm_{Mn}\qmm_{Nm}\nonumber\\
C^{\Kn\sn\nn}&:=&\sgn(\sn)\qmm_{\Kn\nn}\nonumber\\
C^{Mm,\Kn\sn\nn}&:=&\sgn(\sn)\lb(2\qmm_{Mm}\qmm_{\Kn\nn}-\qmm_{\Kn n}\qmm_{N\nn}\rb)
\ea
Then we can rewrite the $\Lambda^{\half}$-expansion as
\ba
\lefteqn{\Lambda^{\half}\lb(\lb\{\xmgenn+\frac{1}{T}\lb(\pmgen\rb)+\frac{T}{4}\lb[\sgn(\snt)\Dtmt-\sgn(\sn)\Dtm\rb]\rb\},\half\sgn(\sn)\e{}{\Kn}{v},\nn\rb)}\nonumber\\
&&\hspace{-0.5cm}
\Lambda^{\half}\lb(\lb\{\xmgenn+\frac{1}{T}\lb(\pmgen\rb)-\frac{T}{4}\lb[\sgn(\snt)\Dtmt-\sgn(\sn)\Dtm\rb]\rb\},\half\sgn(\snt)\e{}{\Knt}{v},\nnt\rb)\nonumber\\
&=&\lb(\frac{a^{\frac{3}{2}}|\det(\pmm)|^{\frac{1}{4}}}{\hbar}\rb)^2\lb(sT\rb)^2\nonumber\\
&&\Big(C^{\Kn\sn\nn}+s\xmm_{Mm}C^{Mm,\Kn\sn\nn}+s^2\xmm_{Mm}\xmm_{Nn}\nonumber\\
\nonumber\\
&&\hspace{1cm}
\lb[\frac{1}{2}C^{Mm,\Kn\sn\nn}C^{Nn}+\frac{1}{3!}\epsilon_{ijk}\lb[\epsilon_{\nn mn}C^{\Kn \sn i}C^{Mj}C^{Nk}
+\epsilon_{\ell\nn n}C^{\Kn \sn j}C^{Li}C^{Nk}
+\epsilon_{\ell m\nn}C^{\Kn \sn k}C^{Li}C^{Mj}\rb]\Big)\rb]\nonumber\\
&&\Big(C^{\Knt\snt\nnt}+s\xmm_{Mm}C^{Mm,\Knt\snt\nnt}+s^2\xmm_{Mm}\xmm_{Nn}\nonumber\\
&&\hspace{1cm}
\lb[\frac{1}{2}C^{Mm,\Knt\snt\nnt}C^{Nn}+\frac{1}{3!}\epsilon_{ijk}\lb[\epsilon_{\nnt mn}C^{\Knt \snt i}C^{Mj}C^{Nk}
+\epsilon_{\ell\nnt n}C^{\Knt \snt j}C^{Li}C^{Nk}
+\epsilon_{\ell m\nnt}C^{\Knt \snt k}C^{Li}C^{Mj}\rb]\Big)\rb]\nonumber\\
&&\lb[\lb(f^{(1)}_{\frac{1}{8}}(1)\rb)^2+ s\xmm_{Mm}C^{Mm}\lb(8f^{(1)}_{\frac{1}{8}}(1)f^{(2)}_{\frac{1}{8}}(1)\rb)\rb.
\nonumber\\
&&\lb.
+s^2\xmm_{Mm}\xmm_{Nn}\lb[C^{Mm,Nn}\lb(4f^{(1)}_{\frac{1}{8}}(1)f^{(2)}_{\frac{1}{8}}(1)\rb)
+C^{Mm}C^{Nn}\lb(40f^{(1)}_{\frac{1}{8}}(1)f^{(3)}_{\frac{1}{8}}(1)+32\lb(f^{(2)}_{\frac{1}{8}}(1)\rb)^2\rb)\rb]\rb]+O(s^2(sT/t)^2)\nonumber\\
\ea
The expansion has the following structure
\ba
\lefteqn{\lb(\frac{sT}{t}\rb)^2\lb(\alpha_0+\alpha_1 s\xmm +\alpha_2 s^2(\xmm)^2\rb)\lb(\beta_0+\beta_1 s\xmm +\beta_2 s^2(\xmm)^2\rb)\lb(\gamma_0+\gamma_1 s\xmm +\gamma_2 s^2(\xmm)^2\rb)}\nonumber\\
&=&\alpha_0\beta_0\gamma_0 + s^2(\xmm)^2\lb[\alpha_2\lb(\beta_0+\gamma_0\rb)+\beta_2\lb(\gamma_2+\alpha_2\rb)+\gamma_2\lb(\alpha_0+\beta_0\rb)+\alpha_1\beta_1\gamma_0+\alpha_1\beta_0\gamma_1+\alpha_0\beta_1\gamma_1\rb]\nonumber\\
&&+\mathrm{lin}(\xmm)+O(s^2(sT/t)^2)
\ea
whereby $\mathrm{lin}(\xmm)$ denotes all terms linear in $\xmm$ which we do not show in detail as they will not contribute to the final result, because they vanish when integrated against the even function $\exp(-2(\xmm)^2)$. Precisely, when considering what the $\alpha,\beta$ and $\gamma$ coefficients actually are, we obtain
\ba
\lefteqn{\Lambda^{\half}\lb(\lb\{\xmgenn+\frac{1}{T}\lb(\pmgen\rb)+\frac{T}{4}\lb[\sgn(\snt)\Dtmt-\sgn(\sn)\Dtm\rb]\rb\},\half\sgn(\sn)\e{}{\Kn}{v},\nn\rb)}\nonumber\\
&&\hspace{-0.5cm}
\Lambda^{\half}\lb(\lb\{\xmgenn+\frac{1}{T}\lb(\pmgen\rb)-\frac{T}{4}\lb[\sgn(\snt)\Dtmt-\sgn(\sn)\Dtm\rb]\rb\},\half\sgn(\snt)\e{}{\Knt}{v},\nnt\rb)\nonumber\\
&=&\lb(\frac{a^{\frac{3}{2}}|\det(\pmm)|^{\frac{1}{4}}}{\hbar}\rb)^2\lb(sT\rb)^2\nonumber\\
&&\lb[C^{\Kn\sn\nn}C^{\Knt\snt\nnt}\lb(f^{(1)}_{\frac{1}{8}}(1)\rb)^2\rb.\nonumber\\
&&+s^2\xmm_{Mm}\xmm_{Nn}\Big\{\nonumber\\
&&+\lb(C^{\Knt\snt\nnt}+\lb(f^{(1)}_{\frac{1}{8}}(1)\rb)^2\rb)\nonumber\\
&&\hspace{0.3cm}
\lb[\frac{1}{2}C^{Mm,\Kn\sn\nn}C^{Nn}+\frac{1}{3!}\epsilon_{ijk}\lb[\epsilon_{\nn mn}C^{\Kn \sn i}C^{Mj}C^{Nk}
+\epsilon_{\ell\nn n}C^{\Kn \sn j}C^{Li}C^{Nk}
+\epsilon_{\ell m\nn}C^{\Kn \sn k}C^{Li}C^{Mj}\rb]\Big)\rb]\nonumber\\
&&+\lb(C^{\Kn\sn\nn}+\lb(f^{(1)}_{\frac{1}{8}}(1)\rb)^2\rb)\nonumber\\
&&\hspace{0.3cm}
\lb[\frac{1}{2}C^{Mm,\Knt\snt\nnt}C^{Nn}+\frac{1}{3!}\epsilon_{ijk}\lb[\epsilon_{\nn mn}C^{\Knt \snt i}C^{Mj}C^{Nk}
+\epsilon_{\ell\nnt n}C^{\Knt \snt j}C^{Li}C^{Nk}
+\epsilon_{\ell m\nnt}C^{\Knt \snt k}C^{Li}C^{Mj}\rb]\Big)\rb]\nonumber\\
&&+\lb(C^{\Kn\sn\nn}+C^{\Knt\snt\nnt}\rb)\nonumber\\
&&\hspace{0.3cm}
\lb[C^{Mm,Nn}\lb(4f^{(1)}_{\frac{1}{8}}(1)f^{(2)}_{\frac{1}{8}}(1)\rb)
+C^{Mm}C^{Nn}\lb(40f^{(1)}_{\frac{1}{8}}(1)f^{(3)}_{\frac{1}{8}}(1)+32\lb(f^{(2)}_{\frac{1}{8}}(1)\rb)^2\rb)\rb]\nonumber\\
&&
+C^{\Kn\sn\nn}C^{Mm,\Knt\snt\nnt}C^{Nn}\lb(8f^{(1)}_{\frac{1}{8}}(1)f^{(2)}_{\frac{1}{8}}(1)\rb)
+C^{\Knt\snt\nnt}C^{Mm,\Kn\sn\nn}C^{Nn}\lb(8f^{(1)}_{\frac{1}{8}}(1)f^{(2)}_{\frac{1}{8}}(1)\rb)\nonumber\\
&&\lb.
+C^{Mm,\Kn\sn\nn}C^{Mm,\Knt\snt\nnt}\lb(f^{(1)}_{\frac{1}{8}}(1)\rb)^2\Big\}\rb] +O(s^2(sT/t)^2)
\ea
When integrating the $\Lambda^{\half}$-functions multiplied with the Gaussian $\exp(-2(\xmm_{Mm})^2)$, the integral is $\sqrt{\pi/2}^9$ and $(9/4)\sqrt{\pi/2}^9$ for the zeroth power and the second power in $\xmm_{Mm}$ respectively. Note that we have a factor $e^{-\frac{t}{4}\sum\limits_{\vt\in V}\sum\limits_{(J,\sigma,j)\atop\in L}\lb(\Del\rb)^2}$ in the expression for the expectation value. Therefore we have to expand this function in powers of $t$. The linear term in $t$ leads to term having a minimal order of $(sT)^2/t$. This order is already smaller than terms of the order $s^2(sT/t)^2$, because
$
 \lb[s^2(sT/t)^2\rb]\lb[t/(sT)^2\rb]=s^2/t=1/t^{2\alpha}\gg 1
$.
Fortunately, we can neglect the linear term in $t$ in the expansion of the exp-function. Consequently, the final expectation value  of $(\Omnt)^{\dagger}\Omn$ is given by
\ba
\label{OmnNLO}
\lefteqn{\MeO}\nonumber\\
&=&e^{+i\sum\limits_{(J,\sigma,j)}\phigen\Del}
\lb(\frac{a^{\frac{3}{2}}|\det(\pmm)|^{\frac{1}{4}}}{\hbar}\rb)^2\lb(sT\rb)^2\nonumber\\
&&\lb[C^{\Kn\sn\nn}C^{\Knt\snt\nnt}\lb(f^{(1)}_{\frac{1}{8}}(1)\rb)^2\rb.\nonumber\\
&&+\frac{9}{4}s^2\Big\{
\lb(C^{\Knt\snt\nnt}+\lb(f^{(1)}_{\frac{1}{8}}(1)\rb)^2\rb)\nonumber\\
&&\hspace{0.3cm}
\lb[\frac{1}{2}C^{Mm,\Kn\sn\nn}C^{Nn}+\frac{1}{3!}\epsilon_{ijk}\lb[\epsilon_{\nn mn}C^{\Kn \sn i}C^{Mj}C^{Nk}
+\epsilon_{\ell\nn n}C^{\Kn \sn j}C^{Li}C^{Nk}
+\epsilon_{\ell m\nn}C^{\Kn \sn k}C^{Li}C^{Mj}\rb]\Big)\rb]\nonumber\\
&&+\lb(C^{\Kn\sn\nn}+\lb(f^{(1)}_{\frac{1}{8}}(1)\rb)^2\rb)\nonumber\\
&&\hspace{0.3cm}
\lb[\frac{1}{2}C^{Mm,\Knt\snt\nnt}C^{Nn}+\frac{1}{3!}\epsilon_{ijk}\lb[\epsilon_{\nn mn}C^{\Knt \snt i}C^{Mj}C^{Nk}
+\epsilon_{\ell\nnt n}C^{\Knt \snt j}C^{Li}C^{Nk}
+\epsilon_{\ell m\nnt}C^{\Knt \snt k}C^{Li}C^{Mj}\rb]\Big)\rb]\nonumber\\
&&+\lb(C^{\Kn\sn\nn}+C^{\Knt\snt\nnt}\rb)\nonumber\\
&&\hspace{0.3cm}
\lb[C^{Mm,Nn}\lb(4f^{(1)}_{\frac{1}{8}}(1)f^{(2)}_{\frac{1}{8}}(1)\rb)
+C^{Mm}C^{Nn}\lb(40f^{(1)}_{\frac{1}{8}}(1)f^{(3)}_{\frac{1}{8}}(1)+32\lb(f^{(2)}_{\frac{1}{8}}(1)\rb)^2\rb)\rb]\nonumber\\
&&\lb.\hspace{-0.5cm}
+\lb(C^{\Kn\sn\nn}C^{Mm,\Knt\snt\nnt}+C^{\Knt\snt\nnt}C^{Mm,\Kn\sn\nn}\rb)C^{Nn}\lb(8f^{(1)}_{\frac{1}{8}}(1)f^{(2)}_{\frac{1}{8}}(1)\rb)
+C^{Mm,\Kn\sn\nn}C^{Mm,\Knt\snt\nnt}\lb(f^{(1)}_{\frac{1}{8}}(1)\rb)^2\Big\}\rb]\nonumber\\
&& +O(s^2(sT/t)^2)
\ea
Recalling the definition of the Master constraint, we have
\ba
\op{C}_{0,\bx}&=&\sum\limits_{\In\Jn\Kn}\sum\limits_{\sn=+,-}\frac{4}{\kappa}\epsilon^{\In\Jn\Kn}\Omn\delta_{\mn,\nn}\nonumber\\
 \op{C}_{\elln,\bx}&=&\sum\limits_{\In\Jn\Kn}\sum\limits_{\sn=+,-}\frac{4}{\kappa}\epsilon^{\In\Jn\Kn}\epsilon_{\elln \mn\nn}\Omn
\ea
Using the fact that the expectation value of $\MCO$ can be expressed in terms of the expectation value of $(\Omnt)^{\dagger}\Omn$ and that we have shown that the leading order agrees with the classical Master constraint, we have our final result given by
\ba
\lefteqn{\frac{\langle\Psigen\,|\,\MCO\,|\,\Psigen\rangle}{||\Psigen||^2}}\nonumber\\
&=&\MC
+\frac{9}{4}s^2\sum\limits_{v\in V(\alpha)}\lb[\sum\limits_{\In\Jn\Kn}\sum\limits_{\Int\Jnt\Knt}\sum\limits_{\sn=+,-}\sum\limits_{\snt=+,-}\rb.\nonumber\\
&&
\Big\{
\epsilon^{\In\Jn\Kn}\epsilon^{\Int\Jnt\Knt}\lb(\delta_{\mn,\nn}\delta_{\mnt,\nnt}
+\sum\limits_{\elln=1}^3 \epsilon_{\elln \mn\nn}\epsilon_{\elln \mnt\nnt}\rb)\nonumber\\
&&\lb(\frac{4a^{\frac{3}{2}}|\det(\pmm)|^{\frac{1}{4}}}{\kappa\hbar}\rb)^2\lb(sT\rb)^2e^{+i\sum\limits_{(J,\sigma,j)}\phigen\Del}\nonumber\\
&&\Big\{
\lb(C^{\Knt\snt\nnt}+\lb(f^{(1)}_{\frac{1}{8}}(1)\rb)^2\rb)\nonumber\\
&&\hspace{0.3cm}
\lb[\frac{1}{2}C^{Mm,\Kn\sn\nn}C^{Nn}+\frac{1}{3!}\epsilon_{ijk}\lb[\epsilon_{\nn mn}C^{\Kn \sn i}C^{Mj}C^{Nk}
+\epsilon_{\ell\nn n}C^{\Kn \sn j}C^{Li}C^{Nk}
+\epsilon_{\ell m\nn}C^{\Kn \sn k}C^{Li}C^{Mj}\rb]\Big)\rb]\nonumber\\
&&+\lb(C^{\Kn\sn\nn}+\lb(f^{(1)}_{\frac{1}{8}}(1)\rb)^2\rb)\nonumber\\
&&\hspace{0.3cm}
\lb[\frac{1}{2}C^{Mm,\Knt\snt\nnt}C^{Nn}+\frac{1}{3!}\epsilon_{ijk}\lb[\epsilon_{\nn mn}C^{\Knt \snt i}C^{Mj}C^{Nk}
+\epsilon_{\ell\nnt n}C^{\Knt \snt j}C^{Li}C^{Nk}
+\epsilon_{\ell m\nnt}C^{\Knt \snt k}C^{Li}C^{Mj}\rb]\Big)\rb]\nonumber\\
&&+\lb(C^{\Kn\sn\nn}+C^{\Knt\snt\nnt}\rb)\nonumber\\
&&\hspace{0.3cm}
\lb[C^{Mm,Nn}\lb(4f^{(1)}_{\frac{1}{8}}(1)f^{(2)}_{\frac{1}{8}}(1)\rb)
+C^{Mm}C^{Nn}\lb(40f^{(1)}_{\frac{1}{8}}(1)f^{(3)}_{\frac{1}{8}}(1)+32\lb(f^{(2)}_{\frac{1}{8}}(1)\rb)^2\rb)\rb]\nonumber\\
&&\lb.\hspace{-0.5cm}
+\lb(C^{\Kn\sn\nn}C^{Mm,\Knt\snt\nnt}+C^{\Knt\snt\nnt}C^{Mm,\Kn\sn\nn}\rb)C^{Nn}\lb(8f^{(1)}_{\frac{1}{8}}(1)f^{(2)}_{\frac{1}{8}}(1)\rb)
+C^{Mm,\Kn\sn\nn}C^{Mm,\Knt\snt\nnt}\lb(f^{(1)}_{\frac{1}{8}}(1)\rb)^2\Big\}\rb]\nonumber\\
&&+O(s^2(sT/t)^2)
\ea
\end{appendix}


\begin{thebibliography}{99}
\parskip -5pt
\bibitem{I}
K. Giesel, T. Thiemann, ``Algebraic Quantum Gravity (AQG).
I. Conceptual Setup'' [gr-qc/0607099]
\bibitem {III}
K. Giesel, T. Thiemann, ``Algebraic Quantum Gravity (AQG).
III. Semiclassical Perturbation Theory'' [gr-qc/0607101]
\bibitem{books} 
C. Rovelli. {\it Quantum Gravity}, (Cambridge University Press, Cambridge,
2004).\\
T. Thiemann. {\it Modern Canonical Quantum General Relativity},
(Cambridge University Press, Cambridge, submitted). Draft available as
[gr-qc/0110034]
\bibitem{reviews} 
C. Rovelli. Loop quantum gravity,
{\it Living  Rev. Rel.} {\bf 1} (1998), 1. [gr-qc/9710008]\\
T. Thiemann. Lectures on loop quantum gravity.
{\it Lect. Notes Phys.} {\bf 631} (2003), 41-135. [gr-qc/0210094]\\
A. Ashtekar and J. Lewandowski. Background independent quantum gravity:
a status report. {\it Class. Quant. Grav.} {\bf 21} (2004), R53.
[gr-qc/0404018]\\
L. Smolin. An invitation to loop quantum gravity. [hep-th/0408048]
\bibitem{QSD} QSD papers
T. Thiemann. Anomaly-free formulation of non-perturbative,
four-dimensional Lorentzian quantum gravity. {\it Physics Letters} {\bf
B380} (1996), 257-264. [gr-qc/9606088]\\
T. Thiemann. Quantum Spin Dynamics (QSD).
{\it Class. Quantum Grav.} {\bf 15} (1998), 839-73. [gr-qc/9606089]\\
T. Thiemann. Quantum Spin Dynamics (QSD): II.
The kernel of the Wheeler-DeWitt constraint operator.
{\it Class. Quantum Grav.} {\bf 15} (1998), 875-905. [gr-qc/9606090]\\
T. Thiemann. Quantum Spin Dynamics (QSD): III.
Quantum constraint algebra and physical scalar product in quantum general
relativity. {\it Class. Quantum Grav.} {\bf 15} (1998), 1207-1247.
[gr-qc/9705017]\\
T. Thiemann. Quantum Spin Dynamics (QSD): IV.
2+1 Euclidean quantum gravity as a model to test 3+1
Lorentzian quantum gravity. {\it Class. Quantum Grav.} {\bf 15} (1998),
1249-1280. [gr-qc/9705018]\\
T. Thiemann. Quantum Spin Dynamics (QSD): V.
Quantum gravity as the natural regulator of the Hamiltonian constraint
of matter quantum field theories.
{\it Class. Quantum Grav.} {\bf 15} (1998), 1281-1314. [gr-qc/9705019]\\
T. Thiemann. Quantum Spin Dynamics (QSD): VI.
Quantum Poincar\'e algebra and a quantum positivity of energy
theorem for canonical quantum gravity.
{\it Class. Quantum Grav.} {\bf 15} (1998), 1463-1485. [gr-qc/9705020]\\
T. Thiemann. Kinematical Hilbert spaces for fermionic and
Higgs quantum field theories.
{\it Class. Quantum Grav.} {\bf 15} (1998), 1487-1512. [gr-qc/9705021]
%
\bibitem{M1} 
T. Thiemann. The Phoenix project: master constraint
programme for loop quantum gravity.
{\it Class. Quant. Grav.} {\bf 23} (2006), 2211-2248.
[gr-qc/0305080]
%
\bibitem{M2} T. Thiemann. Quantum spin dynamics. VIII. The master 
constraint.
{\it Class. Quant.Grav.} {\bf 23} (2006), 2249-2266.
[gr-qc/0510011]
%
\bibitem{Test}
B. Dittrich and T. Thiemann. Testing the master
constraint
programme for loop quantum gravity: I. General framework.
{\it Class. Quant. Grav.} {\bf 23} (2006), 1025-1066.
[gr-qc/0411138]\\
B. Dittrich and T. Thiemann. Testing the master
constraint
programme for loop quantum gravity: II. Finite -- dimensional systems.
{\it Class. Quant. Grav.} {\bf 23} (2006), 1067-1088.
[gr-qc/0411139]\\
B. Dittrich and T. Thiemann. Testing the master
constraint programme for loop quantum gravity: III. SL(2R) models.
{\it Class. Quant. Grav.} {\bf 23} (2006), 1089-1120.
[gr-qc/0411140]\\
B. Dittrich and T. Thiemann. Testing the master
constraint programme for loop quantum gravity:
IV. Free field theories.
{\it Class. Quant. Grav.} {\bf 23} (2006), 1121-1142.
[gr-qc/0411141]\\
B. Dittrich and T. Thiemann. Testing the master
constraint programme for loop quantum gravity:
V. Interacting field theories.
{\it Class. Quant. Grav.} {\bf 23} (2006), 1143-1162.
[gr-qc/0411142]
\bibitem{ALMMT} 
A. Ashtekar, J. Lewandowski, D. Marolf, J. Mour\~ao and T.
Thiemann. Quantization of diffeomorphism invariant theories
of connections with local degrees of freedom. {\it Journ. Math. Phys.}
{\bf 36} (1995), 6456-6493. [gr-qc/9504018]
%
\bibitem{U1hoch3}
T. Thiemann. Quantum Spin Dynamics (QSD): VII.
Symplectic structures and continuum lattice formulations of
gauge field theories. {\it Class. Quant. Grav.} {\bf 18}
(2001) 3293-3338. [hep-th/0005232]\\
T. Thiemann. Gauge Field Theory Coherent States (GCS): I. General
properties. {\it Class. Quant. Grav.} {\bf 18} (2001), 2025-2064.
[hep-th/0005233]\\
T. Thiemann and O. Winkler. Gauge Field Theory Coherent States
(GCS): II. Peakedness properties. {\it Class. Quant.
Grav.} {\bf 18} (2001) 2561-2636. [hep-th/0005237]\\
T. Thiemann and O. Winkler. Gauge Field Theory Coherent States
(GCS): III. Ehrenfest theorems.
{\it Class. Quantum Grav.} {\bf 18} (2001), 4629-4681. [hep-th/0005234]
\bibitem{Complexifier} 
T. Thiemann. Complexifier coherent states for canonical
quantum general relativity.
{\it Class. Quant. Grav.} {\bf 23} (2006), 2063-2118.
[gr-qc/0206037]
\bibitem{STW} H. Sahlmann, T. Thiemann and O. Winkler.
Coherent states for canonical quantum general relativity and the infinite
tensor product extension. {\it Nucl. Phys.} {\bf B606}
(2001) 401-440. [gr-qc/0102038]
\bibitem{Hanno}
H. Sahlmann and T. Thiemann. Towards the QFT on
curved spacetime limit of QGR. 1. A general scheme. 
{\it Class. Quant. Grav.} {\bf 23} (2006), 867-908.
[gr-qc/0207030]\\
H. Sahlmann and T. Thiemann. Towards the QFT on
curved spacetime limit of QGR.
2. A concrete implementation. 
{\it Class. Quant. Grav.} {\bf 23} (2006), 909-954.
[gr-qc/0207031]
\bibitem{ITP}
T. Thiemann and O. Winkler. Gauge field theory coherent
states (GCS): IV. Infinite tensor product and thermodynamic limit.
{\it Class. Quantum Grav.} {\bf 18} (2001), 4997-5033. [hep-th/0005235]
\bibitem{VAL} A. Ashtekar, J. Lewandowski, ``Quantum Theory of Geometry II :
Volume Operators", Adv. Theo. Math. Phys. {\bf 1} (1997) 388-429 
\bibitem{VRS} 
C. Rovelli and L. Smolin.
Discreteness of volume and area in quantum gravity.
{\it Nucl. Phys.} {\bf B442} (1995), 593. Erratum: {\it Nucl. Phys.} {\bf
B456} (1995), 734.\\
\bibitem{GT} 
K. Giesel and T. Thiemann. Consistency check on volume
and
triad operator quantisation in loop quantum gravity. I. [gr-qc/0507036]\\
K. Giesel and T. Thiemann. Consistency check on volume
and triad operator quantisation in loop quantum gravity. II.
[gr-qc/0507037]
\end{thebibliography}
\end{document}